\documentclass[11pt,a4paper]{article}
\pdfoutput=1
\usepackage{jcappub}
\usepackage{aas_macros}

\usepackage{float}
\usepackage{ctable}
\newcommand{\vx}{{\bf x}}
\newcommand{\fnl}{f_{\rm NL}}
\newcommand{\gnl}{g_{\rm NL}}
\newcommand{\vk}{{\bf k}}
\newcommand{\vkp}{{\bf k'}}

\def \MSUN{{\rm M}_{\odot}}

\title{N-body simulations with generic non-Gaussian initial conditions II: halo bias}
\author[a]{Christian Wagner}
\author[a,b]{and Licia Verde}
\affiliation[a]{ICCUB-IEEC, University of Barcelona, Barcelona 08028, Spain.}
\affiliation[b]{ICREA, Instituci\'o Catalana de Recerca i Estudis Avan\c{c}ats.}
\emailAdd{cwagner@icc.ub.edu}
\emailAdd{liciaverde@icc.ub.edu}
\abstract{
We present N-body simulations for generic non-Gaussian initial conditions with the aim of exploring and modelling the scale-dependent  halo bias.   This effect is evident on very large scales requiring large simulation boxes.  In addition,  the previously available prescription to implement generic non-Gaussian initial conditions has been improved to keep under control higher-order terms which were spoiling the  power spectrum on large scales.
We pay particular attention to the differences between physical, inflation-motivated primordial bispectra and their factorizable templates, and to the operational definition of the non-Gaussian halo bias  (which has both a scale-dependent and an approximately scale-independent contributions).
We find that analytic predictions for both the non-Gaussian halo mass function and halo bias work well once a fudge factor (which was introduced before but still lacks convincing physical explanation) is calibrated on simulations.
The halo bias remains  therefore an extremely promising tool to probe primordial non-Gaussianity and thus to give insights into the physical mechanism that generated the primordial perturbations. 
The simulation outputs and tables of the analytic predictions will be made publicly available via the non-Gaussian comparison project web site \url{http://icc.ub.edu/~liciaverde/NGSCP.html}. 
}
\keywords{cosmological simulations, large scale structure, initial conditions}
\begin{document}

\maketitle

%%%%%%%%%%%%%%%%%%%%%%%%%%%%%%%%%%%
\section{Introduction}
\label{introduction}
Inflation, the standard theory for the origin of the primordial density fluctuations, predicts that the primordial density field can be described, to a good approximation, as a Gaussian random field. This nearly Gaussian nature of the primordial fluctuations is well established observationally (for current constraints see, e.g., \cite{komatsu2011}). However, small deviations from Gaussianity are predicted by inflation. The level and form of the non-Gaussianity depend on the inflationary model. The simplest inflationary model, the standard single-field slow-roll inflation, produces non-Gaussianities which are unmeasurably small \cite{maldacena2003,acquaviva2003} (for a thorough review on inflationary non-Gaussianity see \cite{bartolo_review,chen_review}).

Other inflationary models \cite{linde1997,lyth2003,babich2004,chen2007a,chen2007b,holman2008,langlois2008}, which violate one or more assumptions of the standard single-field slow-roll inflation, may potentially cause much larger non-Gaussianities. Hence, measurements of deviations from Gaussianity offer a unique window to distinguish between different inflationary models or to rule out classes of models.

Departures from Gaussianity are described at leading order by the bispectrum of the primordial fluctuations in the gravitational potential, $B_\Phi(k_1,k_2,k_3)$. The bispectrum is the Fourier transform of the 3-point-correlation function and is identical to zero for a Gaussian field. The functional form of the bispectrum specifies the form of the non-Gaussianity, while its amplitude, parametrized by the parameter $\fnl$, gives the level of non-Gaussianity.

Since the Universe is statistically homogeneous and isotropic, the wave vectors $\vk_1$, $\vk_2$, and $\vk_3$ have to form a triangle. The shape of the triangle, for which the bispectrum peaks, is conventionally used to characterize the form of the non-Gaussianity. The local type of non-Gaussianity, $\Phi(x)=\Phi^G(x)+\fnl\left(\Phi^G(x)^2-\langle \Phi^G(x)^2\rangle\right)$, where $\Phi^G(x)$ is gravitational potential in the Gaussian case, yields a bispectrum that peaks for squeezed triangles (e.g., $k_1\ll k_2 \simeq k_3$). Potentially large non-Gaussianities of the local type are generally produced by inflationary models with multiple fields, where non-Gaussianity is created by non-linearities which develop outside the horizon \cite{babich2004}.

Other inflationary models which involve higher-derivative operators (see e.g.~\cite{creminelli2003,alishahiha2004,arkani-hamed2004}) give rise to a bispectrum that has its maximum for equilateral triangles ($k_1=k_2=k_3$). 

Models in which the initial state of the inflaton is different from the Bunch-Davies vacuum \cite{chen2007a,holman2008,meerburg2009,ashoorioon2010}, can produce non-Gaussianity of the enfolded shape, i.e.~the bispectrum is dominated by flattened or enfolded triangle configurations ($k_1\simeq k_2+k_3$). 

Finally, there is the so-called orthogonal shape of non-Gaussianity, which cannot be easily described in terms of a triangle configuration \cite{ssz2010}. This shape is ---with respect to a specific scalar product (see \cite{babich2004} for an explicit definition)--- orthogonal to the equilateral and local shape.

The cosmic microwave background (CMB) offers the cleanest probe of non-Gaussianity, since the temperature perturbations are still very small and well described by linear theory.
If a shape of the primordial bispectrum is assumed\footnote{For an approach without assuming a shape for the bispectrum, see \cite{fergusson2010_cmb}.}, constraints on the corresponding $\fnl$ value can be derived from the observed fluctuations in the CMB. The current constraints for the most commonly used shapes are $-10<\fnl^{\rm local}<74$, $-214<\fnl^{\rm equil}<266$, $-410<\fnl^{\rm orth}<6$ (95\% CL) \cite{komatsu2011}, and $\fnl^{\rm enfl}=114 \pm 72$ (68\% CL) \cite{jorge2011} . The Planck mission \cite{planck} is expected to measure $\fnl$ with an accuracy of $\Delta \fnl \simeq 5$.

Primordial non-Gaussianity leaves also observable signatures in the large-scale structure of the Universe (for recent reviews, see \cite{verde2010,desjacques_review}). The bispectrum of the density field measured through observable tracers like galaxies, however, is dominated by non-linear gravitational evolution and biasing, this makes the measurement of the primordial contribution difficult \cite{verde2000,verde2001a,verde2001b} (see, however, \cite{baldauf2010}).

Another probe of non-Gaussianity from large-scale structure observations, is the abundance of galaxy clusters or large voids in the Universe. Both of these objects originate from the tails of the nearly Gaussian initial density distribution and are therefore sensitive to primordial non-Gaussainity \cite{MVJ,LV}. The signal depends on the skewness of the initial density distribution, which is given by an integral over the bispectrum, and is therefore less sensitive to the shape of the non-Gaussianity. 
This probe measures the primordial non-Gaussianity on scales of the order of $10\, \rm Mpc$, which is smaller than the scales accessible by CMB measurements. Intriguingly, several groups \cite{cayon2010, hoyle2010, enqvist2010} derived estimates of $\fnl$ from the observations of very massive galaxy clusters, which are as large as $\fnl\sim400$ (see, however, \cite{mortonson2011}). 

The clustering of halos on large scales offers another possibility to probe non-Gaussianity \cite{dalal2008,MV}. Non-Gaussianity couples small and large scales. The density peaks on small scales, which are the seeds for halo formation, are modulated by the large scale modes of the gravitational potential. This induces a scale-dependent halo bias on large scales, which can be very different from the scale-independent halo bias in the Gaussian case. 
The amplitude of this effect scales approximately linearly in $\fnl$, while the scale dependence is set by the shape of non-Gaussianity. For example, the local type non-Gaussianity generates a halo bias which scales as $k^{-2}$, whereas the halo bias of equilateral types of non-Gaussianity is expected to have only a very mild scale dependence \cite{VM}.
This technique was already used on observational data of galaxy surveys and quasar catalogues \cite{slosar2008,xia2010a,xia2010b}. The derived  constraints, $-27<\fnl^{\rm local}<70$ (95\% CL) \cite{slosar2008}, are competitive with the CMB measurements. Fisher matrix forecasts for future large-scale structure surveys predict that measurements of the non-Gaussian bias will constrain $\fnl^{\rm local}$ with an uncertainty of $\Delta \fnl\sim 1$ \cite{carbone2008,carbone2010,cunha2010,fedeli2010}.

All of the three aforementioned large-scale structure probes of non-Gaussianity are highly affected by non-linear gravitational evolution. The standard tool for taking non-linearities into account are N-body simulations. However, until recently, only the local type of non-Gaussianity was simulated with N-body simulations and reasonable agreement with analytic predictions was found \cite{dalal2008,desjacques2009,pillepich2010,grossi2009,nishimichi2010}. 

In \cite{wagner2010}, we presented a prescription for setting up initial conditions for N-body simulations with non-local non-Gaussianities. Using this method we were able to simulate the non-linear power spectrum and the cluster mass function for non-local	 shapes, and again found consistency with theoretical expectations. The technique of \cite{wagner2010} was, however, not suitable to study the effect of non-Gaussianity on the halo bias, since  ---depending on the shape of the bispectrum--- it introduced artificial power on large scales, which interferes with the expected signal of the non-Gaussian halo bias.

In this paper, we modify our ansatz used for the generation of non-Gaussian initial conditions in a way that these spurious contributions to the initial power spectrum are always kept under control. 
This enables us to set up and run non-Gaussian simulations with large volumes, which we then use to study ---for the first time by means of N-body simulations--- the non-Gaussian halo bias for different types of non-Gaussianity. As the simulated volume is larger than the one we were able to simulate in \cite{wagner2010} and hence the statistical error on the number of massive halos is smaller, we use these simulations also to revisit the non-Gaussian cluster mass function.

The outline of the paper is the following. First, we review the non-Gaussian halo bias and its theoretical modelling in Sec.~\ref{bias}. In Sec.~\ref{templates}, we compare, in the context of the non-Gaussian halo bias, the bispectra of inflationary models with the commonly used separable templates which approximate these bispectra. In Sec.~\ref{IC}, we review the method of setting up non-Gaussian initial conditions for a given bispectrum and then present the modification of our ansatz.
The practical implementation and numerical settings of our simulation suite are given in Sec.~\ref{sims}. We present and discuss our results on the non-Gaussian mass function and halo bias in Sec.~\ref{results}. Finally, we conclude in Sec.~\ref{conclusions}.

%%%%%%%%%%%%%%%%%%%%%%%%%%%%%%%%%%%
\section{Non-Gaussian halo bias}
\label{bias}
The halo bias describes the relation between the halo density contrast, $\delta_{\rm h}$, and the underlying dark matter density contrast, $\delta_{\rm m}$. In Fourier space we can write the relation as $\delta_{\rm h}(k)=b(k)\delta_{\rm m}(k) + n(k)$, where the random variable $n(k)$ models the stochasticity coming from shot noise and possibly from physical halo formation processes. 
For Gaussian initial conditions the bias is scale-independent on large scales, $b_{M,z}(k)=b_{1;M,z}^{\rm G}$ for $k\lesssim0.1$ \cite{kaiser1984,MW1996,ST2001}. Analytic predictions and results of N-body simulations show that the linear\footnote{The linear bias, $b_1$, is called linear as it corresponds to the linear coefficient in the Taylor expansion of the halo density field: $\delta_{\rm h}(x)=b_0+b_1 \delta_{\rm m}(x)+b_2\delta_{\rm m}^2(x)/2+ \ldots\ $, where both density fields are smoothed on a sufficiently large scale.} Gaussian bias, $b_{1;M,z}^{\rm G}$, increases with redshift $z$ and halo mass $M$. In addition the halo bias may depend on environment and mass accretion history \cite{gao2005,gao2007}.

Primordial non-Gaussianity of the local type induces a scale dependence of the halo bias on large scales, which scales as $\sim k^{-2}$. This particular scaling was first found in N-body simulations by \cite{dalal2008}, who also gave a simple analytic understanding how this effect arises. Later \cite{MV,slosar2008} rederived the analytic prediction in a more rigorous way using two different approaches, the high peak limit and the peak-background split, respectively. In Ref.~\cite{afshordi2008}, the same scale dependence of the non-Gaussian halo bias was found by computing the ellipsoidal collapse threshold in the presence of small non-Gaussianities.  The analytic predictions were tested with further and larger N-body simulations \cite{pillepich2010,desjacques2009,grossi2009,nishimichi2010}. Reasonable agreement between theory and simulations was found if small corrections, which we discuss in detail below, are taken into account. In a number of subsequent papers further theoretical modelling was developed by applying (univariate and multivariate) local biasing in the framework of perturbation theory \cite{mcdonald2008,taruya2008,jeong2009,sefusatti2009,giannantonio2010,baldauf2010}. 
In summary the non-Gaussian bias can be modelled to leading order as
\begin{equation}\label{eq:NG_bias}
\Delta b \equiv b^{\rm NG}_{M,z}(k,\fnl)-b^{\rm G}_{M,z}(k)=\Delta b_{1;M,z}(\fnl)+\fnl\left[b_{1;M,z}^{\rm G} + \Delta b_{1;M,z}(\fnl) -1\right]\frac{q \delta_c}{D(z)}\frac{\mathcal{F}_M(k)}{\mathcal{M}_M(k)} \,,
\end{equation}
where $\Delta b_{1;M,z}(\fnl)$ is the difference in the linear bias, which is non-zero, since a non-vanishing $\fnl$ changes the halo number density and hence the linear halo bias for a fixed halo mass $M$. Mainly, this effect leads to a scale-independent shift relative to the Gaussian ($\fnl=0$) linear bias \cite{desjacques2009,giannantonio2010,desimone2011}. Additionally, it makes the scale-dependent second term in above equation slightly non-linear in $\fnl$ \cite{giannantonio2010,smith2011}. In practice, when fitting observational data, the linear bias is estimated from the data itself.

The amplitude of the non-Gaussian contribution is given by $\fnl$, the linear Lagrangian bias $[b_{1;M,z}(\fnl)-1]$, and the redshift-dependent collapse threshold $q\delta_c/D(z)$, where $\delta_z=1.686$ is the spherical collapse threshold in an Einstein-de Sitter universe, $D(z)$ is the growth function normalized to $(1+z)^{-1}$ at high redshift, and $q$ can either be regarded as a fudge factor introduced to improve the fitting to N-body simulations \cite{grossi2009} or as a correction of the spherical collapse threshold \cite{afshordi2008,valageas2010}. The scale dependence of the non-Gaussian bias is hidden in $\mathcal{M}_M(k)$ and $\mathcal{F}_M(k)$. $\mathcal{M}_M(k)$ connects the density with the primordial potential. The linear matter density contrast smoothed on a mass scale $M$ is related to the primordial potential by the Poisson equation,
\begin{equation}
\delta_{{\rm m};M,z}(\vk)=\frac{2}{3}\frac{T(k)k^2}{\Omega_m H_0^2}W_M(k)D(z)\Phi(\vk)\equiv\mathcal{M}_M(k)D(z)\Phi(\vk)\,.
\end{equation}
Here, $\Omega_m$ and $H_0$ are the present day matter fraction and the Hubble constant, respectively. The transfer function $T(k)$ models the physics until the end of recombination and is normalised to unity on large scales. The function $W_M(k)$ is the Fourier transform of the top-hat filter with a smoothing scale given by the mass $M$. As $W_M(k \rightarrow 0)=1$, we see that $\mathcal{M}_M(k)$ scales as $k^2$ on large scales.

The ``form factor'', $\mathcal{F}_M(k)$, depends on the shape of the non-Gaussianity through the bispectrum of the primordial potential $B_\Phi(k_1,k_2,k_3)$, \cite{MV}: 
\begin{equation}\label{eq:formfactor}
\mathcal{F}_M(k)=\frac{1}{8\pi^2\sigma_M^2 \fnl}\int dk' {k'}^2\mathcal{M}_M(k')\int_{-1}^{1} d\mu \mathcal{M}_M(|\vk+\vkp |)\frac{{B}_{\Phi}(k,k',|\vk+\vkp|)}{P_\Phi(k)}\,,
\end{equation}
where $\mu$ is defined by $\vk \cdot \vkp = k k' \mu$, $P_\Phi(k)$ is the primordial power spectrum, and $\sigma_M^2$ denotes the variance of linear density fluctuations on the mass scale $M$ at redshiht $z=0$.
Considering the limit of $k\rightarrow 0$, we see that on large scales the form factor depends primarily on the squeezed limit of the bispectrum.
For the local type of non-Gaussianity, the ``form factor'' becomes scale and mass independent on large scales, ${F}_M(k\ll1)=2$. Hence, we obtain the aforementioned $k^{-2}$ scale dependence of the halo bias in the case of local non-Gaussianity. As we discuss below in Sec.~\ref{templates}, other shapes of non-Gaussianity lead to different and in general weaker scale dependences.

One higher-order correction to $\Delta b$, which was introduced in \cite{desjacques2009}, is a relatively small contribution coming from the fractional difference in the Gaussian and non-Gaussian non-linear matter power spectrum.
This effect becomes larger at smaller scales but remains below a few percent on all scales for observationally allowed values of $\fnl$.

Some additional effects are also missed in Eq.~(\ref{eq:NG_bias}): The impact of the mass accretion history of the halos ---in particular recent major mergers--- on the non-Gaussian halo bias was pointed out in \cite{slosar2008}, and further investigated and tested with N-body simulations in \cite{reid2010}. The formation redshift of the halos, $z_f$, changes the amplitude of the non-Gaussian correction by $\left[b_{1;M,z}^{\rm G}(\fnl)-1\right] \longrightarrow \left[b_{1;M,z}^{\rm G}(\fnl)-\mu(z_f)\right]$, where $\mu(z_f)$ varies roughly between $-2$ and $+2$ from very young to very old halos \cite{reid2010}. For recent major mergers, $\mu$ takes the value $-1-1/\delta_c\approx -1.6$ \cite{slosar2008,reid2010}.

In addition, higher-order effects in the biasing description (e.g., terms of the order $\fnl^2$ \cite{mcdonald2008,desjacques2010,giannantonio2010}) and in the primordial non-Gaussianity (e.g., non-Gaussianity at the cubic order quantified by $\gnl$ \cite{desjacques2010,giannantonio2010}) are neglected in Eq.~(\ref{eq:NG_bias}). 

Finally, general relativistic effects are not taken into account, which might become important on very large scales \cite{bartolo2005,yoo2009,wands2009,VM,yoo2010,bartolo2010}.

So far we discussed the scale-dependent halo bias caused by the local type of non-Gaussianity with a scale-independent $\fnl$. The effect of a scale-dependent $\fnl$ were investigated analytically and by means of N-body simulations in \cite{shandera2010,becker2010}. In this paper we are mainly interested in the scale-dependent halo bias of non-local types of non-Gaussianity. The prescription presented above is sufficiently general to model the effects of other shapes of non-Gaussianity. As discussed above, the shape dependence is hidden in the form factor $\mathcal{F}_M(k)$. The expression for the form factor, see Eq.~(\ref{eq:formfactor}),  was derived by \cite{MV} in the framework of local biasing of density peaks and was evaluated for different types on non-Gaussianity in \cite{VM}. Recently, \cite{schmidt2010} generalized the peak-background split approach to non-local types of non-Gaussianities and derived a very similar but not identical expression for the form factor. 
In the next section we discuss the analytic predictions of the non-Gaussian halo bias for non-local shapes using the high peak formulation of \cite{MV}. 

\begin{figure}[htb]
\begin{center}
\includegraphics[angle=0,width=0.8\textwidth]{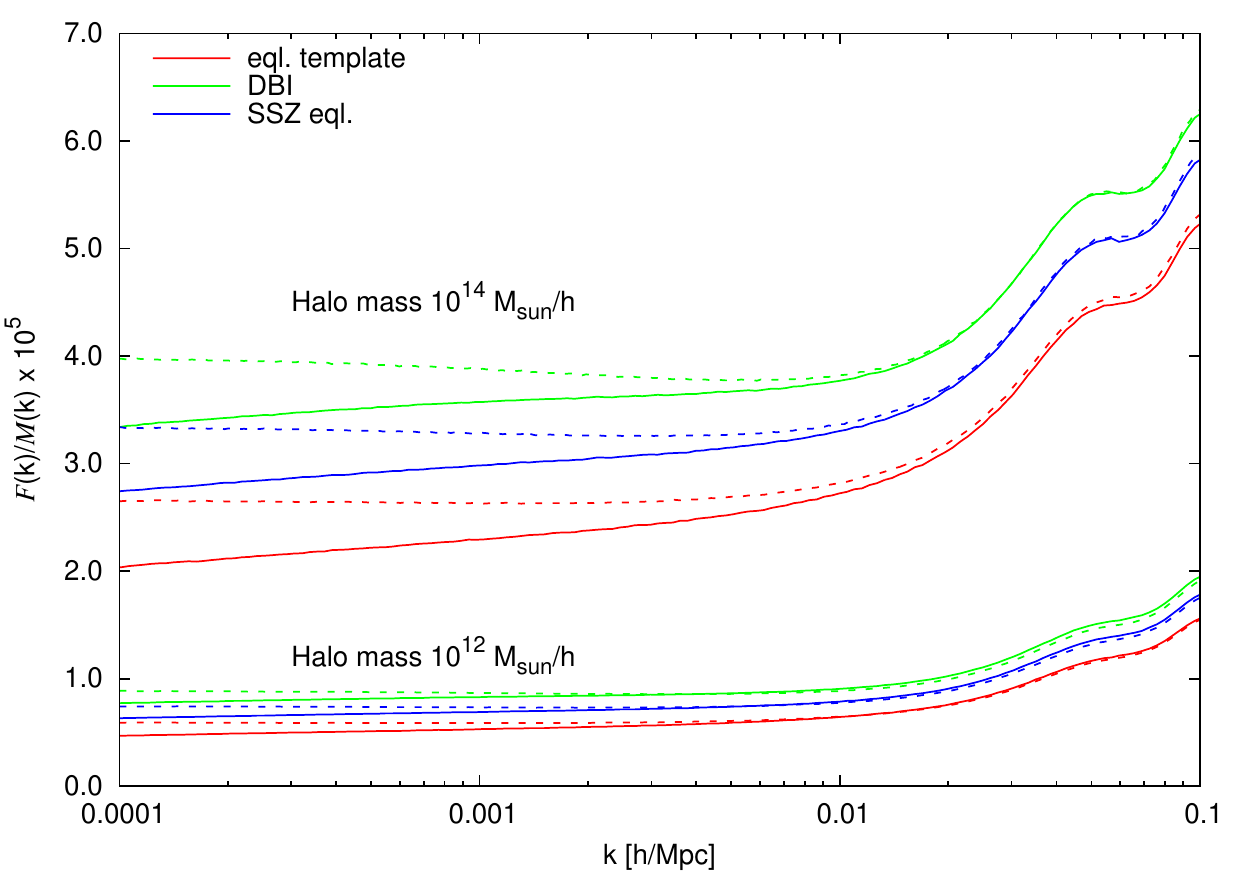}
\end{center}
\caption
{Scale-dependent part of the non-Gaussian halo bias, $\mathcal{F}_M(k)/\mathcal{M}_M(k)$, for the equilateral types of non-Gaussianity. Solid and dashed lines correspond to a spectral index of $n_s=0.95$ and $n_s=1$, respectively. In both cases, the amplitude of the power spectrum is set by keeping the mass variance fixed, $\sigma_8=0.7913$.
}
\label{fig:beta_eql}
\end{figure}

\section{Physical shapes vs. templates}
\label{templates}
The functional form of the primordial bispectrum, $B_\Phi(k_1,k_2,k_3)$, predicted by inflationary models can be quite complex. In particular, the bispectrum is usually not separable, i.e.~it cannot be written as a finite sum of terms which are factorizable as a product of functions of $k_1$, $k_2$ and $k_3$. 
Separability is important for computationally efficient analysis and simulation of both CMB maps \cite{creminelli2003,fergusson2010_cmb} and large-scale structure \cite{wagner2010,fergusson2010_lss}.
Hence, most physical shapes have been approximated by so-called templates, which are 
separable and constructed to effectively maximize the correlation  between  the physical shape and the template across all triangle configurations. In practice a so-called ``cosine" shape  correlator is used \cite{babich2004}, which  absolute value is  always $\le1$ with 1 indicating perfect correlation.
 This measure of similarity is useful for constraints on the bispectrum coming from CMB or large-scale structure analysis, as it probes the entire range of possible triangle configurations. However, these templates can be misleading for predictions of the non-Gaussian halo bias, as in this case the shape is probed mostly in the squeezed limit. In this limit the templates do not necessarily need to be good approximations of the physical shapes to yield a cosine close to unity \cite{jorge2011}. For the non-Gaussian halo bias, however, the correct scaling of the templates in this regime is crucial for sensible predictions. Let us consider several templates in this regard. We start with the equilateral\footnote{This template peaks when the wavenumbers $k$ form a equilateral triangle.} template, which was introduced in \citep{creminelli2006} as a separable approximation to the shape functions of DBI inflation \cite{alishahiha2004}, inflation with higher-derivative terms \cite{creminelli2003}, and ghost inflation \cite{arkani-hamed2004}:
\begin{equation}
B_\Phi^{\rm eql}(k_1,k_2,k_3) = 6 \fnl^{\rm eql}\left(-F^{\rm loc}-2F^A+F^B\right)\,,
\end{equation}
with\footnote{The original expression of the equilateral template was proposed for a scale-invariant power spectrum and given in terms of the wavenumbers $k$. Here we generalize it to non scale-invariant power spectra by substituting $k$ with $P^{-1/3}(k)$ with proper normalization. We use the same substitution ansatz in the case of other bispectra.}
\begin{eqnarray}
F^{\rm loc} &=& P_\Phi(k_1)P_\Phi(k_2)+ P_\Phi(k_2)P_\Phi(k_3) + P_\Phi(k_1)P_\Phi(k_3) \\
F^{A}\ &=& \left[P_\Phi(k_1)P_\Phi(k_2)P_\Phi(k_3)\right]^{2/3} \\
F^{B}\ &=& P_\Phi^{1/3}(k_1)P_\Phi^{2/3}(k_2)P_\Phi(k_3) + {\rm 5\,perm.}
\end{eqnarray}
and the dependence of $F$ on the three wavenumbers $k$ is understood. The primordial power spectrum is specified by the amplitude $\mathcal{A}_0$ at $k_0$ and the spectral index $n_s$:
\begin{equation}
P_\Phi(k)=\mathcal{A}_0\,(k/k_0)^{n_s-1}k^{-3}\,.
\end{equation}

The equilateral template was also adopted in other models. 
More generally, the form of the bispectrum is interesting because of its direct connection to the (self) interactions of the inflaton.   
In the  effective field theory of inflation \cite{EFTI}, there is a direct connection between the  coefficients of  operators  of the Lagrangian for the inflationary fluctuations and the shapes of primordial non-Gaussianity.
In \cite{ssz2010}, it was shown that the bispectrum of single-field inflation with an approximate shift symmetry protecting the Goldstone boson is a linear combination of two different shapes.
The two shapes, $B_\Phi^{\dot\pi(\partial_i\pi)^2}$ and $B_\Phi^{\dot\pi^3}$, which correspond to the interaction operators $\dot\pi(\partial_i\pi)^2$ and $\dot\pi^3$ (see \cite{ssz2010} for details), are both close to the equilateral template. Especially, the cosine of $B_\Phi^{\dot\pi(\partial_i\pi)^2}$ with the equilateral template is very close to one. For this reason, we denote this shape also by ``SSZ equilateral''.\footnote{Previously, in the framework of general single field inflation, \cite{chen2007a} obtained two independent shapes which span the same space as the two shapes of \cite{ssz2010}.}

We now turn to the question how well the equilateral template captures the behaviour of these physical shapes in  the squeezed limit. Choosing $k_1\rightarrow 0$ and using the following notation, ${\vk}\equiv \vk_1$, $\vk'\equiv \vk_2$, and $k_3 = |\vk+\vk'|=\sqrt{k^2+k'^2+2k'k\mu}$, 
we obtain for the equilateral template
\begin{eqnarray}
{B_\Phi^{\rm eql}}_{\rm squeezed}&=& 12 \fnl^{\rm eql}\left[P_\Phi^{1/3}(k) P_\Phi^{5/3}(k') 
	- \left(\frac{n_s-4}{3}\right)^2\mu^2 P_\Phi(k)k^2 P_\Phi(k')k'^{-2}\right]\\
		&=& 12 \mathcal{A}_0^2\fnl^{\rm eql}\left[\left(\frac{k^{\frac{1}{3}} k'^{\frac{5}{3}}}{k_0^2}\right)^{n_s-1}
	-  \left(\frac{n_s-4}{3}\right)^2\mu^2 \left(\frac{kk}{k_0^2}'\right)^{n_s-1}\right]
		k^{-1}k'^{-5}\\
		&=&12 \mathcal{A}_0^2 \fnl^{\rm eql}\left(1-\mu^2\right) k^{-1} k'^{-5} \sim k^{-1}\,,
\end{eqnarray}
where in the last line a scale-invariant power spectrum was assumed, i.e. $n_s=1$. We find a similar  scaling of the bispectrum in the squeezed limit also for the physical shapes generated by the aforementioned inflationary models. However, the exact scaling and its amplitude differ slightly from case to case. In the case of a scale-invariant power spectrum, the scaling reduces for all models to $k^{-1}$ with a model-dependent amplitude. For example for the bispectrum of DBI inflation (see \cite{babich2004} for an explicit expression) we get
\begin{equation}
{B_\Phi^{\rm DBI}}_{\rm squeezed}=\frac{99}{7}\mathcal{A}_0^2 \fnl^{\rm eql}\left(1-\frac{5}{11}\mu^2\right) k^{-1} k'^{-5}\,,
\end{equation}
while the scaling of the bispectrum corresponding to $\dot\pi(\partial_i\pi)^2$ \cite{ssz2010} is 
\begin{equation}
{B_\Phi^{\rm SSZ\ eql.}}_{\rm squeezed}=\frac{216}{17} \mathcal{A}_0^2 \fnl^{\rm eql}\left(1-\frac{5}{8}\mu^2\right) k^{-1} k'^{-5} \,.
\end{equation}
Here, we used the equilateral normalization convention for all shapes, i.e.~$B_\Phi(k_0,k_0,k_0)=6\mathcal{A}_0^2 k_0^{-6}\fnl^{\rm eql}$.

In Fig.~\ref{fig:beta_eql}, the scale-dependent part of the non-Gaussian halo bias, $\mathcal{F}_M(k)/\mathcal{M}_M(k)$, is shown for different bispectra of the equilateral type. The dashed lines depict $\mathcal{F}_M(k)/\mathcal{M}_M(k)$ for a scale-invariant power spectrum. Note that on very large scales the amplitude of the effect for the equilateral template and the physical models, namely DBI and SSZ eql., is different by a factor of $1.5$ and $1.256$, respectively, in agreement with the above expressions for the squeezed limit. The solid lines show $\mathcal{F}_M(k)/\mathcal{M}_M(k)$ computed with a power spectrum with $n_s=0.95$.
The upturn and the wiggles on smaller scales stem from the transfer function included in $\mathcal{M}_M(k)$.
Note, in contrast to the local type of non-Gaussianity, $\mathcal{F}_M(k)/\mathcal{M}_M(k)$ for the equilateral shapes is mass-dependent even on large scale. 

In summary, predicting the non-Gaussian halo bias with the equilateral template instead of the physical shapes results in approximately the correct scaling of the effect but a slightly smaller amplitude. The amplitude derived from the physical shapes of DBI and SSZ eql.~is larger by $50\%$ and $26\%$, respectively. 

\begin{figure}[htb]
\begin{center}
\includegraphics[angle=0,width=0.8\textwidth]{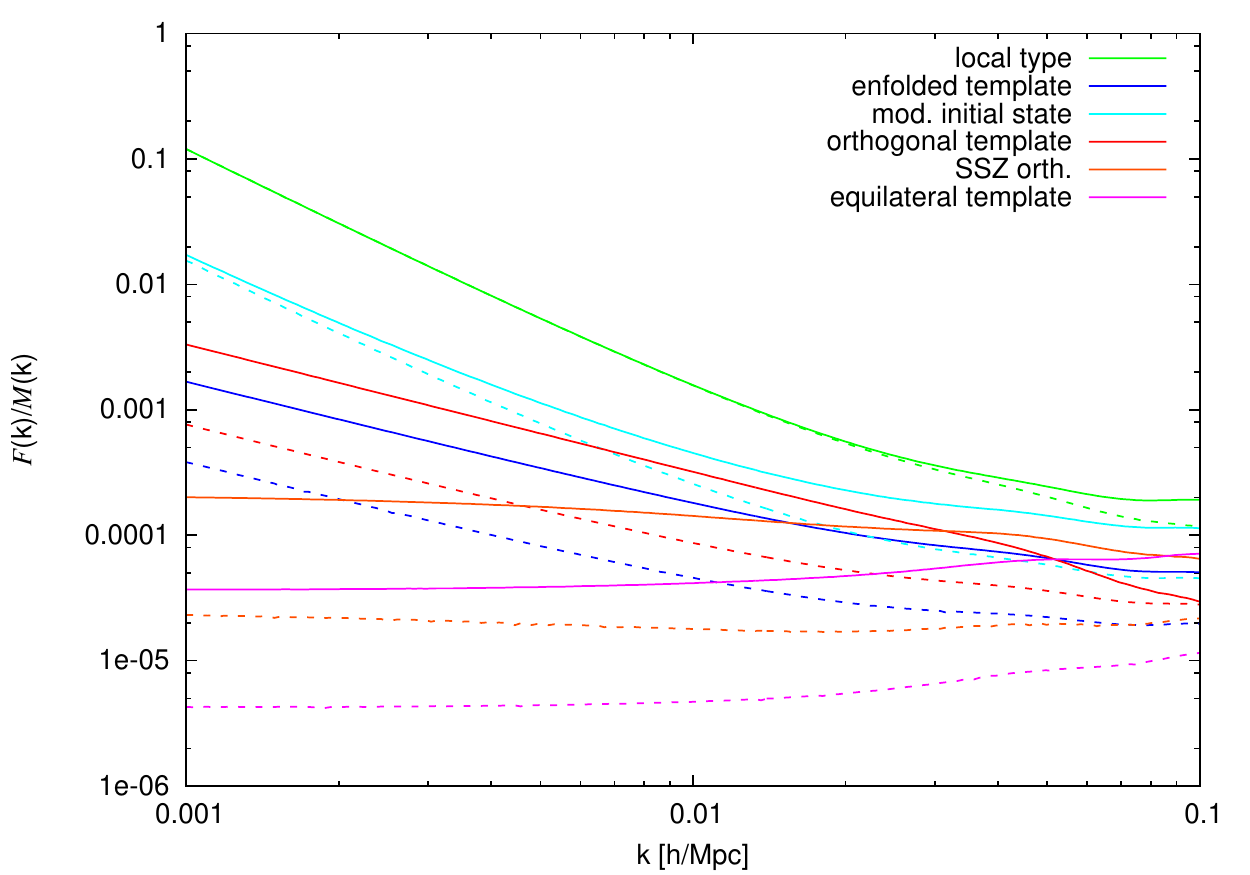}
\end{center}
\caption
{Scale-dependent part of the non-Gaussian halo bias, $\mathcal{F}_M(k)/\mathcal{M}_M(k)$, for different inflationary models and their corresponding templates. The solid and dashed lines show the results for halos of mass $3\times 10^{14}\,\MSUN/h$ and $3\times 10^{11}\,\MSUN/h$, respectively.
}
\label{fig:NG_bias}
\end{figure}

For other templates, as we discuss now, the situation becomes more problematic.
The orthogonal\footnote{The orthogonal template was introduced as a separable approximation to the linear combination of the shapes $B_\Phi^{\dot\pi(\partial_i\pi)^2}$ and $B_\Phi^{\dot\pi^3}$ that is orthogonal (i.e. has a vanishing cosine) to $B_\Phi^{\dot\pi(\partial_i\pi)^2}$} template \cite{ssz2010}
\begin{equation}
B_\Phi^{\rm orth}(k_1,k_2,k_3) = 6 \fnl^{\rm orth}\left(-3F^{\rm loc}-8F^A+3F^B\right)\,,
\end{equation}
and the enfolded\footnote{This template approximates the shape function produced by inflationary models with a modifications of the initial-state of the inflaton field and has its maximum for flattened triangles, i.e. $k_1=k_2+k_3$.} template \cite{meerburg2009}
\begin{equation}
B_\Phi^{\rm enfl}(k_1,k_2,k_3) = 6 \fnl^{\rm enfl}\left(F^{\rm loc}+3F^A-F^B\right)=\frac{1}{2}\left(B_\Phi^{\rm eql}-B_\Phi^{\rm orth}\right)\,,
\end{equation}
scale both as $\sim k^{-2}$ in the squeezed limit, while the corresponding physical shapes scale as $\sim k^{-1}$ and $\sim k^{-3}$, respectively. 
In Fig.~\ref{fig:NG_bias} we show again the scale-dependent part of the non-Gaussian bias, $\mathcal{F}_M(k)/\mathcal{M}_M(k)$, computed from these commonly used templates and the corresponding physical shapes. For comparison, the scale dependence obtained from the local type of non-Gaussianity, $B_\Phi^{\rm loc}=2\fnl^{\rm loc} F^{\rm loc}$, and from the equilateral template are also shown. Here, we used in all cases a scale-invariant power spectrum. The solid and dashed lines correspond to halos of mass $3\times 10^{14}\,\MSUN/h$ and $3\times 10^{11}\,\MSUN/h$, respectively. 
The scale dependence of the halo bias caused by modified-initial-state type of non-Gaussianity \cite{meerburg2009} is almost degenerate with the one caused by the local type of non-Gaussianity, i.e.~both predictions are very similar when  $\fnl^{\rm loc}$ is rescaled by $1/8$. However, on scales of $\sim 0.01 h/{\rm Mpc}$ the different halo mass dependence lifts this degeneracy to some extent.
The orthogonal shape of \cite{ssz2010} (SSZ orth.) ---having the same scaling in the squeezed limit as the equilateral shapes--- does not lead to a scale-dependent halo bias on large scales and is almost indistinguishable from the equilateral type. Only on smaller scales, the ``orthogonal'' halo bias has a slightly different scale dependence than the equilateral one.

In conclusion, in respect of the halo bias, the commonly used enfolded and orthogonal templates are not adequate to capture the behaviour of the underlying physical models. The physical shapes scale as $\sim k^{-3}$ or as $\sim k^{-1}$ in the squeezed limit, while the scaling of the enfolded and orthogonal templates is $\sim k^{-2}$. 

In \cite{fergusson2010_cmb} a mode expansion of the bispectrum in separable functions was introduced. This expansion can be applied to arbitrary non-separable bispectra. In many cases already a small number of modes is sufficient to yield a large cosine with the original bispectrum. While this method of constructing separable templates for arbitrary bispectra is useful for efficient measurements of the bispectrum from the CMB or the large-scale structure, it fails to reproduce the correct scaling of the bispectrum in the squeezed limit. By construction of the mode expansion, all templates derived in this way have a scaling of $\sim k^{-2}$ in the squeezed limit, irrespectively of the scaling of the original bispectrum.

New separable templates which are not only overall good approximations to the original bispectrum but also faithfully capture the scaling in the squeezed limit can most probably be constructed (e.g \cite{jorge2011}). However, we will see in the next section that separability of the bispectrum does not speed up our current implementation of generating the initial conditions for N-body simulation.

As a concluding remark, we note that similar considerations apply for the skewness specified by the bispectrum. The skewness is the relevant parameter used for the predictions of the non-Gaussian halo mass function. Although having a large mean correlation over all triangle configurations, templates and corresponding physical shapes may lead to different 
values of the skewness. However, the differences in the skewness will be, in general, much smaller than the aforementioned differences in the scaling behaviour of the squeezed limit.

%%%%%%%%%%%%%%%%%%%%%%%%%%%%%%%%%%%
\section{Initial conditions}
\label{IC}
In \cite{wagner2010} we introduced a prescription how to generate non-Gaussian initial conditions for a given bispectrum, i.e.~how we find a realization of the potential field $\Phi$, which has the desired bispectrum
\begin{equation}\label{eq:bispectrum}
\langle \Phi_{\vk_1}\Phi_{\vk_2}\Phi_{\vk_3}\rangle = (2\pi)^3 \delta^{\rm D}(\vk_1+\vk_2+\vk_3)B_\Phi(\vk_1,\vk_2,\vk_3)\,.
\end{equation}
We shortly review our method.
First, we decompose the potential field in Fourier space, $\Phi_\vk$, in a Gaussian and a non-Gaussian part
\begin{equation}
\Phi_\vk=\Phi^{\rm G}_\vk+ \Phi^{\rm NG}_\vk\,.
\end{equation}
The Gaussian field $\Phi^{\rm G}_\vk$ is statistically completely described by the power spectrum
\begin{equation}
\langle\Phi^{\rm G}_{\vk_1}\Phi^{\rm G}_{\vk_2}\rangle=(2\pi)^3 \delta^D(\vk_1+\vk_2) P_\Phi(k_1)\,.
\end{equation}
As the non-Gaussian contribution is small, the leading order of the bispectrum is given by
\begin{equation}\label{eq:leadingorder}
\langle \Phi_{\vk_1}\Phi_{\vk_2}\Phi_{\vk_3}\rangle = \langle \Phi^G_{k_1} \Phi^G_{k_2} \Phi^{NG}_{k_3}\rangle + \langle \Phi^G_{k_1} \Phi^{NG}_{k_2} \Phi^{G}_{k_3}\rangle +\langle \Phi^{NG}_{k_1} \Phi^G_{k_2} \Phi^{G}_{k_3}\rangle \,.
\end{equation}
Using the following ansatz for $\Phi^{NG}$ 
\begin{eqnarray}
\label{eq:ansatz}
\!\!\!\Phi^{NG}_{\bf k}\!&=&\!\frac{1}{6(2\pi)^3}\!\!\int\! d^3k_2 d^3 k_3 B_\Phi(k, k_2, k_3)\delta^D({\bf k}+{\bf k}_2+{\bf k}_3)\frac{\Phi^{*G}_{\bf k_2}\Phi^{*G}_{\bf k_3}}{P_\Phi(k_2)P_\Phi(k_3)}  \nonumber  \\
&=& \!\frac{1}{6(2\pi)^3}\!\!\int\! d^3k_2 B_\Phi(k, k_2, |{\bf k}+{\bf k}_2|)\frac{\Phi^{*G}_{\bf k_2}}{P_\Phi(k_2)}
\frac{\Phi^{G}_{{\bf k}+{\bf k}_2}}{P_\Phi( |{\bf k}+{\bf k}_2|)}\,,
\end{eqnarray}
with $\Phi^*_{\bf k}$  the complex conjugate of $\Phi_{\bf k}$, one can easily show that 
\begin{equation}
 \langle \Phi^G_{k_1} \Phi^G_{k_2} \Phi^{NG}_{k_3}\rangle=\frac{1}{3}(2\pi)^3  B_\Phi(k_1, k_2, k_3)\delta^D({\bf k}_1+{\bf k}_2+{\bf k}_3)\,.
\end{equation}
Finally, we recover at leading order Eq.~(\ref{eq:bispectrum}) by virtue of the symmetry property of the bispectrum.

One problem of the above ansatz is that the non-Gaussian contributions to the power spectrum can become very large on large scales. 
\footnote{On large scales, we require the non-Gaussian contributions to the power spectrum to be negligible (or renormalizable), in order to be in agreement with CMB measurements.}
Naively, one expects $\langle \Phi^{NG}\Phi^{NG}\rangle$ to be much smaller than the Gaussian contribution $\langle \Phi^{G}\Phi^{G}\rangle$. However in the case of certain bispectra, $\langle \Phi^{NG}\Phi^{NG}\rangle$ diverges more strongly for small wavenumbers than the Gaussian power spectrum. With the above ansatz for $\Phi_\vk^{NG}$ the non-Gaussian contribution to the power spectrum reads
\begin{equation}
\langle \Phi^{NG}_\vk \Phi^{NG}_{\bf q}\rangle=\frac{1}{18}\delta^D(\vk+{\bf q})\int{d^3k^\prime \frac{ B_\Phi^2(k,k^\prime,|\vk+{\bf k^\prime}|)}{P_\Phi(k^\prime)P_\Phi(|\vk+{\bf k^\prime}|)}}\,.
\end{equation}
In practice, there is a minimum wavenumber $k_{\rm min}$, the largest mode which fits in the simulation box, and a maximum wavenumber $k_{\rm max}$, which corresponds to the smallest wavelength still resolved by the used grid size. These wavenumbers provide large and small-scale cutoffs for the above integral. Nevertheless, this spurious contribution to the power spectrum can be large enough to affect substantially the results of the N-body simulations.  In particular, simulations with large box sizes, which are needed to predict the halo bias on large scales, suffer from these divergences. 
On large scales, $k\ll1$, the integral is dominated by the bispectrum in the squeezed limit. Using the scalings presented in Sec.~\ref{templates}, we see that the non-Gaussian contribution scales as $\fnl^2 k P_\Phi(k)$, $\fnl^2 k^{-1}P_\Phi(k)$, and $\fnl^2 k^{-3}P_\Phi(k)$ for the equilateral, orthogonal/enfolded templates, and the local type of non-Gaussianity, respectively. Hence, in the case of the equilateral type the contribution is small, whereas in the other cases it starts to dominate the power spectrum on large scales.\footnote{In \cite{fergusson2010_lss} the mode expansion technique of \cite{fergusson2010_cmb} was used to develop a computationally efficient method to generate initial conditions for N-body simulations for a given bispectrum. However, due to the $k^{-2}$ scaling in the squeezed limit of the expanded bispectrum, also in this case, non-Gaussian contributions to the power spectrum of the form $\fnl^2 k^{-1}P_\Phi(k)$ arise on large scales.}

For the local type of non-Gaussianity, we found in \cite{wagner2010} that our ansatz becomes identical to the real space formulation $\Phi(\vx)=\Phi^G(\vx)+\fnl\left[ (\Phi^G(\vx))^2-\langle(\Phi^G(\vx))^2\rangle\right]$, which does not lead to the aforementioned divergences, when we use $B_\Phi(k_1,k_2,k_3)=6\fnl P\Phi(k_2)P_\Phi(k_3)$ in Eq.~(\ref{eq:ansatz}) instead of $B_\Phi^{\rm loc}$. Here we generalize this modification by writing it in the following way
\begin{equation}
B_\Phi(k_1,k_2,k_3)\longrightarrow B_\Phi(k_1,k_2,k_3)\frac{3P_\Phi(k_2)P_\Phi(k_3)}{P_\Phi(k_1)P_\Phi(k_2)+P_\Phi(k_2)P_\Phi(k_3)+P_\Phi(k_1)P_\Phi(k_3)}\,.
\end{equation}
Inserting this expression in Eq.~(\ref{eq:ansatz}) yields the modified ansatz for $\Phi_\vk^{NG}$ (see also \cite{schmidt2010})
\begin{eqnarray}
\label{eq:new_ansatz}
\!\!\!\Phi^{NG}_{\bf k}\!&=& \!\frac{1}{2(2\pi)^3}\!\!\int\! d^3k' \frac{B_\Phi(k, k', |{\bf k}+{\bf k'}|)\Phi^{*G}_{\bf k'}
\Phi^{G}_{{\bf k}+{\bf k'}}}{P_\Phi(k)P_\Phi(k')+P_\Phi(k')P_\Phi(|\vk+\vkp|)+P_\Phi(k)P_\Phi(|\vk+\vkp|)}\,,  
\end{eqnarray}
which by using Eq.~(\ref{eq:leadingorder}) reproduces Eq.~(\ref{eq:bispectrum}). 

Note that the integrand in the modified ansatz is in general not separable. This means that the integration cannot be written in a simple way as a sum of convolutions, which can be efficiently computed by Fast Fourier Transformations (FFTs). However, a possible avenue to still factorize our modified ansatz is the use of Schwinger parameters (see \cite{smith2006} for details).
This is in contrast to our original ansatz, which is separable as long as the bispectrum is separable. 
However, the modified ansatz does not lead to contributions to the power spectrum, which diverge for $k \rightarrow 0$, since they are suppressed by $P_\Phi^{-2}(k)$:\footnote{Nevertheless, without cutoffs the integral can still be divergent, e.g.~in the local case this would lead to a renormalization of the amplitude. In practice, the cutoffs given by the box size and grid size render this contribution negligible.}
\begin{equation}
\langle \Phi^{NG}_\vk \Phi^{NG}_{\bf q}\rangle=\frac{1}{2}\delta^D(\vk+{\bf q})\int{d^3k^\prime \frac{ B_\Phi^2(k,k^\prime,|\vk+{\bf k^\prime}|)P_\Phi(k^\prime)P_\Phi(|\vk+{\bf k^\prime}|)}{\left[P_\Phi(k)P_\Phi(k')+P_\Phi(k')P_\Phi(|\vk+\vkp|)+P_\Phi(k)P_\Phi(|\vk+\vkp|)\right]^2}}\,.
\end{equation}

Note that neither Eq.~(\ref{eq:ansatz}) nor Eq.~(\ref{eq:new_ansatz}) specify the trispectrum of the non-Gaussian field at the leading order. A possible inclusion of the trispectrum in the generation of the initial conditions was proposed in \cite{fergusson2010_lss}. In this paper, however, we neglect contributions from higher-order correlations.

Next, we can convert the potential $\Phi_\vk$ into the linear density field $\delta_\vk$ with the help of the transfer function, which we compute with CAMB \cite{CAMB}, and the Poisson equation:
\begin{equation}
\label{eq:poisson}
\delta_{\vk}=\frac{2}{3}\frac{k^2T(k)D(z)}{\Omega_m H_0}\,.
\end{equation}
Finally, the density is sampled with particles. The positions and velocities of the particles are derived from the displacement field at the initial redshift using the Zel'dovich Approximation or second-order Lagrangian perturbation theory \cite{zeldovich,2LPT,sirko}.

%%%%%%%%%%%%%%%%%%%%%%%%%%%%%%%%%%%
\section{Simulations}
\label{sims}
In this paper we are mainly interested in the scale-dependent halo bias of different types of non-Gaussianity. As this effect, in general, is larger on large scales, we need to simulate large volumes but still have sufficient mass and force resolution to resolve halos. Since we are not interested in halos of a particular mass, a reasonable trade-off between size and resolution is a box size of about $2\,{\rm Gpc/h}$ and a particle load of one billion particles. This enables us to probe large scales ($k\sim 0.003\, h/{\rm Mpc}$) and resolve halos with masses above  $10^{13}\, \MSUN/h$ ($\sim 20$ particles).

The downside of our modified ansatz for the initial non-Gaussian part of the potential, $\Phi_\vk^{NG}$ given in Eq.~(\ref{eq:new_ansatz}), is the non-separability of the integrand. This precludes us from using FFTs to compute the integral. Instead, we perform the integration in Fourier space by direct summation, i.e.~for each $\vk$ mode the following sum has to be computed
\begin{equation}
\label{eq:ansatz_sum}
\Phi_\vk^{NG}= \frac{1}{2}\ \!\!\sum_\vkp\! \frac{B_\Phi(k, k', |{\bf k}+{\bf k'}|)\Phi^{*G}_{\bf k'}
\Phi^{G}_{{\bf k}+{\bf k'}}}{P_\Phi(k)P_\Phi(k')+P_\Phi(k')P_\Phi(|\vk+\vkp|)+P_\Phi(k)P_\Phi(|\vk+\vkp|)}\,,
\end{equation}
where $\vkp$ runs over all grid points in Fourier space. If $N_g$ denotes the number of grid points in one dimension, then the computational cost of computing $\Phi_\vk^{NG}$ for all $\vk$ modes scales as $N_g^6$. This scaling renders the usage of large grid sizes infeasible. Even for a moderately small grid size of $400$, the generation of the non-Gaussian initial conditions takes two days on 256 cores of present-day CPUs. In order to still use a large particle load, we use two different grid sizes: One grid being as large as the particle grid, $N_p$, which samples only the Gaussian contribution to the potential, and one smaller grid for the non-Gaussian part.\footnote{To compute the non-Gaussian part, we use Eq.~(\ref{eq:ansatz_sum}) where we only sum over modes which lie within the smaller grid.}
By doing this, we neglect non-Gaussianities on scales smaller than the grid spacing of the non-Gaussian grid. This not only changes the clustering on these scales, but may also affect the halo clustering on large scales, as the formation of the halos occurs on small scales. This effect will depend on the halo mass; halos which form on scales larger than the resolution of the non-Gaussian grid will, presumably, hardly be affected. The choice of the non-Gaussian grid size is therefore a trade-off between computational resources and the lowest halo mass, which is still unaffected by the spatially unresolved non-Gaussianity.

\begin{figure}[htb]
\begin{center}
\includegraphics[angle=0,width=0.49\textwidth]{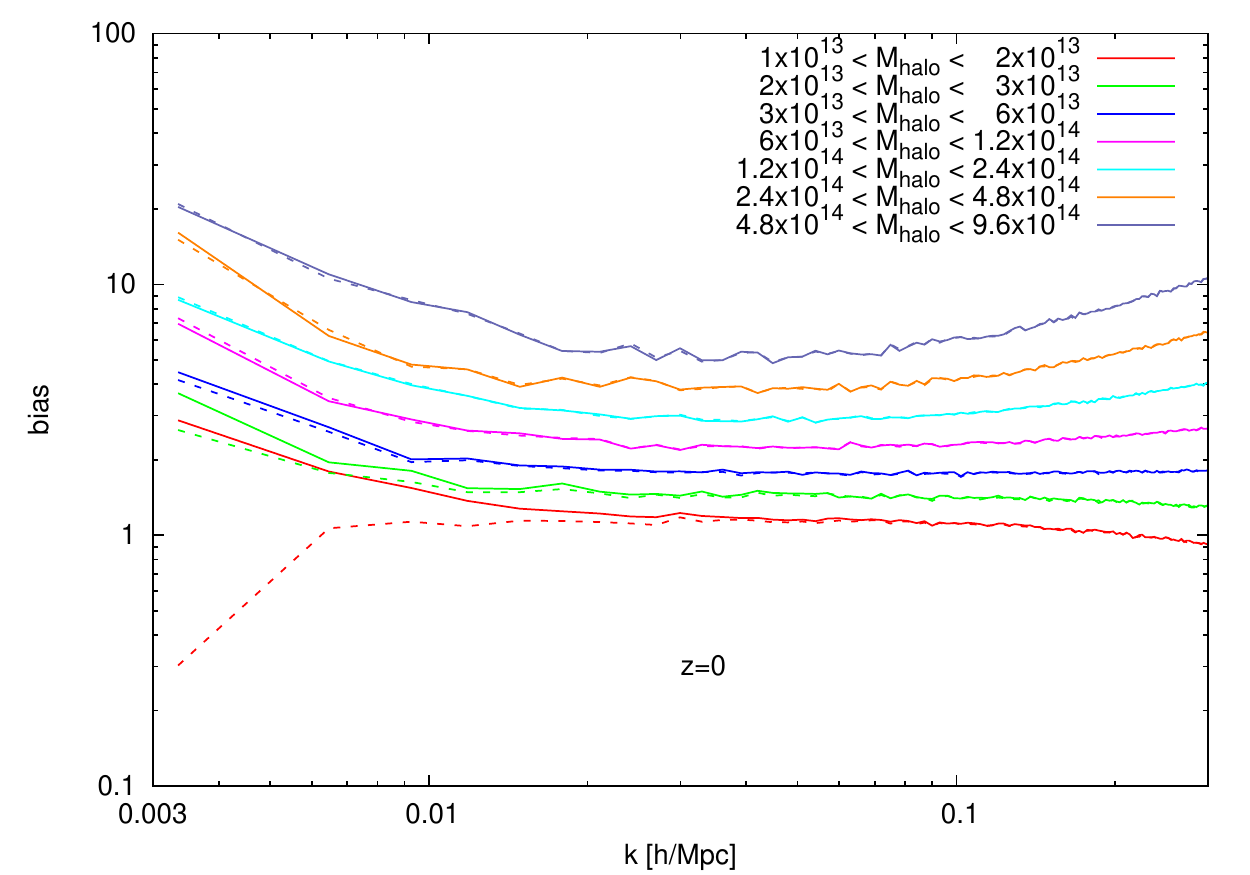}
\includegraphics[angle=0,width=0.49\textwidth]{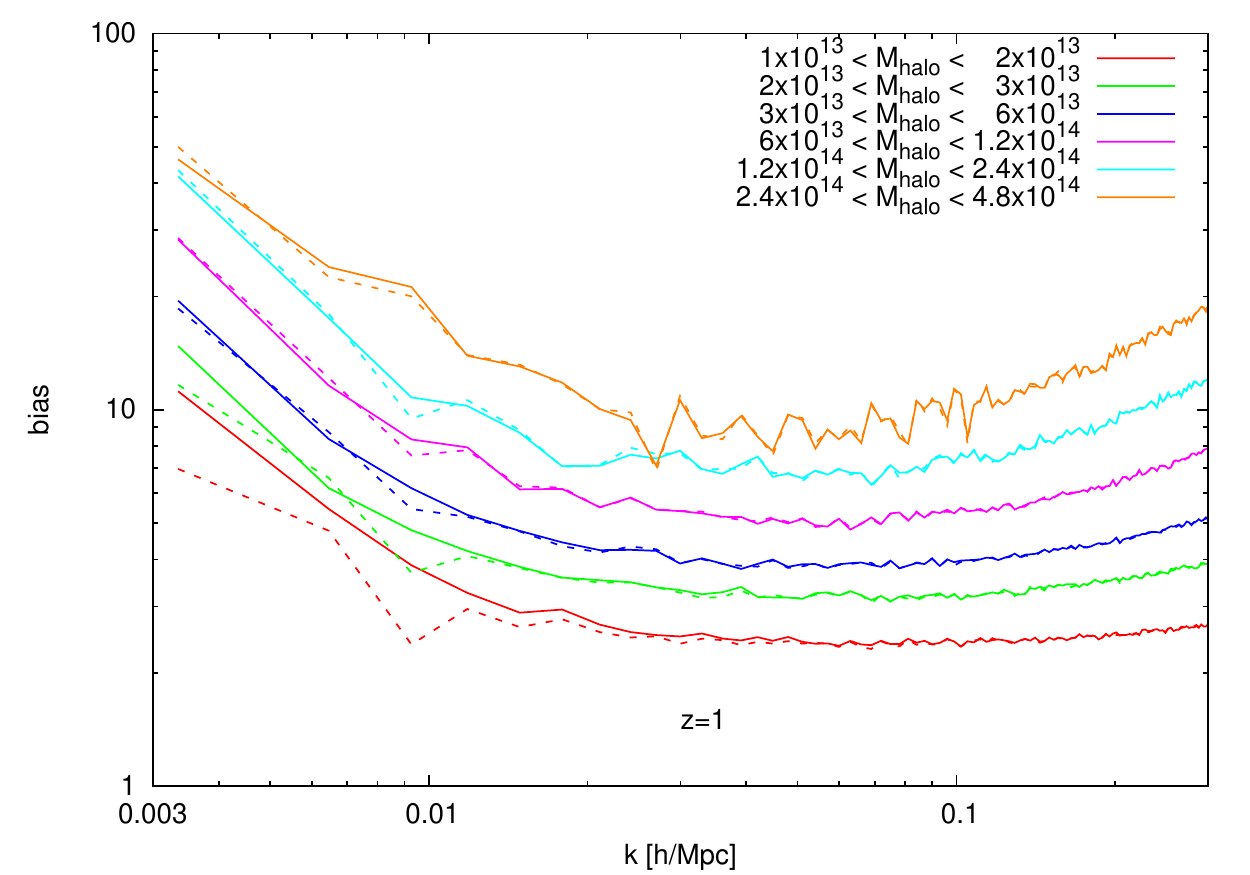}
\end{center}
\caption
{Halo bias for different halo masses computed by the ratio of the halo-matter cross power spectrum to the matter power spectrum obtained from N-body simulations with local non-Gaussianities with $\fnl=250$. The solid and dashed lines correspond to \texttt{local-1024-1024} and \texttt{local-1024-400}, respectively. The masses are given in units of $\MSUN/h$.
}
\label{fig:bias_loc_cut}
\end{figure}

To quantify this effect in more detail, we run two simulations of the local type of non-Gaussianity with $\fnl=250$. The settings of the simulations are identical except for the way we set up the initial conditions. In one case, we use a Gaussian grid of size 1024 and a non-Gaussian grid of size 400 (\texttt{local-1024-400}), in the other, both grids are of size 1024 (\texttt{local-1024-1024}). This large non-Gaussian grid size of 1024 is possible because in the case of local non-Gaussianity the computation can be performed efficiently in real space, $\Phi(x)=\Phi^G(x)+\fnl\left(\Phi^G(x)^2-\langle \Phi^{G}(x)^2 \rangle\right)$.

\begin{figure}[htb]
\begin{center}
\includegraphics[angle=0,width=0.8\textwidth]{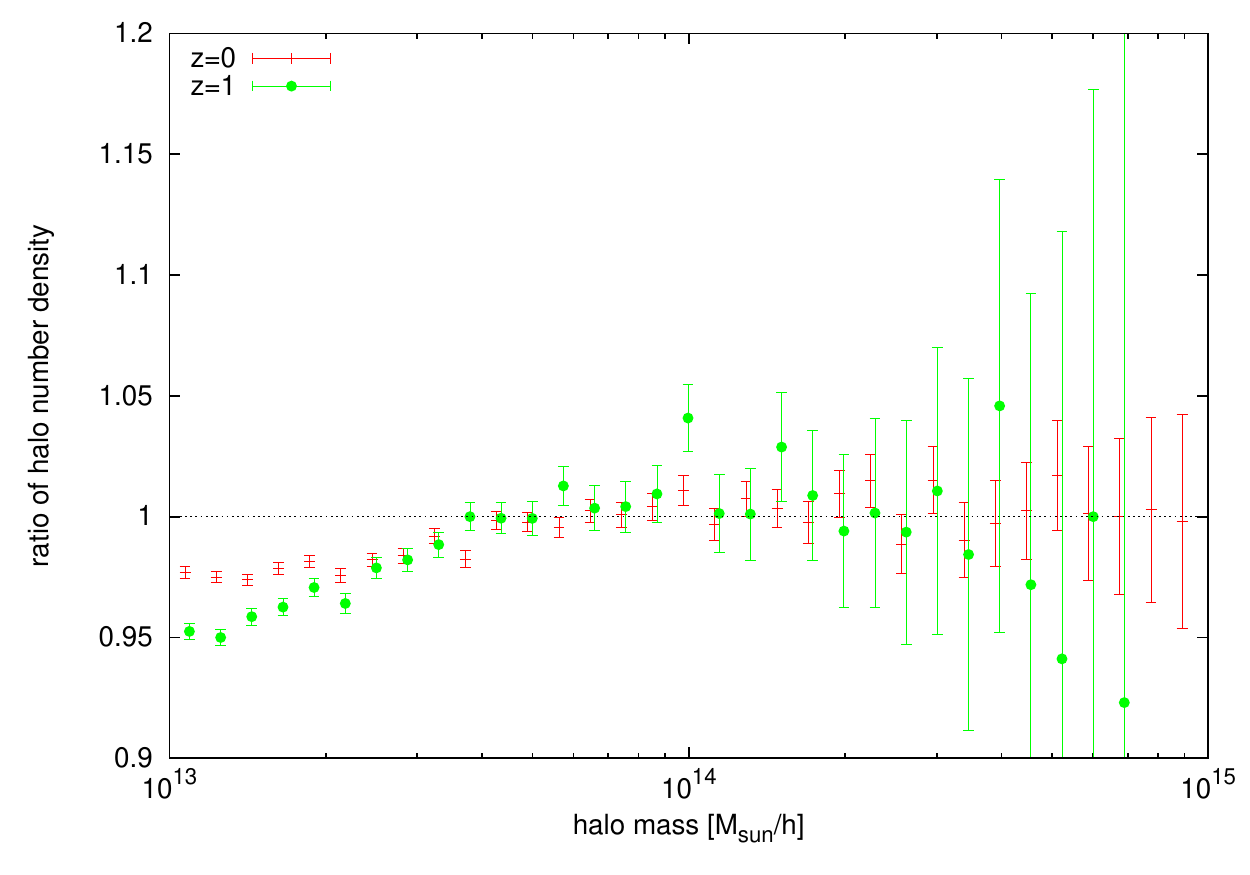}
\end{center}
\caption
{The ratio of the number density of halos found in the simulation \texttt{local-1024-400} and in the simulation \texttt{local-1024-1024} as a function of halo mass. The error bars show the Poisson error. For clarity, the green circles are shifted slightly along the x-axis.
}
\label{fig:mf_loc_cut}
\end{figure}

We adopt a flat $\Lambda$CDM cosmology with cosmological parameters consistent with current observational constraints \cite{komatsu2011}, namely $\Omega_m=0.27$ (the present-day matter fraction), $\Omega_b=0.047$ (present-day baryon fraction), $H_0=70\,{\rm km/s\,Mpc^{-1}}$ (the Hubble constant), $n_s=0.95$ (the spectral index of the primordial power spectrum), and  $\sigma_8=0.7913$ (the \textit{rms} of the linear density fluctuations at $z=0$ in spheres of $8\,{\rm Mpc}/h$).

The simulations are carried out with Gadget-2 \cite{springel2005} taking only the gravitational interaction into account. The box size is $1875\,{\rm Mpc}/h$, and $1024^3$ particles with a softening length of $40\,{\rm kpc}/h$ are used to sample the density field. The simulations are started at an initial redshift of $z_{\rm initial}=49$ using second-order Lagrangian perturbation theory to displace the particles from a regular grid.

Halos are found in the particle outputs of the N-body simulations by the publicly available halo finder AHF \cite{knollmann2009}. This halo finder defines halos as gravitationally bound objects with a spherical overdensity equal to the redshift-dependent virial overdensity. 
With the halo catalogues and particle snapshots at hand, we can compute the Fourier modes $\delta_\vk^{\rm h}$ and $\delta_\vk$ of the halo and matter density field, respectively. This is achieved by first assigning the halos or particles to a regular grid ($512^3$) using the cloud-in-cell (CIC) scheme and then performing an FFT.

In Fig.~\ref{fig:bias_loc_cut}, we show the halo bias obtained from the two simulations \texttt{local-1024-1024} (solid lines) and \texttt{local-1024-400} (dashed lines) for different halo masses at redshift $z=0$ and $z=1$. We compute the halo bias by the ratio of the halo-matter cross power spectrum to the matter power spectrum
\begin{equation}\label{eq:comp_bias}
b(k)=\frac{P_{\rm hm}(k)}{P_{\rm m}(k)}\,,
\end{equation}
with $P_{\rm m}(k)=\langle|\delta_\vk|^2\rangle$ and $P_{\rm hm}(k)=\langle {\operatorname{Re}}(\delta_\vk^{\rm h}\delta_\vk^*)\rangle$, where the average is taken over all modes which lie in a shell of radius $k$ and thickness $\Delta k=0.003\,h/{
\rm Mpc}$. 
As expected, the clustering of halos with a mass below $\sim 3\times 10^{13}\, \MSUN/h$ is affected by the smaller grid size used for the non-Gaussian $\Phi_\vk^{NG}$. The grid spacing of the non-Gaussian field is $4.7\,{\rm Mpc}/h$. This corresponds approximately to the Lagrangian radius of halos of mass $3\times 10^{13}\, \MSUN/h$. The small difference in the halo clustering for more massive halos is consistent with shot noise. At higher redshift, the noise increases due the smaller number of halos per mass bin. 

The other quantity of interest, which is also affected by the smaller non-Gaussian grid, is the halo number density. In Fig.~\ref{fig:mf_loc_cut}, we show the ratio of the mass functions derived from the two simulations \texttt{local-1024-400} and \texttt{local-1024-1024}. We find that the smaller non-Gaussian grid size decreases the halo number density at the low-mass end by a few percent. For halos more massive than $\sim 4\times 10^{13}\, \MSUN/h$, the halo number densities of both simulations agree within the errors.

In App.~\ref{app:eql_comp}, we compare the results of two N-body simulations of the equilateral type, for which we used the two different formulations for the non-Gaussian part of the potential, Eq.~(\ref{eq:ansatz}) and Eq.~(\ref{eq:new_ansatz}). This comparison gives us another estimate of the minimum halo mass, $\sim 5 \times 10^{13}\, \MSUN/h$, above which the results are not affected by the low spatial resolution of the non-Gaussianity. 

We conclude that, for halos more massive than $5\times 10^{13}\,\MSUN/h$, even the simulations which are set up with a small non-Gaussian grid ($N_g=400$) model correctly the characteristics of non-Gaussianity, in particular the halo bias and the halo number density. As a rule of thumb, the non-Gaussian grid spacing should be smaller than the Lagrangian radius of the lowest-mass halo of interest.

\begin{ctable}
  [caption={N-body simulations. $N_g$ denotes the size of the grid used for the non-Gaussian part of the potential. The size of the particle grid is identical to the size of the grid used for the Gaussian part of the potential and is given by $N_p$.}, label={tab:sims}]
  {ccccc}
	{\tnote{In this case, we use ansatz Eq.~(\ref{eq:ansatz}) for the non-Gaussian $\Phi_\vk^{NG}$. The convolution is computed with an FFT in real space.}}
{\FL
  type of non-Gaussianity		 	& $\fnl$&$N_g$ &$N_p$ & \# realizations \ML  		  
local 				& 250 	&  400 & 1024 & 1 \NN
local 				& 250 	& 1024 & 1024 & 2 \NN
local 				& 60 	& 1024 & 1024 & 3 \ML
equilateral template\phantom{\tmark} 		& 1000	&  400 & 1024 & 2 \NN
equilateral template\tmark 	& 1000	& 1024 & 1024 & 1 \ML
orthogonal template			& -1000 & 400 &  1024 & 2 \NN  
orthogonal template		& -250 & 400  &  1024 & 3 \ML
Gaussian			&  -	& -	 & 1024 & 3 \LL}
\end{ctable}

For the main study of the non-Gaussian halo bias, we perform a suite of simulations of different shapes of non-Gaussianity and different $\fnl$ values, which are summarized in Tab.~\ref{tab:sims}. The cosmology and simulation settings, like box size, particle load, etc., are the same as given above.

Although the orthogonal template is not a good approximation of the physical shape (SSZ orth.) in the context of halo bias (see Sec.~\ref{templates}) and although our ansatz is applicable to non-separable bispectra, we still use this template for some of our simulations. This has the following reasons. First, the corresponding physical shape (SSZ orth.) causes a halo bias which is very similar to the equilateral type and, hence, hard to measure. Second, the scale dependence of the non-Gaussian halo bias of the orthogonal template lies in between the local and the equilateral one, which makes it an interesting toy model to test the analytic predictions of the non-Gaussian halo bias against N-body simulations. Lastly, other groups (e.g.~\cite{scoccimarro2011}) are running N-body simulations of this type, too, which makes a future comparison easier.

The chosen values for $\fnl$ are partly larger than allowed by current constraints coming from CMB data \cite{komatsu2011}. Here, however, we want primarily to compare theoretical predictions with the results of N-body simulations. To do this accurately, we need high signal-to-noise and therefore a relatively large value of $\fnl$. This is in particular the case for the equilateral type of non-Gaussianity, for which the non-Gaussian halo bias has only a very weak scale dependence. 
Note that if we find that the analytic modelling of the non-Gaussian bias works for relatively large non-Gaussianity, it will also be valid for small values of $\fnl$.

%%%%%%%%%%%%%%%%%%%%%%%%%%%%%%%%%%%
\section{Results}\label{results}
In this section we present the outcome of our N-body simulations of different types on non-Gaussianity. 
Here, we used the standard templates (namely the local, equilateral and orthogonal templates) to create
the initial conditions for the suite of N-body simulations.
The templates are  useful  as  toy models to test  and calibrate the
analytical predictions. If the analytical expressions can correctly
reproduce the bias behaviour for shapes that have such different
$k$-dependence in the squeezed limit, this  lends  support to their
validity.

Although the simulations were targeted to study the non-Gaussian halo bias, we first consider the effect of non-Gaussianity on the halo mass function. Afterwards, we analyse in detail the scale-dependent halo bias induced by the different types of non-Gaussianity.

\subsection{Halo mass function}
Primordial non-Gaussianity affects significantly the number density of objects which correspond to the tails of the probability distribution function (PDF) of the initial density field \cite{MVJ} (for recent reviews, see \cite{verde2010,desjacques_review}). Galaxy cluster-sized halos originate from rare high peaks in the density field smoothed on the scale of the cluster mass. If the PDF has a positive skewness on this mass scale, more of these high peaks are initially present and give rise to more galaxy clusters. 

\begin{figure}[htb]
\begin{center}
\includegraphics[angle=0,width=0.49\textwidth]{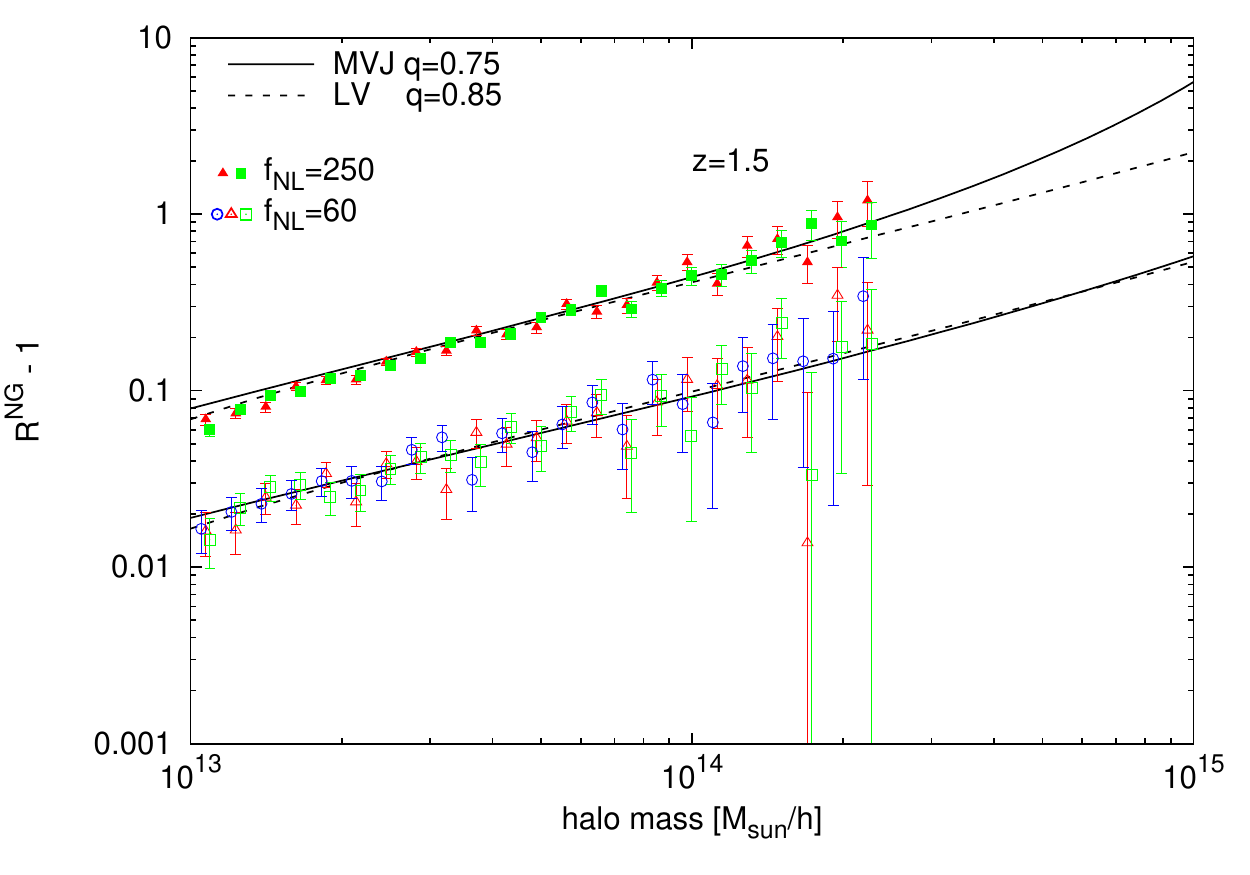}
\includegraphics[angle=0,width=0.49\textwidth]{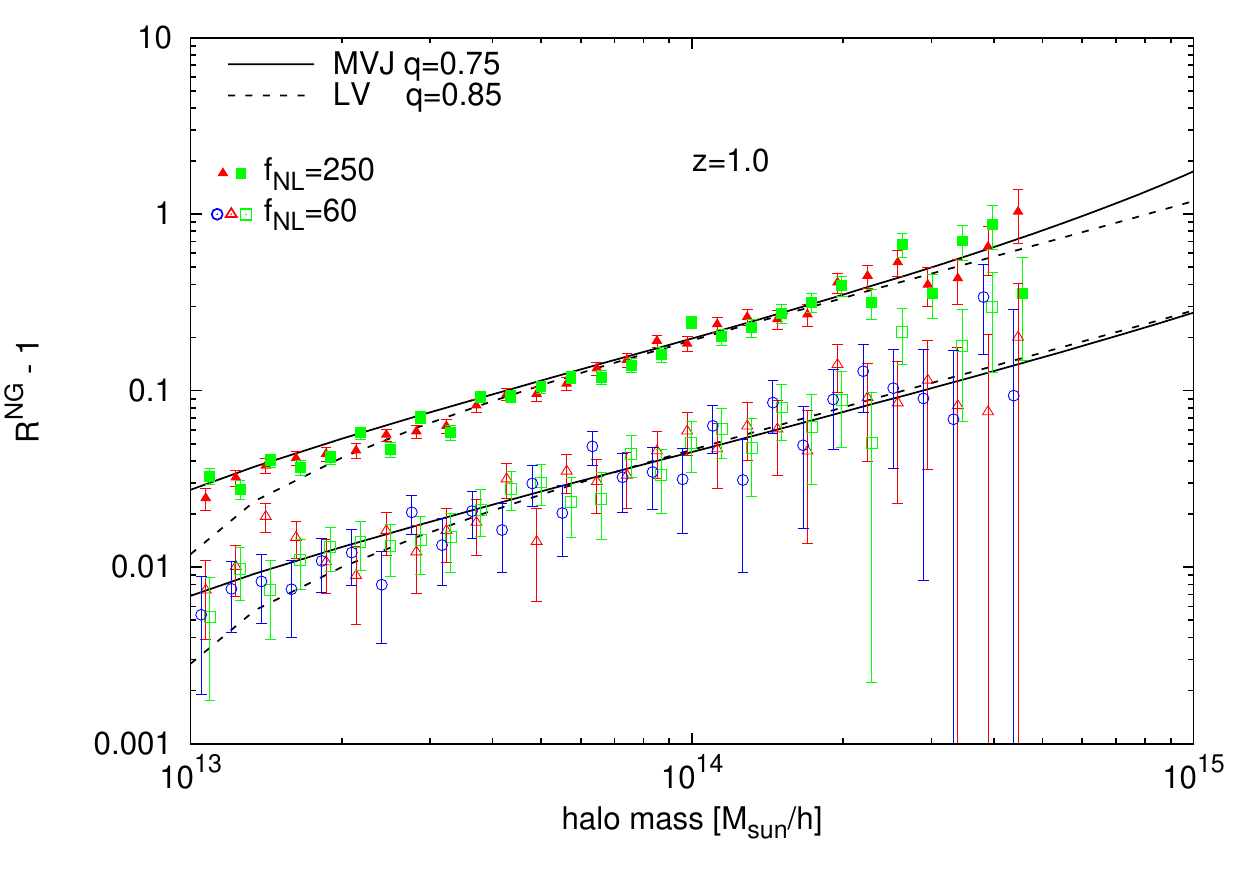}
\includegraphics[angle=0,width=0.49\textwidth]{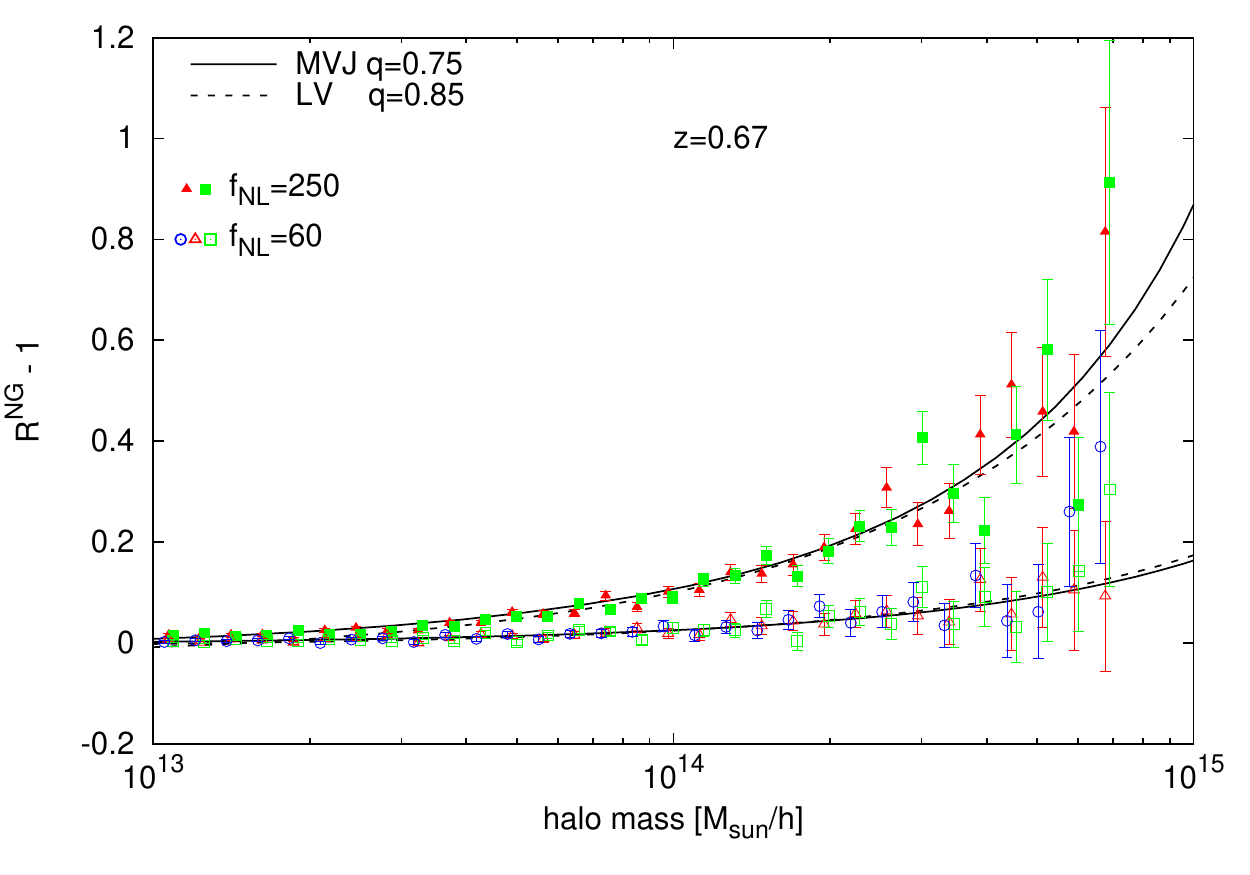}
\includegraphics[angle=0,width=0.49\textwidth]{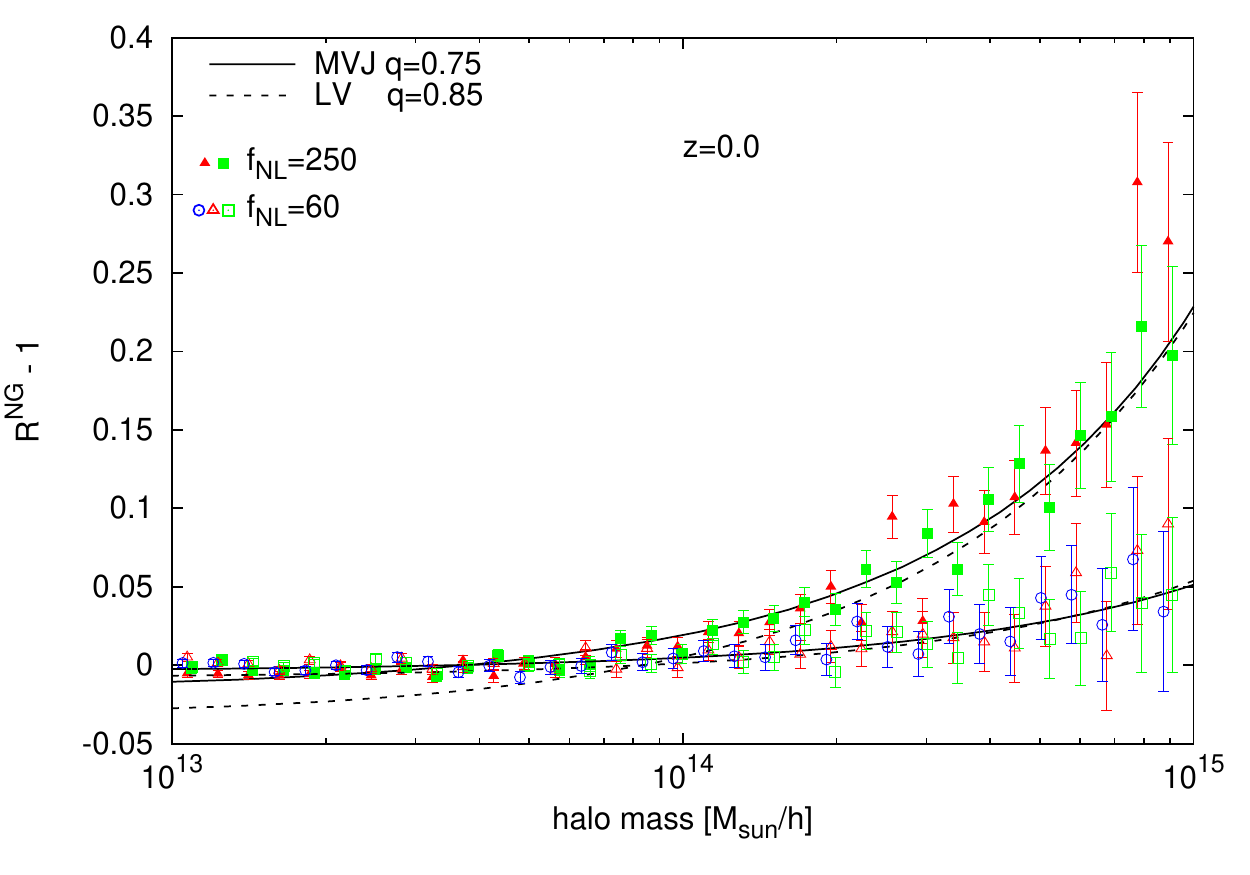}
\end{center}
\caption
{Fractional difference in the mass function derived from non-Gaussian simulations of the local type and from Gaussian simulations. Different realizations are depicted with symbols of different shapes. For clarity, the data points of different realizations are shifted slightly along the x-axis. The solid and dashed lines show the predictions of \cite{MVJ} (MVJ) and \cite{LV} (LV) using the best-fit fudge factor of $q=0.75$ and $q=0.9$, respectively.
}
\label{fig:mf_loc}
\end{figure}

The ratio of the non-Gaussian to the Gaussian halo mass function, $R^{NG}(M,z)$, can be modelled analytically in the framework of  the Press-Schechter formalism \cite{MVJ,LV} and the excursion set approach \cite{MR,norena2011}. For the comparison with our N-body results, we use the formulas of \cite{MVJ} (MVJ) and \cite{LV} (LV), who used the saddle-point technique and the Edgeworth expansion, respectively, to derive the following expressions
\begin{equation}
R^{NG}_{\rm MVJ}(M,z)=\exp\left(\frac{\Delta_c^3(z) S_{3,M}}{6\sigma^2_M}\right)\left|\frac{\Delta_c(z)}{6\sqrt{1-\Delta_c(z)S_{3,M}/3}}\frac{{\rm d}S_{3,M}}{{\rm d}\ln \sigma_M}+\sqrt{1-\Delta_c(z)S_{3,M}/3}\right|
\end{equation}
and
\begin{equation}
R^{NG}_{\rm LV}(M,z)=1+\frac{\sigma_M^2}{6\Delta_c(z)}\left[S_{3,M}\left(\frac{\Delta_c^4(z)}{\sigma_M^4}-2 \frac{\Delta_c^2(z)}{\sigma_M^2}-1\right)+\frac{{\rm d}S_{3,M}}{{\rm d}\ln \sigma_M}\left(\frac{\Delta_c^2(z)}{\sigma_M^2}-1\right)\right]\,,
\end{equation}
where $\sigma_M^2$ is the linear mass variance on a mass scale $M$ at $z=0$. The redshift dependence is modelled by the collapse threshold, $\Delta_c(z)=\sqrt{q}\delta_c D(z=0)/D(z)$, which includes a fudge factor, $\sqrt{q}$\footnote{This fudge factor was introduced in \cite{grossi2009} and interpreted as a modification to the collapse threshold which also applies to the Gaussian mass function. Here, however, we do not adopt this interpretation and consider $q$ only as a fudge factor in the ratio of the non-Gaussian to the Gaussian mass function, which we calibrate using N-body simulations.}. The non-Gaussianity is described by the normalized skewness of the linear density field, $\delta$, on a mass scale $M$ at $z=0$, $S_{3,M}=\langle \delta_{M}^3\rangle/\sigma_M^4$. The normalized skewness scales linearly in $\fnl$. The magnitude and sign of $S_{3,M}$ depend on the shape of non-Gaussianity, while the mass dependence is only weakly dependent on the type of non-Gaussianity.

\begin{figure}[htb]
\begin{center}
\includegraphics[angle=0,width=0.49\textwidth]{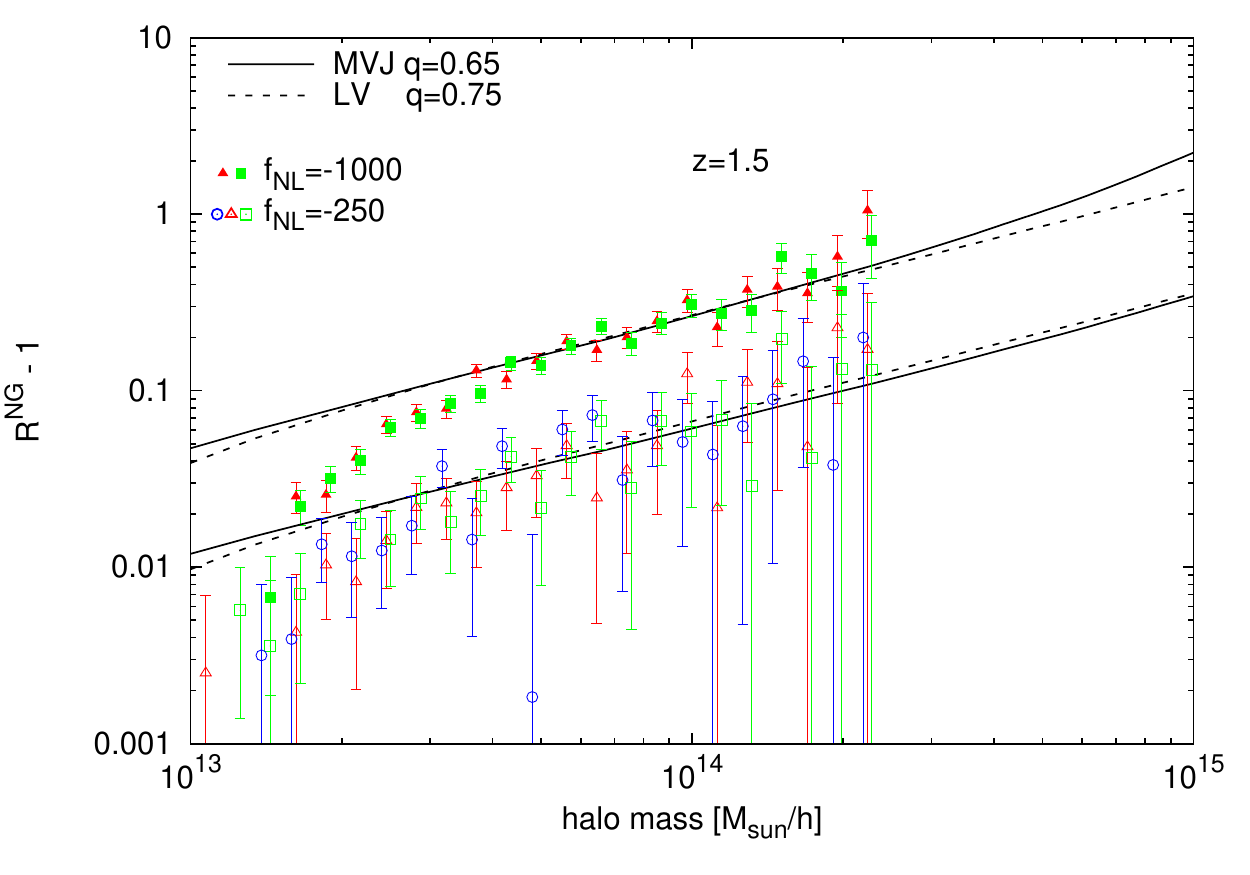}
\includegraphics[angle=0,width=0.49\textwidth]{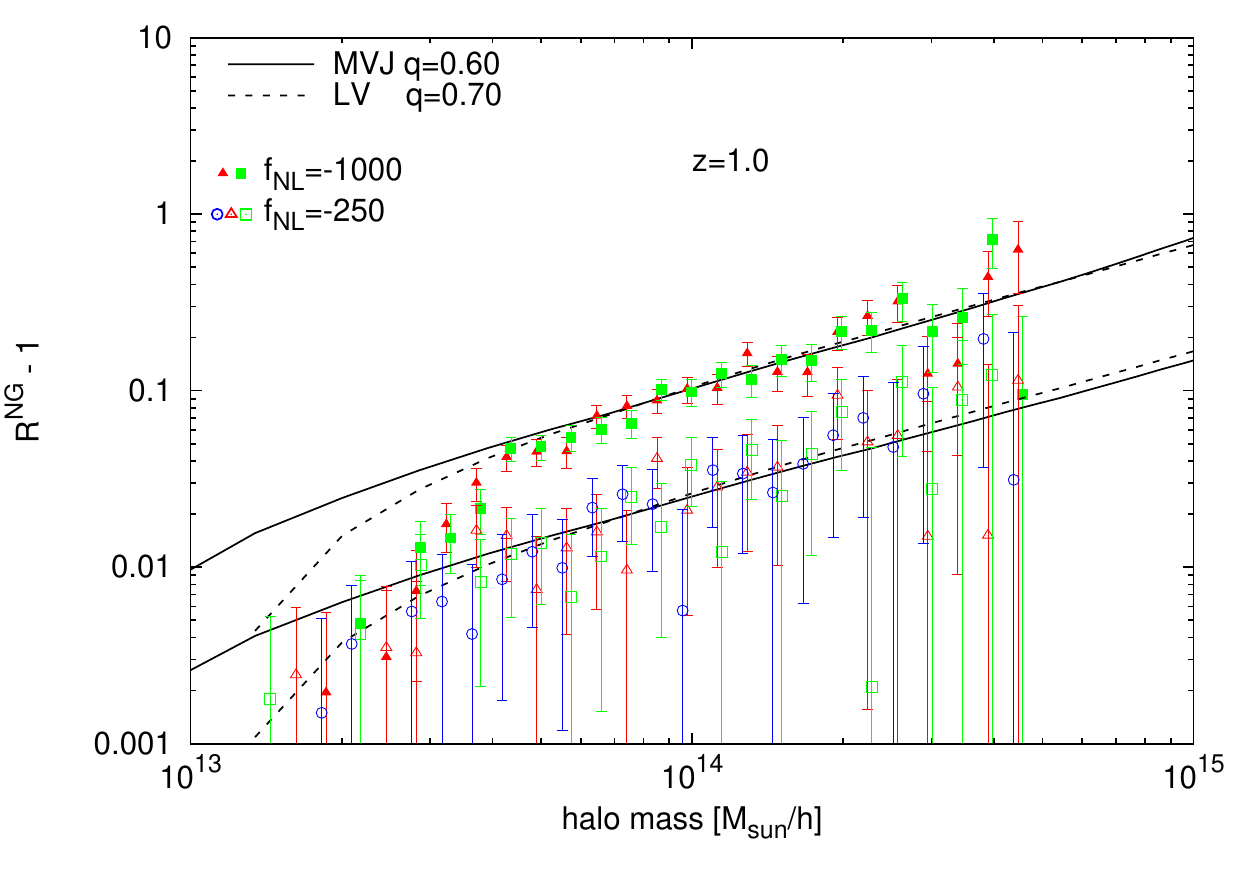}
\includegraphics[angle=0,width=0.49\textwidth]{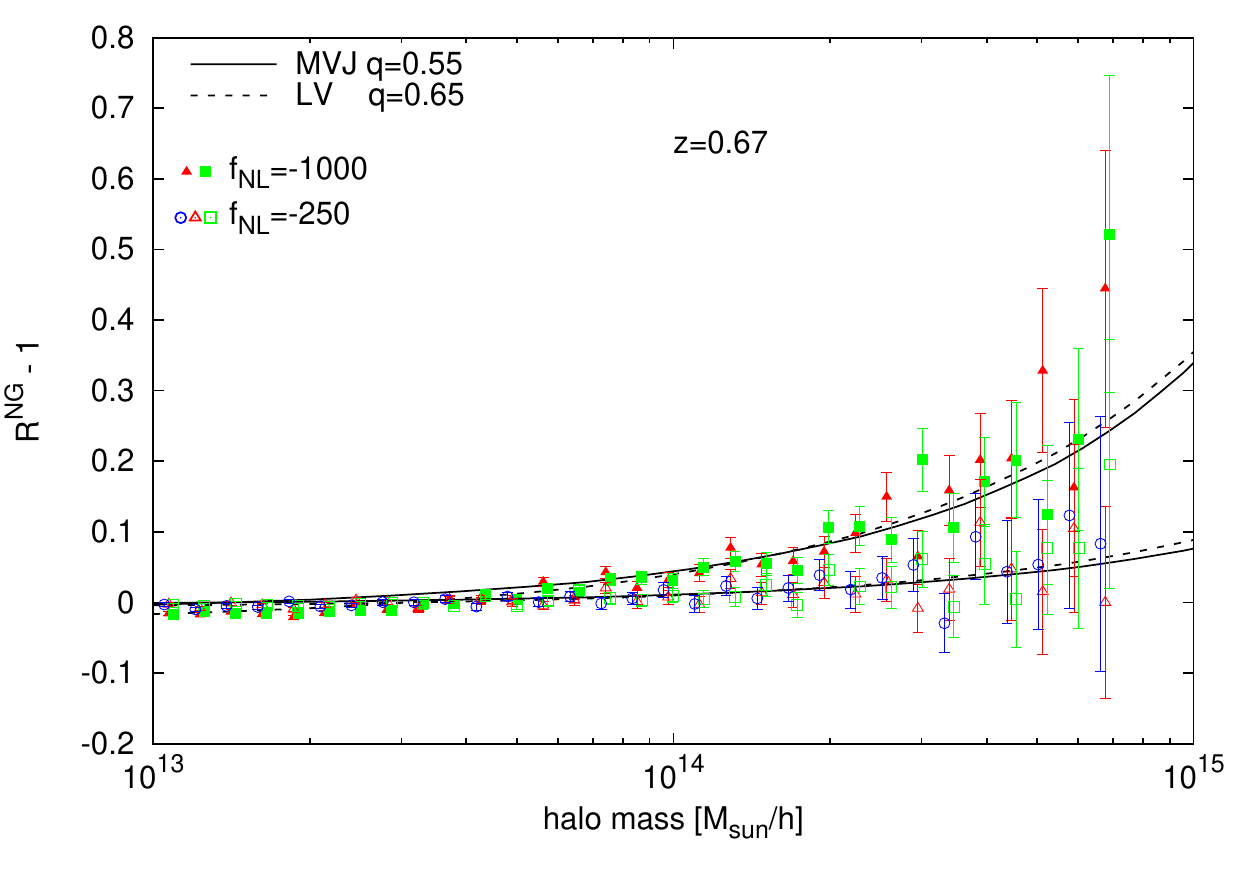}
\includegraphics[angle=0,width=0.49\textwidth]{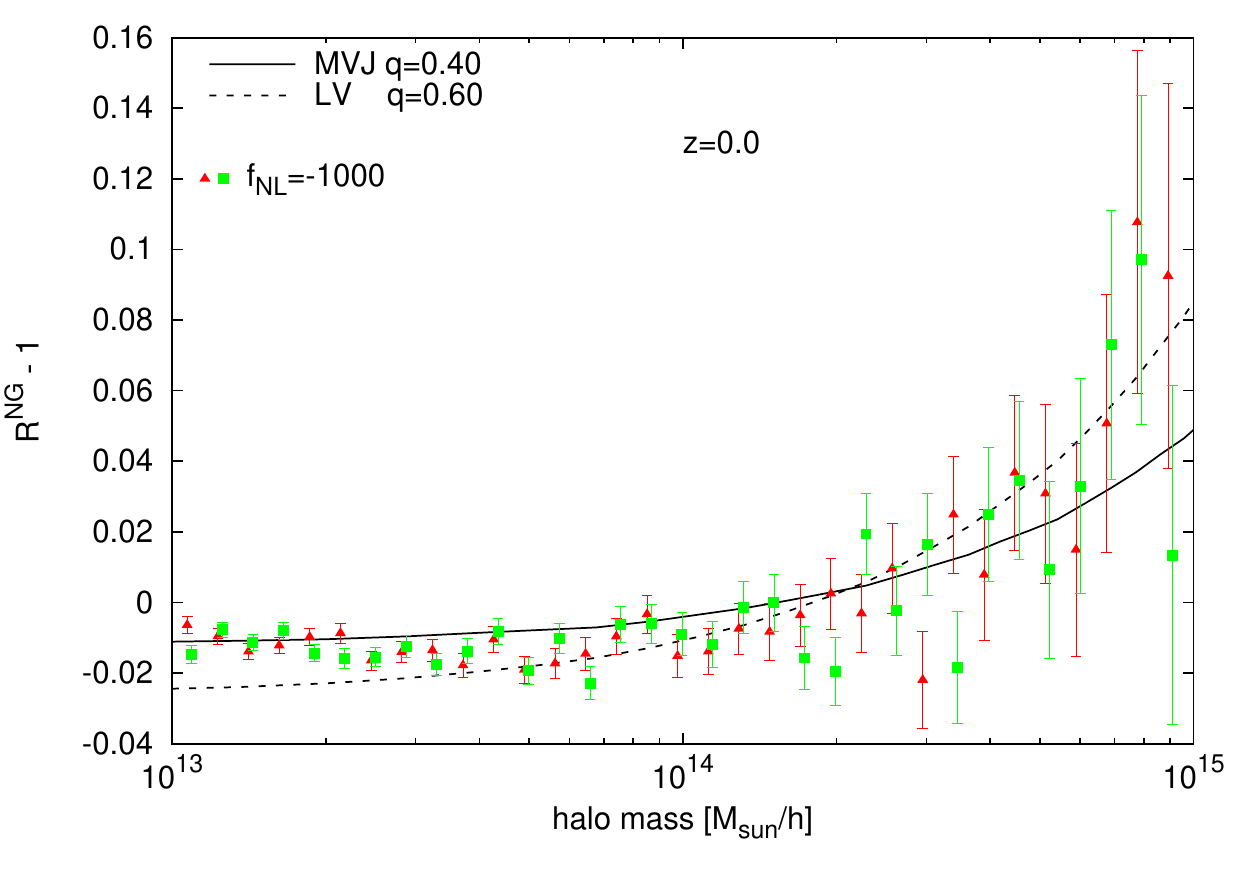}
\end{center}
\caption
{Effect on the halo mass function induced by the orthogonal shape of non-Gaussianity. Symbols and lines have the same meaning as in Fig.~\ref{fig:mf_loc}.
Due to small signal to noise, the best-fit value for q at $z=0$ is only determined to about 20\%.
}
\label{fig:mf_ort}
\end{figure}

In Fig.~\ref{fig:mf_loc}, we show the fractional difference in the non-Gaussian and Gaussian mass functions derived from the simulations of the local type of non-Gaussianity. The four panels correspond to different redshifts, $z=1.5$ (top left), $z=1$ (top right), $z=0.67$ (bottom left), and $z=0$ (bottom right). The filled and open symbols represent the results for $\fnl=250$ and $\fnl=60$, respectively. The corresponding analytic predictions of MVJ and LV (with the best-fit value for $q$) are depicted by the solid and dashed lines, respectively.
Overall, the agreement between simulations and predictions is very good. Only at the low-mass end at $z=0$, the data points do not follow the theory lines very closely. This is not of practical importance, as the difference is much smaller than observationally uncertainties in halo mass measurements. The applied fudge factor of $q=0.75$ for MVJ and $q=0.9$ for LV is in agreement with our findings in \cite{wagner2010}. Other groups \cite{grossi2009,pillepich2010}, who worked with friends-of-friends (FOF) halos, used $q\approx 0.75$ for both predictions, while \cite{desjacques2009}, who applied a spherical-overdensity (SO) halo finder as we do, found $q\approx 1$ for LV and $q\approx 0.75$ for MVJ (V. Desjacques, private communication). For the standard definitions of FOF and SO halos, i.e. using a linking length of $0.2$ times the mean particle distance (FOF) and the redshift-dependend virial overdensity (SO), \cite{desjacques_review} found that for a given halo mass the effect of non-Gaussianity on the number density of FOF halos is smaller than on the number density of SO halos.
Note however that it remains to be seen whether FOF or SO halos most closely match observationally-selected halos (e.g., SZ-selected or X-ray selected clusters). Nevertheless, the fact that $q$ is quite close to unity in both cases, suggests that this will not introduce a dominant systematic  effect.

\begin{figure}[htb]
\begin{center}
\includegraphics[angle=0,width=0.49\textwidth]{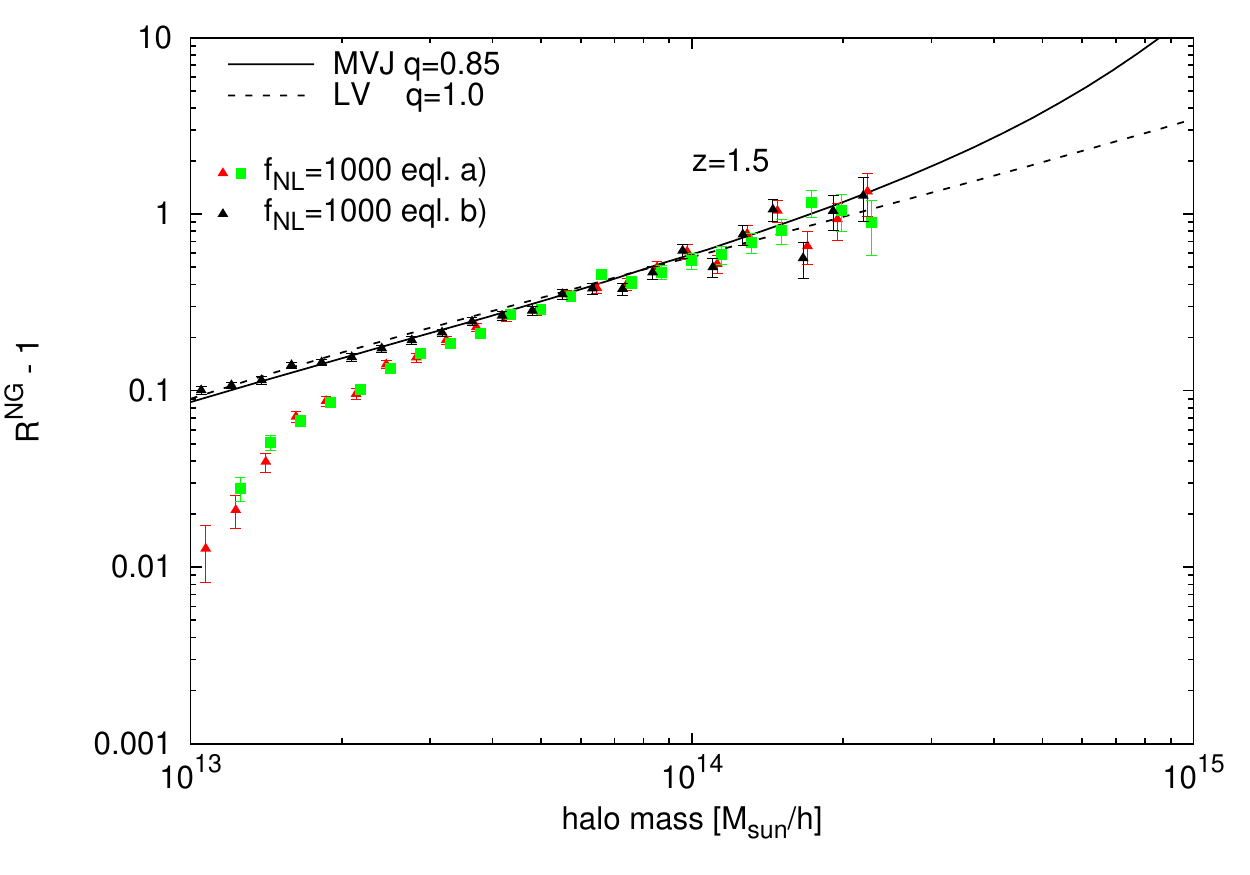}
\includegraphics[angle=0,width=0.49\textwidth]{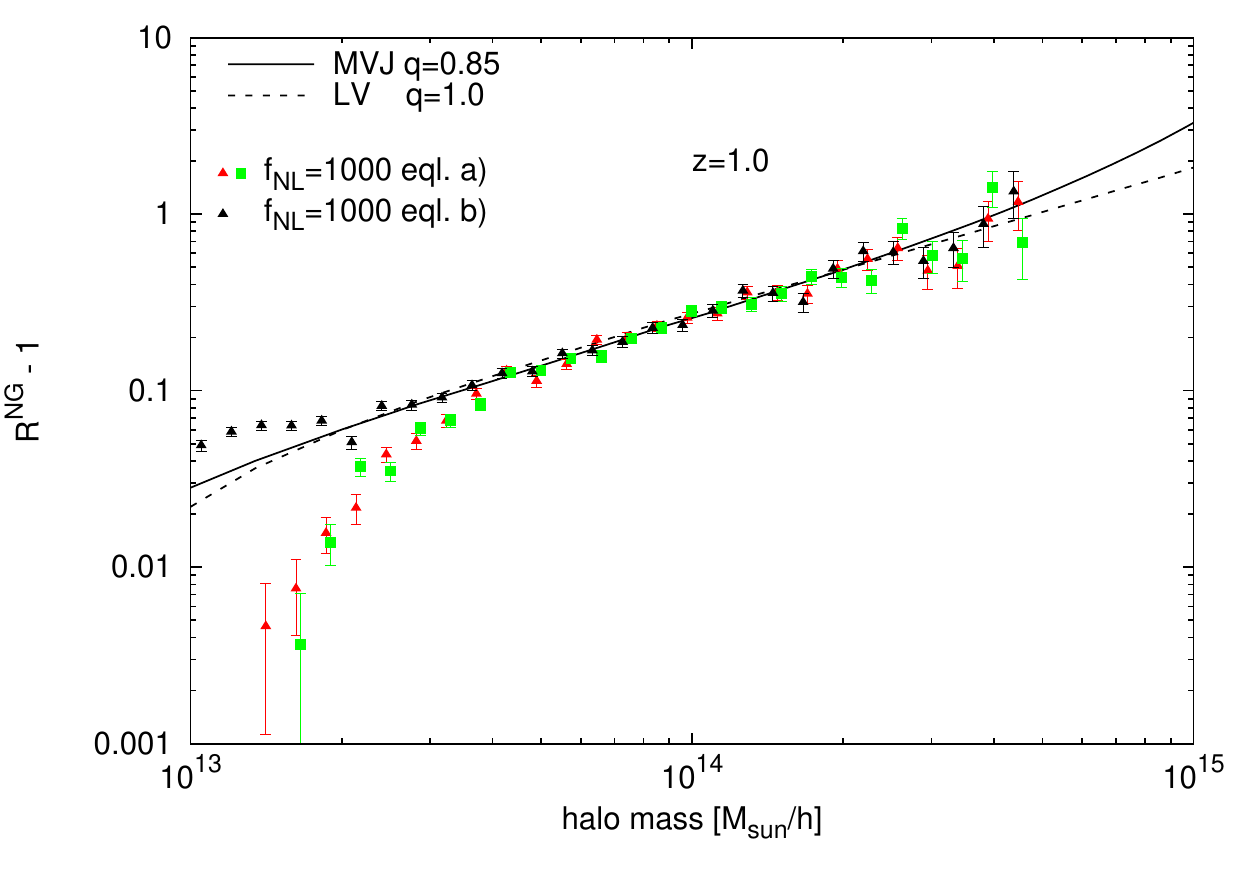}
\includegraphics[angle=0,width=0.49\textwidth]{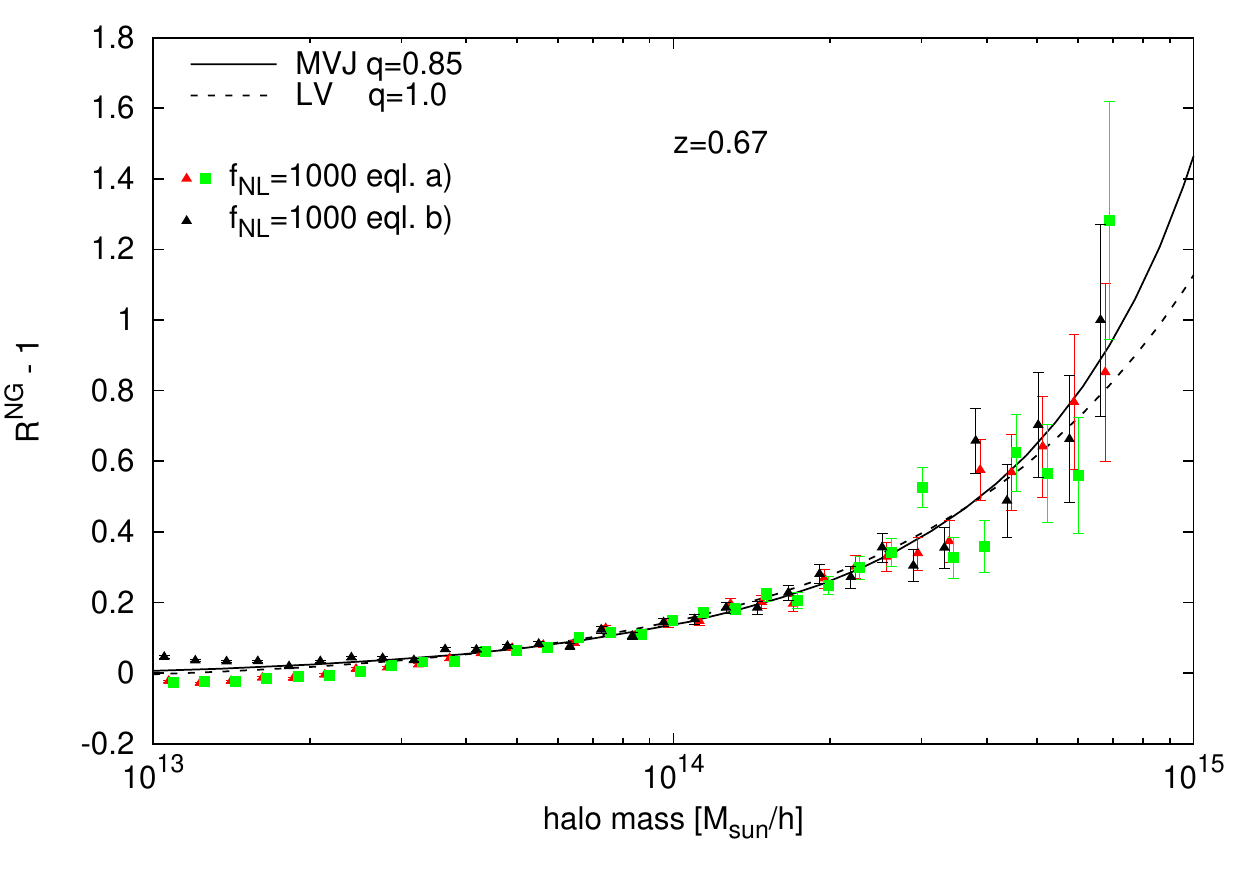}
\includegraphics[angle=0,width=0.49\textwidth]{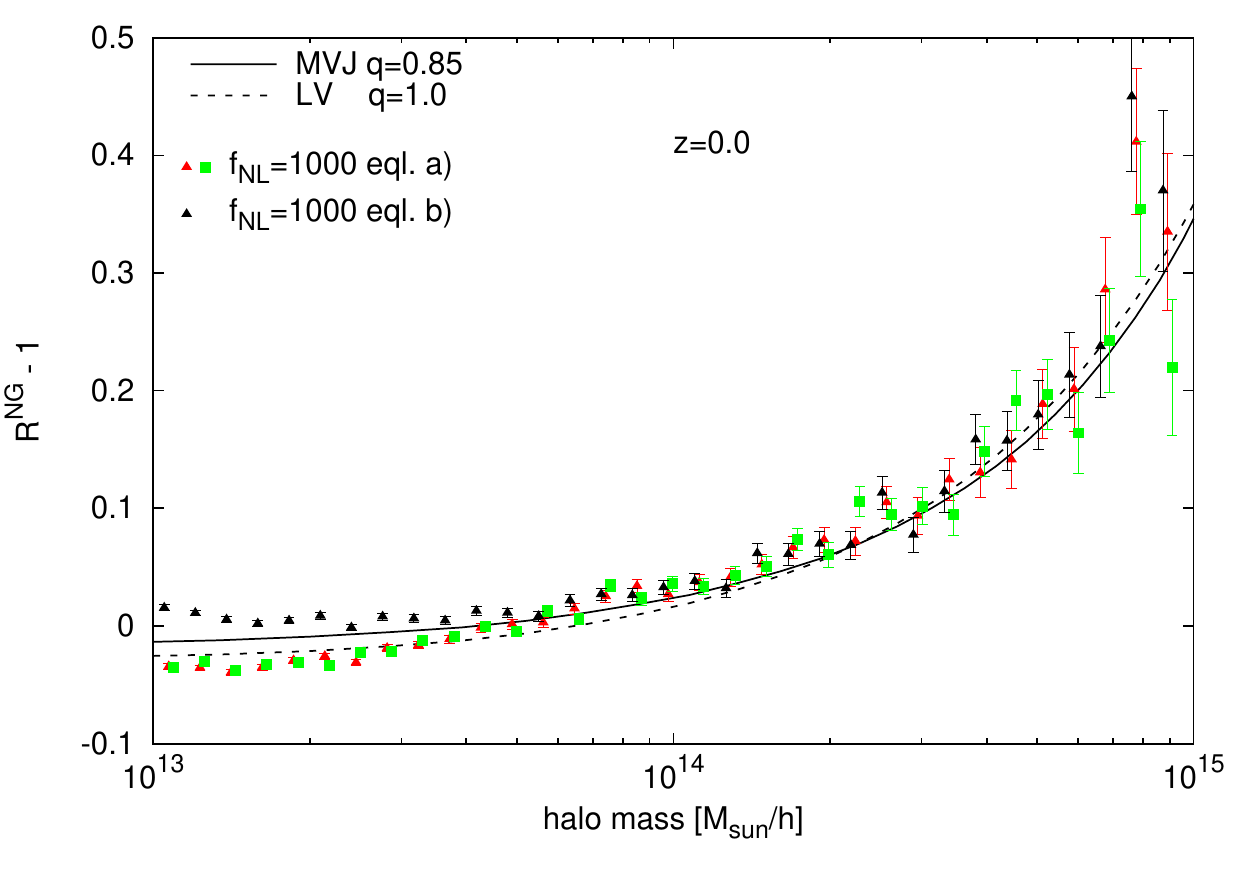}
\end{center}
\caption
{Fractional difference in the mass function caused by non-Gaussianity of the equilateral type. 
The triangles correspond to two simulations with the same realization of the initial Gaussian field but differing in the method how the initial non-Gaussian contribution was computed (Eq.~\ref{eq:ansatz} (black) and Eq.~\ref{eq:new_ansatz} (red)). The green squares depict the results of a second simulation with a different realization of the initial Gaussian field. Dashed and solid lines show the analytic predictions of \cite{LV} and \cite{MVJ}.
}
\label{fig:mf_eql}
\end{figure}

The mass function results from our non-Gaussian N-body simulations of the orthogonal type are presented in Fig.~\ref{fig:mf_ort}. Here, the $\fnl$ values are $-1000$ (filled symbols) and $-250$ (open symbols).
Note that these simulations used a rather small grid size ($N_g=400$) for the non-Gaussian initial density field. Hence, halos with masses below $~5\times 10^{13}\,\MSUN/h$ are expected to show a smaller non-Gaussian effect than predicted by the theory. This is indeed the case, as can be seen in the figure. More interestingly, even at the high-mass end the non-Gaussian effect is smaller than one could expect. To bring the theory in agreement with the N-body data, the fudge factor has to be quite small, decreasing from $q\approx0.75$ to $q\approx0.5$ with decreasing redshift.

Lastly, we consider the increase in the halo mass function due to non-Gaussianity of the equilateral type. In Fig.~\ref{fig:mf_eql}, the results of three simulations with $\fnl=1000$ are shown. Two of them, ``eql.~a)'', started from initial conditions computed with our modified ansatz, Eq.~(\ref{eq:new_ansatz}), using a non-Gaussian grid size of $400$. The third simulation, ``eql.~b)'' (black triangles), used initial conditions generated with our original method, Eq.~(\ref{eq:ansatz}), with $N_g=1024$ (see also App.~\ref{app:eql_comp}).  
Remember that in the case of the equilateral type, artificial contributions to the power spectrum are kept under control also when our original ansatz is used.
As expected, the N-body results agree with each other for halo masses above $~5\times 10^{13}\,\MSUN/h$. The fall-off of ``eql.~a)'' at the low-mass end is caused by the small non-Gaussian grid. Interestingly, we find that, in the case of the equilateral shape of non-Gaussianity, $q$ is very close to one and does not change significantly with redshift.

In conclusion, on the mass scales probed, the non-Gaussian effect on the halo mass function can reasonably well modelled by the analytic predictions, if a shape-dependent fudge factor is taken into account. In the case of the orthogonal type, there is a hint that $q$ is in addition redshift dependent.

\subsection{Halo bias}
In order to explicitly explore the dependence on the halo mass, we divide the halos into five mass bins: 
\begin{eqnarray}
3\times 10^{13}\,\MSUN/h<& M_{\rm halo} &< \ \ \, 6\times 10^{13}\,\MSUN/h  \nonumber \\
6\times 10^{13}\,\MSUN/h<& M_{\rm halo} &< 1.2\times 10^{14}\,\MSUN/h  \nonumber \\
1.2\times 10^{14}\,\MSUN/h<& M_{\rm halo} &< 2.4\times 10^{14}\,\MSUN/h \nonumber \\
2.4\times 10^{14}\,\MSUN/h<& M_{\rm halo} &< 4.8\times 10^{14}\,\MSUN/h \nonumber \\
4.8\times 10^{14}\,\MSUN/h<& M_{\rm halo} &< 9.6\times 10^{14}\,\MSUN/h \nonumber \,.
\end{eqnarray}
The number of halos per mass bin decreases with mass. In particular, at high redshift and for large masses the exponential fall-off in the mass function reduces the number of halos per bin quickly. In the following, we only consider mass bins with more than $250$ halos.

In order to keep the statistical error as small as possible in our analysis, we compute and fit the effect of the non-Gaussian halo bias in the following way. First, we compute the difference between the non-Gaussian and Gaussian bias from the corresponding N-body simulations:
\begin{equation}
\Delta b(k) = \frac{P_{\rm hm}^{NG}(k)}{P_{\rm m}^{NG}(k)} - \frac{P_{\rm hm}^{G}(k)}{P_{\rm m}^G(k)}\,.
\end{equation}

\begin{figure}[htb]
\begin{center}
\includegraphics[angle=0,width=0.8\textwidth]{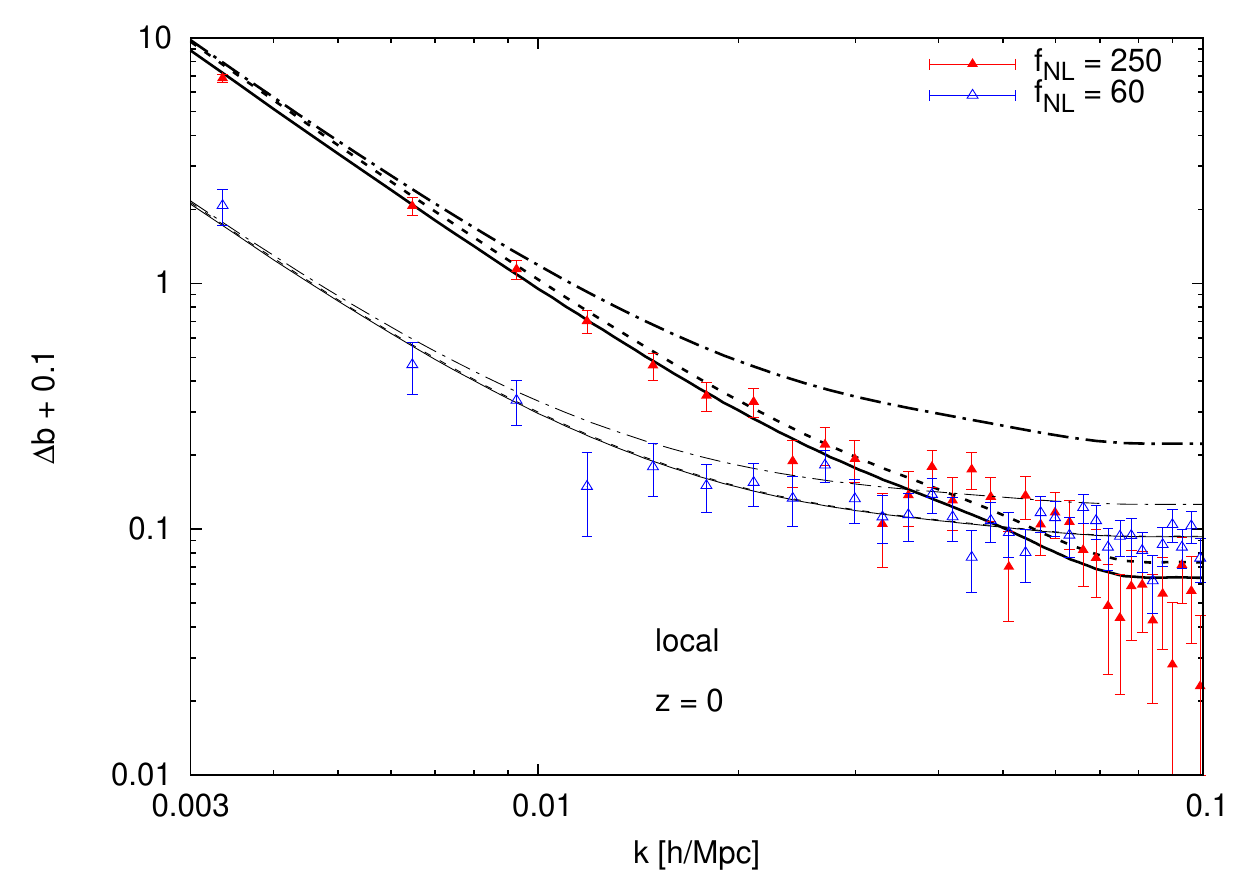}
\end{center}
\caption
{Local non-Gaussian bias of halos with mass $1.2\times 10^{14}\,{\rm Mpc}/h< M <2.4\times 10^{14}\,{\rm Mpc}/h$ at $z=0$. The difference of the bias measured from the non-Gaussian simulations and Gaussian simulations is depicted by the data points (for computation of the error bars, see App.~\ref{app:fitting}). The solid lines show the best fit using the model given in Eq.~\ref{eq:bias_fit}. The dot-dashed lines show the model predictions when the scale-independent bias shift, $\Delta b_I$, is neglected. The short-dashed lines neglect the term which is non-linear in $\fnl$ (see text for details). Thick lines and red symbols correspond to $\fnl=250$, while thin lines and blue symbols show the results for $\fnl=60$. Note that we actually show $\Delta b + 0.1$ to allow for a logarithmic scale.
}
\label{fig:ex_bias_loc}
\end{figure}

As the non-Gaussian and the Gaussian simulation share the same realization of the initial Gaussian field, $\Delta b(k)$ is almost free of sample variance and, in addition, has smaller shot noise than $b^{NG}(k)$ and $b^{G}(k)$ individually. We estimate the error on $\Delta b(k)$ directly from the distribution of $\Delta b(\vk)$ in each $k$ bin (for details, see App.~\ref{app:fitting}).

Following Eq.~\ref{eq:NG_bias}, we model $\Delta b(k)$  by
\begin{equation}\label{eq:bias_fit}
\Delta b(k) = \Delta b_I + \fnl\left[b_1^G+\Delta b_I-1\right]\frac{q \delta_c}{D(z)}\frac{\mathcal{F}_M(k)}{\mathcal{M}_M(k)}\,,
\end{equation}
where $b_1^G$ is the linear halo bias obtained from the Gaussian simulation on large scales (see App.~\ref{app:fitting}). The scale-independent shift, $\Delta b_I$, can be predicted by the difference in the linear bias derived from the non-Gaussian and Gaussian mass functions using the peak-background split approach \cite{desjacques2009}
\begin{equation}\label{eq:bI}
 \Delta b_I = b_1^{NG}-b_1^G = - \frac{\partial \ln R^{NG}(M)} {\partial \delta_c}\,,
\end{equation}
with $R^{NG}(M)$ being the ratio of the non-Gaussian to the Gaussian mass function. 
Here, however, we treat $\Delta b_I$ as a free parameter and compare it later on with the prediction derived from the mass functions.
We choose $q$ as the second free parameter. All other quantities in Eq.~(\ref{eq:bias_fit}) are computed from the theory and are kept fixed.

In Fig.~\ref{fig:ex_bias_loc}, we show as an example the effect of local non-Gaussianity on the halo bias for halos of mass $1.2-2.4\times 10^{14}\,\MSUN/h$ at $z=0$. 
Note that we plot $\Delta b(k) + 0.1$. As $\Delta b_I$ is negative, this addition of $0.1$ is needed to still make use of the logarithmic scale.

The different line types visualize the effect of the different terms in Eq.~(\ref{eq:bias_fit}). The solid lines show the best fit to the data (using all modes up to $k_{\rm max}=0.1\,{\rm Mpc}/h$) and include all terms given above. 
The short-dashed lines neglect $\Delta b_I$ appearing inside the square brackets in Eq.~(\ref{eq:bias_fit}). The inclusion of this term makes the non-Gaussian bias non-linear in $\fnl$ \cite{giannantonio2010,smith2011}, since $\Delta b_I$ depends on $\fnl$. 
The dot-dashed lines neglect $\Delta b_I$ completely. This scale-independent bias shift becomes important on smaller scales ($k>0.02$), for which the scale-dependent part becomes small.

\begin{figure}[htb]
\begin{center}
\includegraphics[angle=0,width=0.8\textwidth]{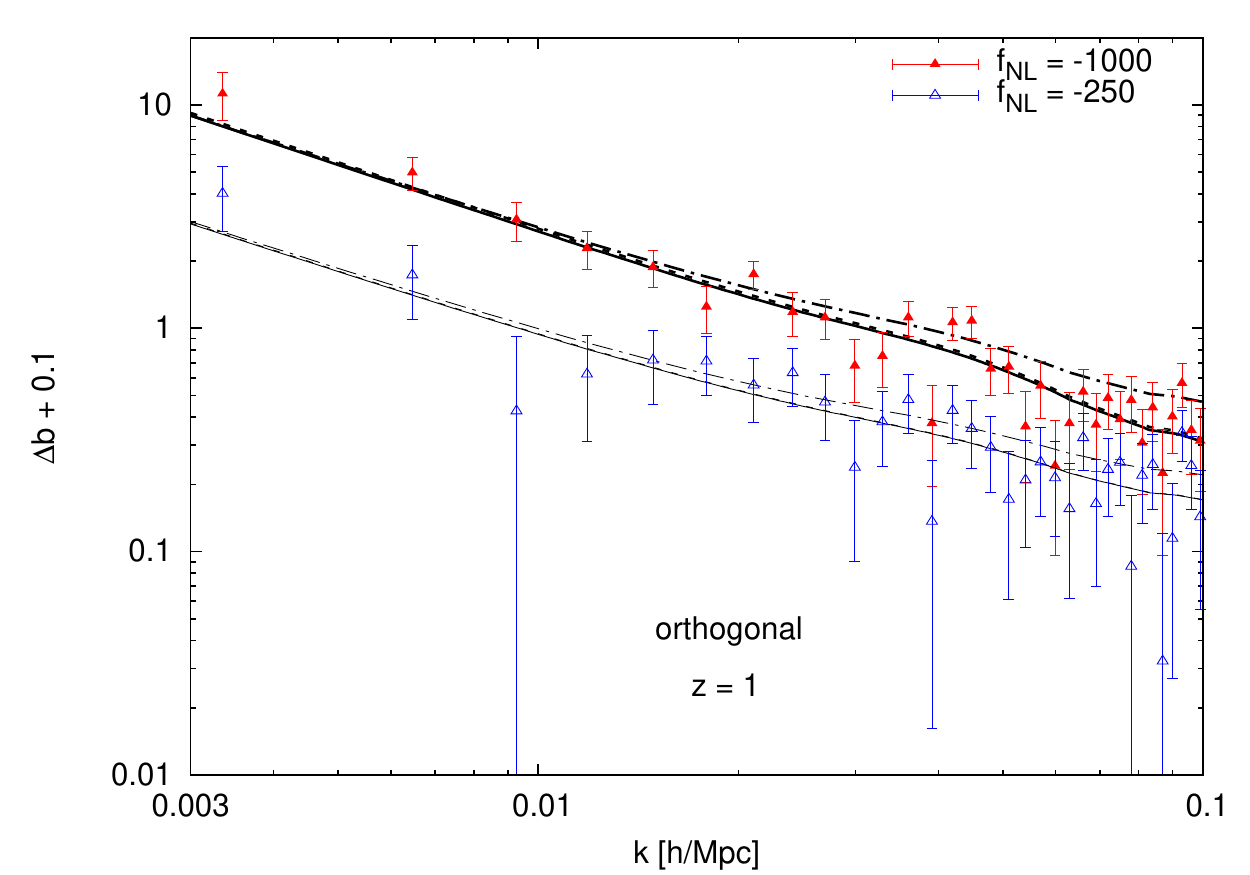}
\end{center}
\caption
{Same as in Fig.~\ref{fig:ex_bias_loc}, but for the orthogonal shape of non-Gaussianity. Note, however, that here the redshift of the halos with mass $1.2\times 10^{14}\,{\rm Mpc}/h< M <2.4\times 10^{14}\,{\rm Mpc}/h$ is $z=1$.
}
\label{fig:ex_bias_ort}
\end{figure}

An example of the measured non-Gaussian bias from the simulations of the orthogonal type is given in Fig.~\ref{fig:ex_bias_ort}. 
Here, the halos have again a mass of $1.2-2.4\times 10^{14}\,\MSUN/h$, but were found in simulation outputs at $z=1$.
Consequently, the number of halos is smaller than in the previous figure and the residual shot noise is larger.
The line types have the same meaning as before. 
On large scales, the halo bias scales as $\sim k^{-1}$ as predicted by the theory. Hence, with increasing wavenumber, the effect does not drop as rapidly as in the local case and extends to smaller scales.

\begin{figure}[htb]
\begin{center}
\includegraphics[angle=0,width=0.8\textwidth]{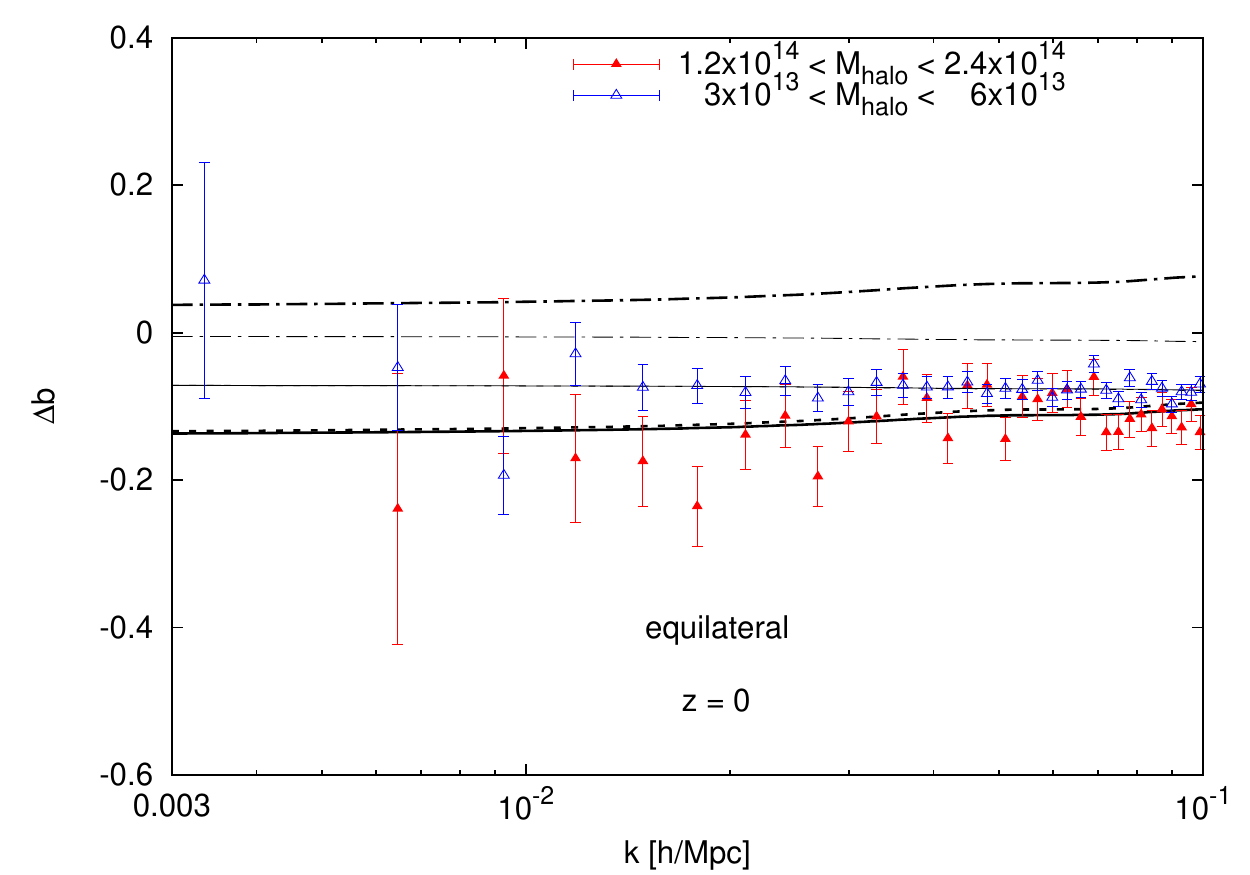}
\end{center}
\caption
{Same as in Fig.~\ref{fig:ex_bias_loc}, but for the equilateral shape of non-Gaussianity. Note the linear scale of the y-axis.
}
\label{fig:ex_bias_eql}
\end{figure}

Next, we present an example of the non-Gaussian halo bias induced by the equilateral shape. In Fig.~\ref{fig:ex_bias_eql}, the halo bias corresponding to two different mass bins at $z=0$ is shown. As expected, the scale dependence is very weak and in agreement with the theoretical predictions. In particular, the observed mass dependence of the effect is consistent with the model predictions.

\begin{figure}[htb]
\begin{center}
\includegraphics[angle=0,width=0.49\textwidth]{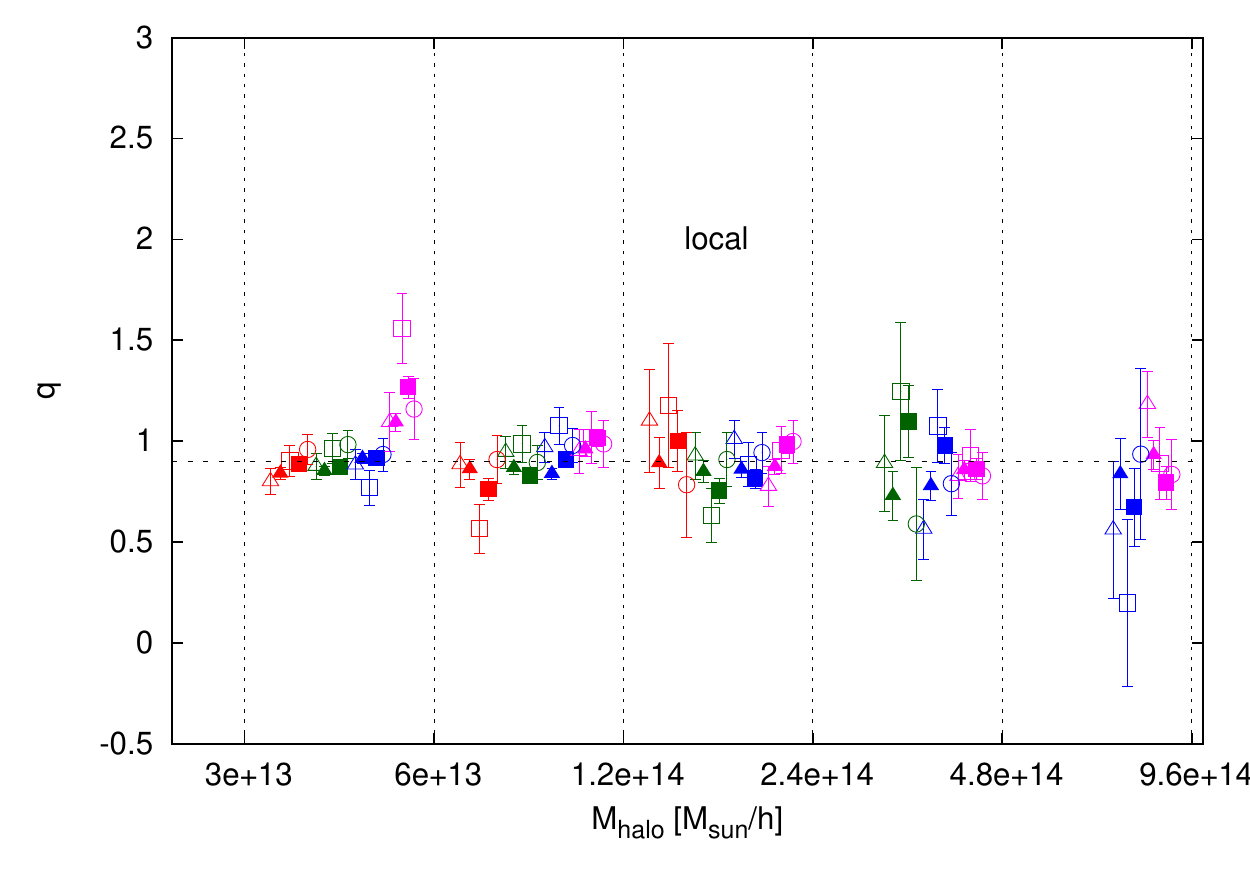}
\includegraphics[angle=0,width=0.49\textwidth]{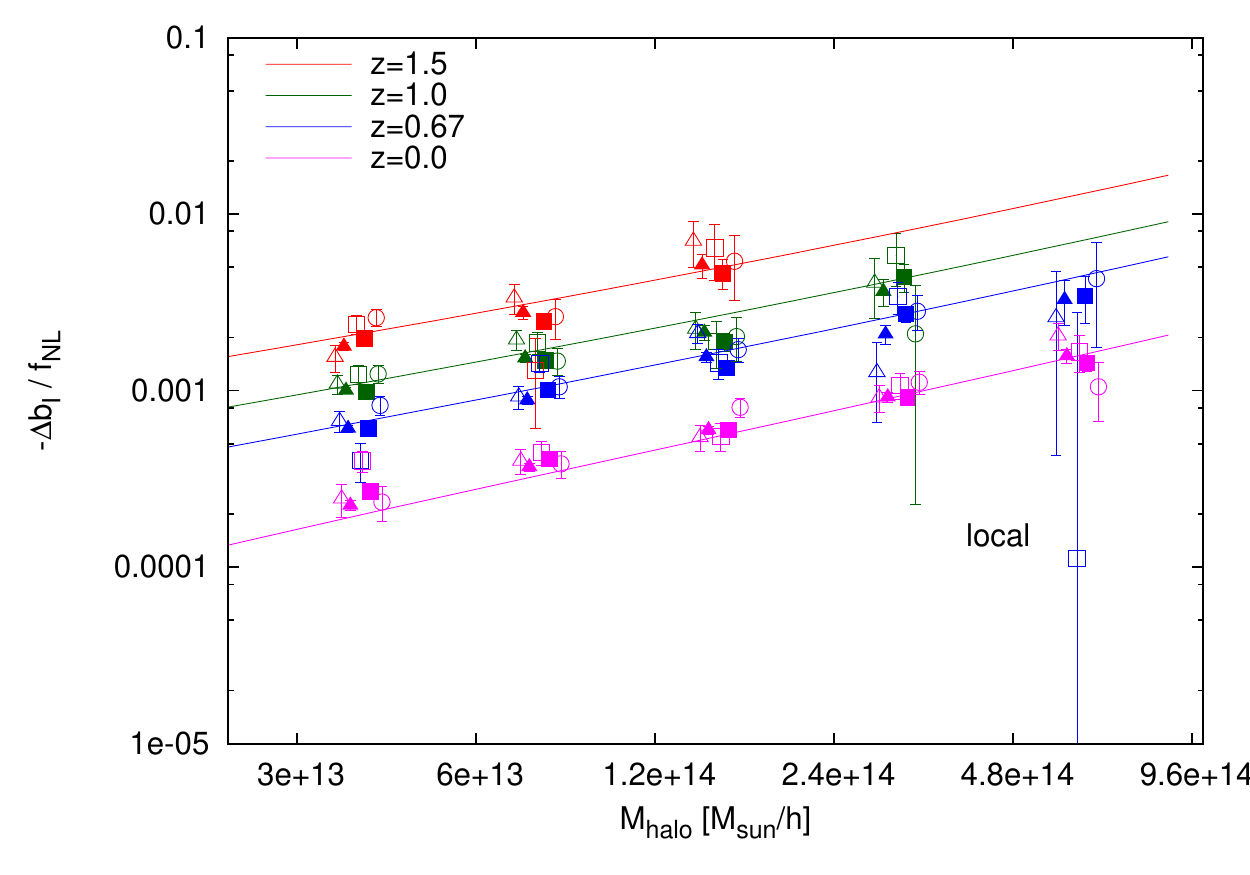}
\includegraphics[angle=0,width=0.49\textwidth]{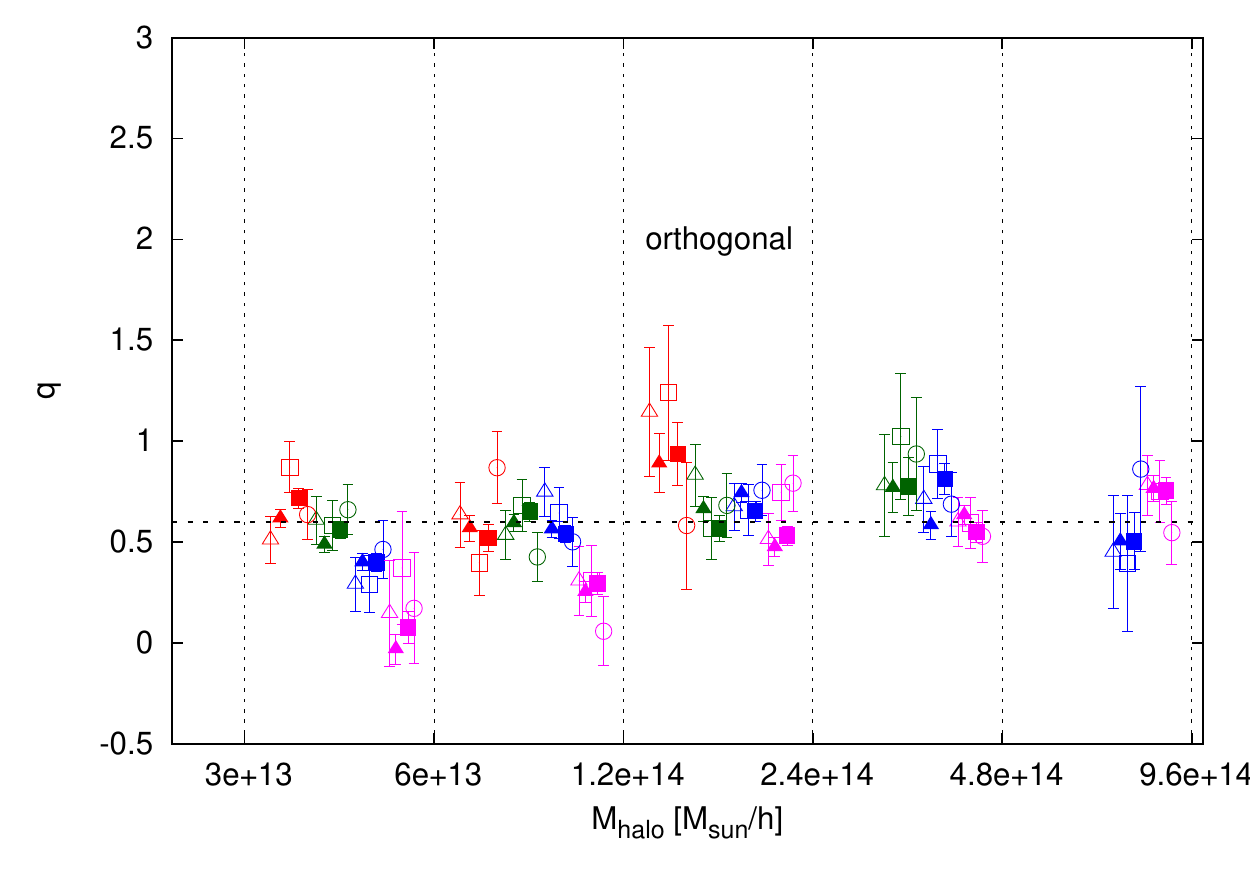}
\includegraphics[angle=0,width=0.49\textwidth]{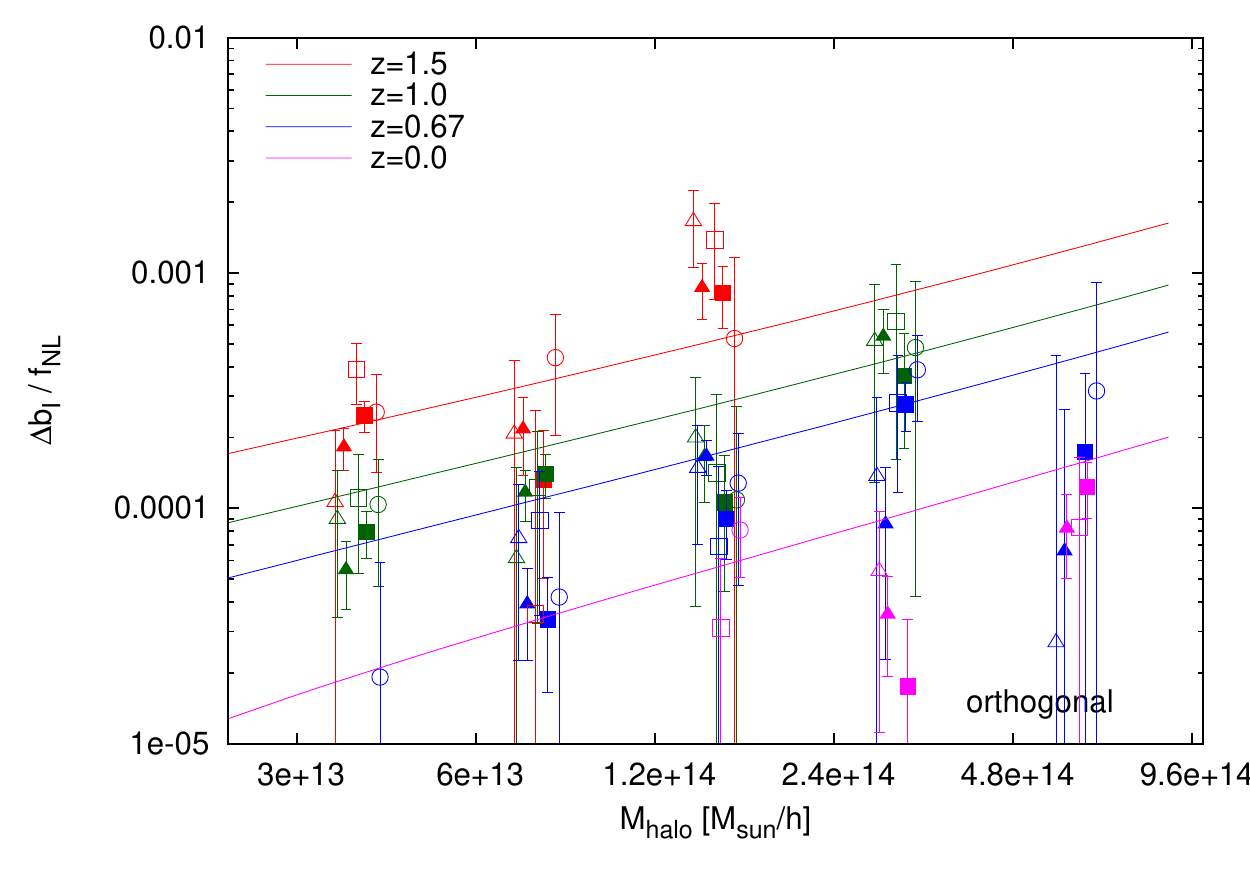}
\includegraphics[angle=0,width=0.49\textwidth]{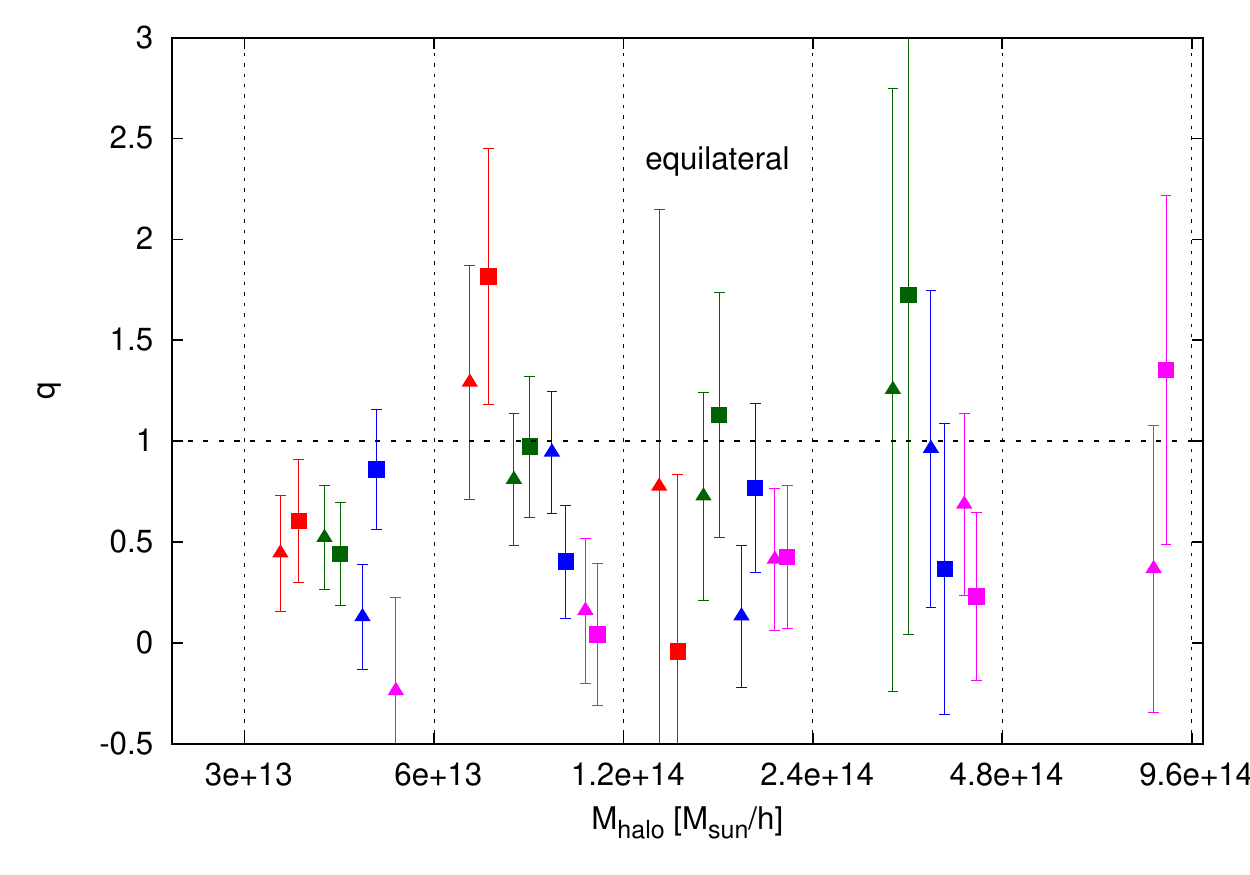}
\includegraphics[angle=0,width=0.49\textwidth]{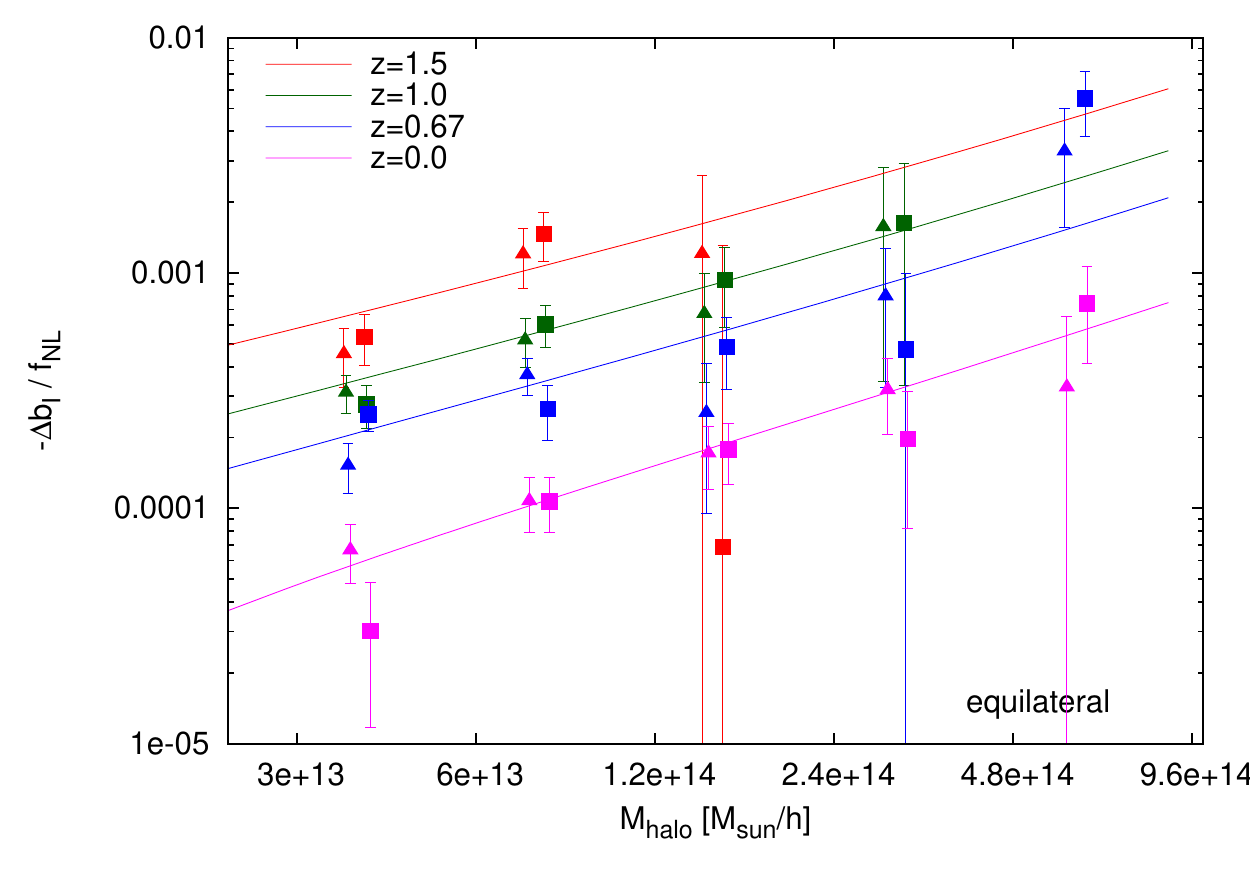}
\end{center}
\caption
{Best-fit values (fitting all modes with $k<0.1\,{\rm Mpc}/h$) of the fudge factor $q$ (left) and the scale-independent shift, $\Delta b_I$, normalized by $\fnl$ (right). Different colours correspond to different redshifts (see key in right panels). The symbol coding is the same we used in the mass function plots (Fig. \ref{fig:mf_loc}, \ref{fig:mf_ort}, and \ref{fig:mf_eql}), see text for details. 
On the left-hand side, all points in between two vertical dashed lines correspond to the same halo mass bin, but show different redshifts (coulor), realizations (symbol shape), and $\fnl$ values (open/filled symbol). The horizontal dashed lines show the $q$ values estimated from the LV non-Gaussian mass function.
The coloured solid lines on the right-hand side show the predictions for the scale-independent bias shift using Eq.~\ref{eq:bI}.
}
\label{fig:fit_q_a0}
\end{figure}

After having discussed for each type of non-Gaussianity typical examples, we show the complete set of best-fit values of the fitting parameters in Fig.~\ref{fig:fit_q_a0}. On the left-hand side, the best-fit values of the fudge factor $q$ are presented. Different colours correspond to different redshifts: $z=1.5$ (red), $z=1$ (green), $z=0.67$ (blue), and $z=0$ (magenta). Triangles, boxes, and circles depict the three different realizations of the initial Gaussian random field used for the generation of the initial conditions. In the case of the local and orthogonal type, the open symbols correspond to $\fnl=60$ and $\fnl=-250$, respectively. Filled symbols show the results for $\fnl=250$ (local), $\fnl=-1000$ (orth.), and $\fnl=1000$ (eql.). For clarity, the points of each mass bin are spread over the range of each mass bin.

For the local type of non-Gaussianity, we recover within the error bars the same $q$ value, which was needed to bring the mass functions in agreement with analytic predictions. 
For the orthogonal shape, the effect is suppressed by a factor which is similar to the one we found for the non-Gaussian mass function, $q\sim 0.6$. This suppression of the non-Gaussian halo bias effect is clearly redshift and halo mass dependent. 
In the case of the equilateral type of non-Gaussianity, the error bars are as expected very large. Nevertheless, also in this case there is a hint of a redshift and halo mass dependence of the effect. Note, however, that this suppression visible for the non-Gaussian halo bias effect for the equilateral shape was not found for the effect on the mass function, for which the measured fudge factor was $q\sim 1$. 
These findings are very interesting and ---if solidified by larger simulations--- may help to lead to a better theoretical understanding of the halo biasing (see discussion in \cite{schmidt2010}).

On the right-hand side of Fig.~\ref{fig:fit_q_a0}, the best-fit values of $\Delta b_I$ normalized by $\fnl$ are shown. The colour and symbol coding is the same as before. As open and filled symbols (corresponding to different $\fnl$ values) are consistent with each other, we can infer that the scale-independent shift is linear in $\fnl$ for the $\fnl$ values probed.
The solid lines represent the predictions from Eq.~(\ref{eq:bI}) using the LV mass function ratio, $R_{\rm LV}^{NG}(M)$, and taking the measured fudge factor $q$ from the mass function into account. Keeping in mind that the LV mass function is not in all cases a good fit to the mass functions derived from our simulations (see for example bottom right panel of Fig.~\ref{fig:mf_ort}), the agreement between the best-fit values and the predicted bias shift is very reasonable.

%%%%%%%%%%%%%%%%%%%%%%%%%%%%%%%%%%%
\section{Conclusions}
\label{conclusions}
N-body simulations are an indispensable tool to test, develop and
calibrate any statistical analysis of large-scale structure.
In this paper, we have further explored the issue of setting up generic
non-Gaussian initial conditions for N-body simulations.  In a previous
paper \cite{wagner2010} we began addressing this issue, focussing ---as
we do here--- on non-Gaussianities specified by a non-zero bispectrum.
In \cite{wagner2010} we focussed  on the non-Gaussian corrections to
the  halo mass function and to the non-linear matter power
spectrum, which, for the  scales and redshfits  of interest, could be
reliably computed from simulation boxes of $600\,{\rm Mpc}/h$ on a side. Here we
concentrate on the large-scale non-Gaussian halo bias. The
scale-dependent bias is evident on very large scales thus needs
multiple simulations of  much larger boxes to be measured reliably for
$\fnl$ values not too large as to push beyond the regime of validity
of the analytical description of the effect.

The scale-dependent non-Gaussian halo bias has been recognized to be a
very  competitive approach to constrain primordial non-Gaussianity
(e.g., \cite{slosar2008,carbone2008,carbone2010,giannantonio2010})
yielding forecasted error bars on the local non-Gaussian parameter of
$\Delta \fnl^{\rm loc} \sim 1$ which makes this approach formally better than an ideal CMB
experiment \cite{yadav2010}.
So far this effect has been explored with  N-body simulations  only
for the so-called local type of non-Gaussianity. However, the shape of
non-Gaussianity is a window into the physical mechanism of inflation
and the generation of primordial perturbations, and it is of value to
be able to simulate initial conditions for general bispectrum shapes.

The technique presented in \cite{wagner2010} was not suitable to study
the halo bias because, in general, higher-order contributions would leave
an artificial signature on the large-scale power spectrum (which are the scales
where the scale dependence of the halo bias is evident). These
components could be kept under control only in specific cases. Here
the issue is solved and these components are always under control.
This however comes at the expense of computational cost: the new
expressions are not separable (even if the bispectrum itself is)
meaning that computational short-cuts that could be implemented  for
separable bispectra before, now cannot be applied in the same way.

Nevertheless we find  a speed-up solution which involves  sampling
only the Gaussian contribution to the potential  with a full
resolution grid
 and using a smaller grid for the non-Gaussian part. We assess the
performance  of this approach for the local case, where the
computation
can be performed in real space and is not computationally intensive.
By using a smaller grid size for the non-Gaussian part of the initial conditions, 
we can keep the computational time taken by the calculation of the initial condition comparable 
to the time needed to run the simulation.

The form of the bispectrum predicted by inflationary models can be
complicated and in general is not-factorizable. As factorization is
very important for an efficient analysis of CMB data, physical
bispectrum shapes have been approximated by   factorizable templates.
We highlight here that, since the halo bias is dominated by squeezed
configurations,  what is a good template for CMB analysis may not be
correct for the halo bias.
We show that the equilateral template results in roughly the correct
scale dependence of the halo bias  for the corresponding physical
models, but  with a  shift in amplitude; if unaccounted for, it could
introduce  a bias on the estimated $f_{\rm NL}$ of up to 50\%.
Fortunately, this shift can be quantified well.
Enfolded and orthogonal templates on the other hand do not work as well. They both lead to a non-Guassian bias that scales as $k^{-1}$ on large scales. This is different from the halo bias induced by the corresponding physical models: 
  The scale dependence of the halo bias caused by
modified-initial-state type of non-Gaussianity  is almost degenerate
with the one caused by the local type of non-Gaussianity. The halo
bias effect of the  orthogonal physical shape is very similar to the 
effect of the equilateral shape.

We performed several simulations with non-Gaussian initial conditions
of different shapes to further test and calibrate the analytic
predictions.
For this purpose we used the standard  templates: these are useful toy
models as they span a range of scalings in the squeezed regime. If the
analytical expressions can correctly reproduce the bias behaviour for
shapes that have such different $k$-dependence in the squeezed limit,
this  lends  support to their validity.

Revisiting the  mass function predictions, we find that  the
non-Gaussian effect on the halo mass
function can be well modelled by the analytic predictions, if a
shape-dependent fudge factor, $q$, is taken into account. In the case
of the orthogonal type, there is a hint that this factor depends
 on redshift and halo mass.

For the halo bias we also find that the analytic predictions work well
once  a  a shape-dependent normalization factor $q$ is included: $q$
effectively re-scales $f_{\rm NL}$.
Interestingly, for the non-local types of non-Gaussianity, this fudge factor varies with redshift and halo mass.

Throughout we pay particular attention to the operational definition
of non-Gaussian bias, taking into account that a non-zero $\fnl$
modifies the mass function and thus modifies also the linear bias. However, a
Gaussian distribution with the same mass function would have a linear halo
bias modified in the same way.

In most inflationary models considered so far, the bispectrum scales either as the equilateral shape ($\sim k^{-1}$)
or as the local shape ($\sim k^{-3}$) in the squeezed limit (see, however, the quasi-single field models of \cite{chen2010a,chen2010b}, which have intermediate scalings). 
The equilateral shape  leads to
a small and almost scale-independent halo bias and is thus not easily
accessible with this technique.
Other shapes that, on large scales, have the same scale dependence as
the local one, may in principle be distinguished through the halo bias
dependence on halo mass on intermediate scales.
On large scales, however, the limits for the local $f_{\rm NL}$ can be
re-interpreted as limits on e.g., modified-initial-state type of
non-Gaussianity, by an appropriate rescaling of $\fnl$: 
$\fnl^{\rm modin.}\simeq 8 \fnl^{\rm loc}$.

\section*{Acknowledgements}
We would like to thank Lotfi Boubekeur for useful discussions. 
The simulations and computations were performed on the cluster Hipatia (UB computing facilities and ERC grant FP7- IDEAS Phys.LSS 240117).
We are grateful to the Centro de Ciencias de Benasque Pedro Pascual where the very last stages of the work were carried out. 
CW is supported by  MICINN grant AYA2008-03531. 
LV acknowledges support from FP7-PEOPLE-2007-4-3-IRG n. 202182 and FP7- 
IDEAS Phys.LSS 240117 and  MICINN grant AYA2008-03531.

\appendix
\section{Simulations of the equilateral type using two different approaches}
\label{app:eql_comp}
In this section we compare the results of two N-body simulations with $\fnl=1000$ non-Gaussian initial conditions of the equilateral type. Both simulations have a box size of $1875\,{\rm Mpc}/h$ and a particle load of $1024^3$. The simulations were started at $z_{\rm initial}=49$ and carried out with Gadget-2 using the same numerical settings. In both cases, halos were identified with AHF \cite{knollmann2009}. The only difference between the simulations is the different prescription and implementation how the non-Gaussian part of the initial gravitational potential was generated. In the first case, ``equilateral a)'', we used our modified ansatz for $\Phi_\vk^{NG}$, Eq.~(\ref{eq:new_ansatz}), and a grid size of $400$ and $1024$ for the non-Gaussian and Gaussian part, respectively. The initial conditions of the second simulation, ``equilateral b)'', were generated using our original ansatz, Eq.~(\ref{eq:ansatz}). The integral appearing in this ansatz is a sum of convolutions and can be computed efficiently with the help of FFTs. Hence, in this case, we were able to use an equal large grid size of $1024$ for the non-Gaussian part, too. In both cases, we use the same realization of the Gaussian random field.
Note although the two approaches are different at the level of $\Phi_\vk$, they both produce a non-Gaussian field, which has, at the leading order, the same power spectrum and bispectrum. However, high-order spectra, e.g. the trispectrum, are different.

\begin{figure}[htb]
\begin{center}
\includegraphics[angle=0,width=0.8\textwidth]{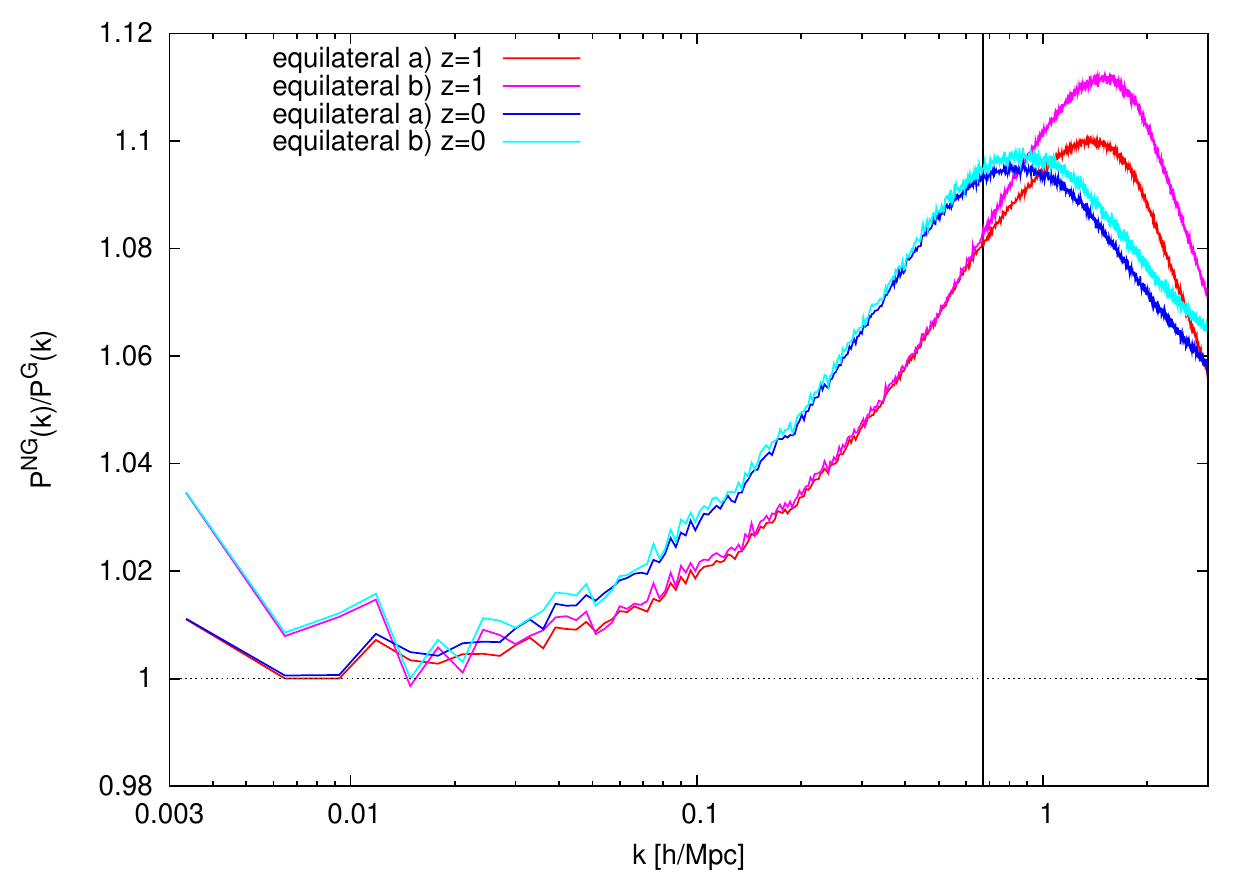}
\end{center}
\caption
{The ratio of the non-linear power spectrum obtained from the two non-Gaussian simulations to the Gaussian power spectrum at redshift $z=1$ and $z=0$. The black vertical line shows the Nyquist frequency, up to which the primordial non-Gaussian field $\Phi_\vk^{NG}$ is sampled in ``equilateral a)''.
}
\label{fig:Pk_eql}
\end{figure}

In the remainder of this section, we consider observables which are affected by primordial non-Gaussianity, and analyse if the different ways of setting up the initial conditions lead to differences in these quantities. We start with the non-linear matter power spectrum. In Fig.~\ref{fig:Pk_eql}, we show the ratio of the power spectrum measured from the two non-Gaussian simulations to the power spectrum obtained from a Gaussian simulation with the same realization. The cyan/blue and magenta/red lines correspond to redshift $z=0$ and $z=1$, respectively. The vertical black line shows the Nyquist frequency of the $400^3$ grid used for the non-Gaussian part in ``equilateral a)'', which corresponds to the smallest scale still resolved by the grid.
For both redshifts, significant differences between the non-Gaussian simulations only appear on scales smaller than the smallest scale sampled by the non-Gaussian grid used in ``equilateral a)'', i.e.~on the right hand side of the vertical line. On the largest scales, the small number of modes leads to differences in the power spectra, since the term $\langle \Phi^G \Phi^{NG} \rangle \propto \langle \Phi^G \Phi^G \Phi^G \rangle$ does not completely vanish.
The very good agreement on the intermediate scales confirms the theoretical expectation, that the non-Gaussian amplification of the power spectrum is at leading order caused by the primordial bispectrum.

\begin{figure}[htb]
\begin{center}
\includegraphics[angle=0,width=0.8\textwidth]{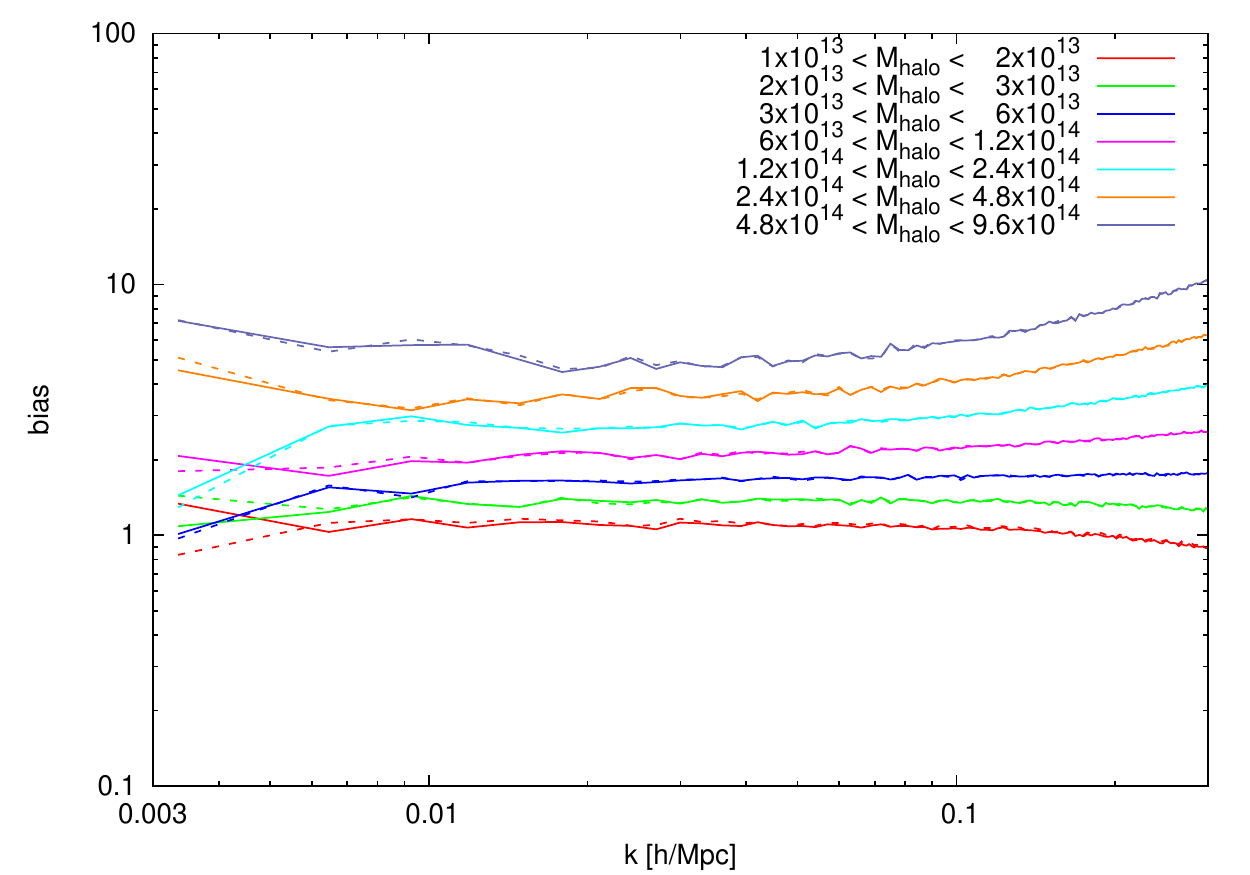}
\end{center}
\caption
{Halo bias for different halo masses in units of $\MSUN/h$ at redshift $z=0$. The dashed and solid lines show the results obtained from  the non-Gaussian simulations ``equilateral a)'' and ``equilateral b)'', respectively.
}
\label{fig:bias_eql}
\end{figure}

Next, we show in Fig.~\ref{fig:bias_eql} the halo bias for different halo masses at redshift $z=0$. The halo bias is computed from the ratio of the halo-matter cross power spectrum to the matter power spectrum, see Eq.~(\ref{eq:comp_bias}). The dashed and solid lines correspond to the simulations ``equilateral a)'' and ``equilateral b)'', respectively. As the effect of non-Gaussianity of the equilateral type on the halo bias is very weak (see Fig.~\ref{fig:NG_bias}), no effect of spatially unresolved non-Gaussianity, like we observed in the local case (see Fig.~\ref{fig:bias_loc_cut}), is visible here. The small differences between the two simulations can be ascribed to the differences in the shot noise.

\begin{figure}[htb]
\begin{center}
\includegraphics[angle=0,width=0.8\textwidth]{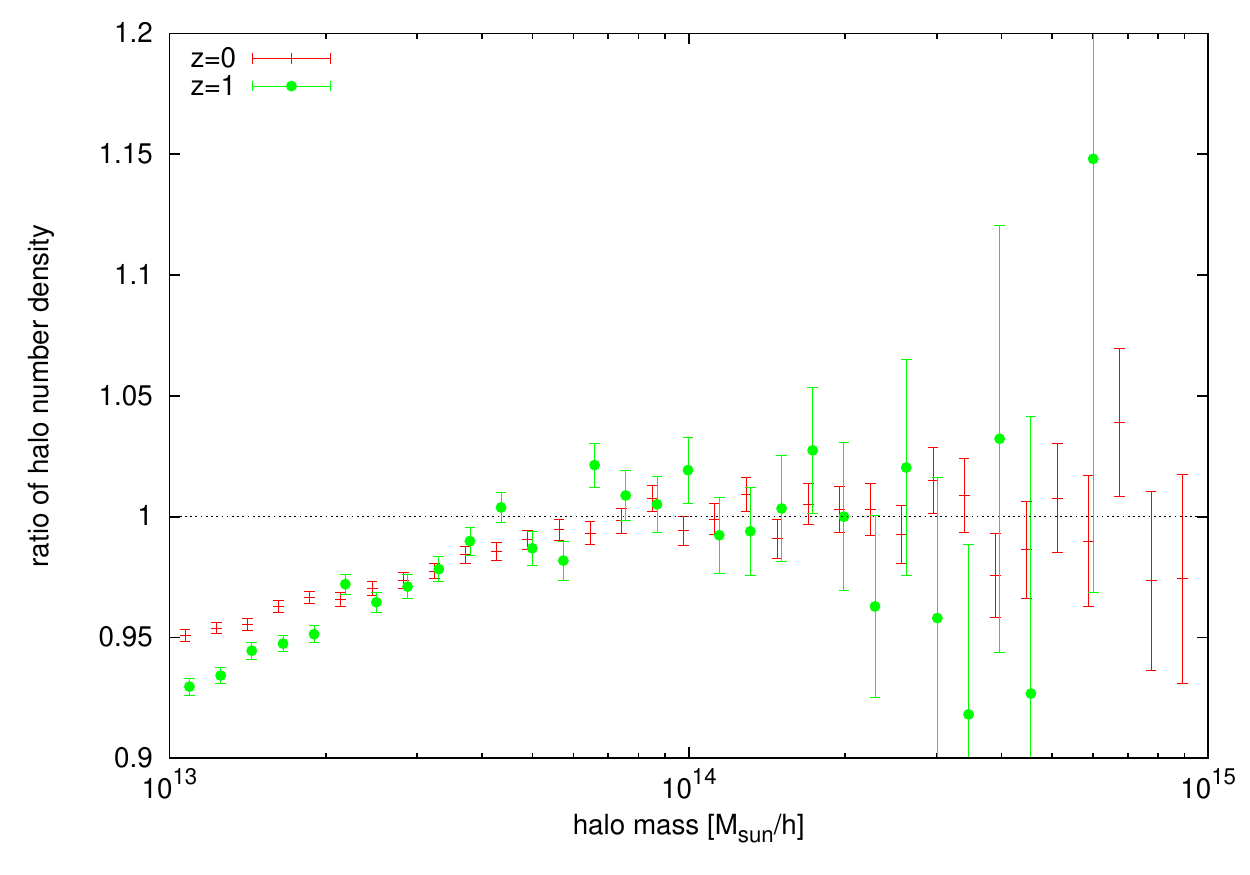}
\end{center}
\caption
{The ratio of mass function of halos found in the simulation ``equilateral a)'' and in the simulation ``euilateral b)'' at redshift $z=1$ and $z=0$. The error bars show the Poisson error. For clarity, the green circles are slightly shifted along the x-axis.
}
\label{fig:mf_eql_app}
\end{figure}

Lastly, we consider the halo mass function derived from the two non-Gaussian simulations. The ratio of the halo number densities measured from the simulations ``equilateral a)'' and ``equilateral b)'' as a function of halo mass is shown in Fig.~\ref{fig:mf_eql_app}. Similar to our test with the local type of non-Gaussianity (see Fig.~\ref{fig:mf_loc_cut}), the number of halos with mass below $~3\times10^{13}\,\MSUN/h$ is reduced by a few percent in the simulation for which a smaller non-Gaussian grid was used to set up the initial conditions. Halos with masses larger than $\sim 5\times 10^{13}\,\MSUN/h$ are not affected by the smaller grid size. Both, the magnitude of the effect and the affected mass range are a bit larger than in the local case, which is probably due to the larger effect on the halo mass function caused by the primordial non-Gaussianity with $\fnl^{\rm eql}=1000$ instead of $\fnl^{\rm loc}=250$ (see Fig.~\ref{fig:mf_loc} and Fig.~\ref{fig:mf_eql}).

As in the case of the non-linear matter power spectrum, no additional effects due to the differences in the initial non-Gaussian fields of ``equilateral a)'' and ``equilateral b)'' are noticeable. This strengthens the analytic modelling that the effects on the mass function due to non-Gaussianity are primarily caused by the primordial bispectrum. 

\section{Details on the fitting of the non-Gaussian bias}
\label{app:fitting}
We assume that the modes of the halo density field are related to the matter density by $\delta_{\rm h}=b(k)\delta_{\rm m}+n(k)$, where $n(k)$ is a noise variable which models the stochasticity in the halo formation process and the discrete nature of halos. The quantity which we want to measure from the simulations is the bias $b(k)$. 
Assuming that the noise is uncorrelated with the density the bias is given by 
\begin{equation}\label{eq:bias_est1}
b(k)=\frac{\left<\operatorname{Re}(\delta_{\rm h}\delta_{\rm m}^*)\right>} {\langle\left|\delta_{\rm m}\right|^2\rangle}\,.
\end{equation}
In practice, the average is taken over the finite number of modes per $k$ bin. 
The term $\langle \operatorname{Re}(n \delta_m^*)\rangle$ is therefore not exactly zero. Assuming $\operatorname{Re}(n \delta_m^*)$ has a Gaussian distribution with variance $\sigma_{n\delta}^2$, the error of $b(k)$ is then given by $\sigma_{n\delta}/\sqrt{N_{\rm modes}}/\langle|\delta_{\rm m}|^2\rangle$, where $N_{\rm modes}$ is the number of independent modes. 

\begin{figure}[htb]
\begin{center}
\includegraphics[angle=0,width=0.8\textwidth]{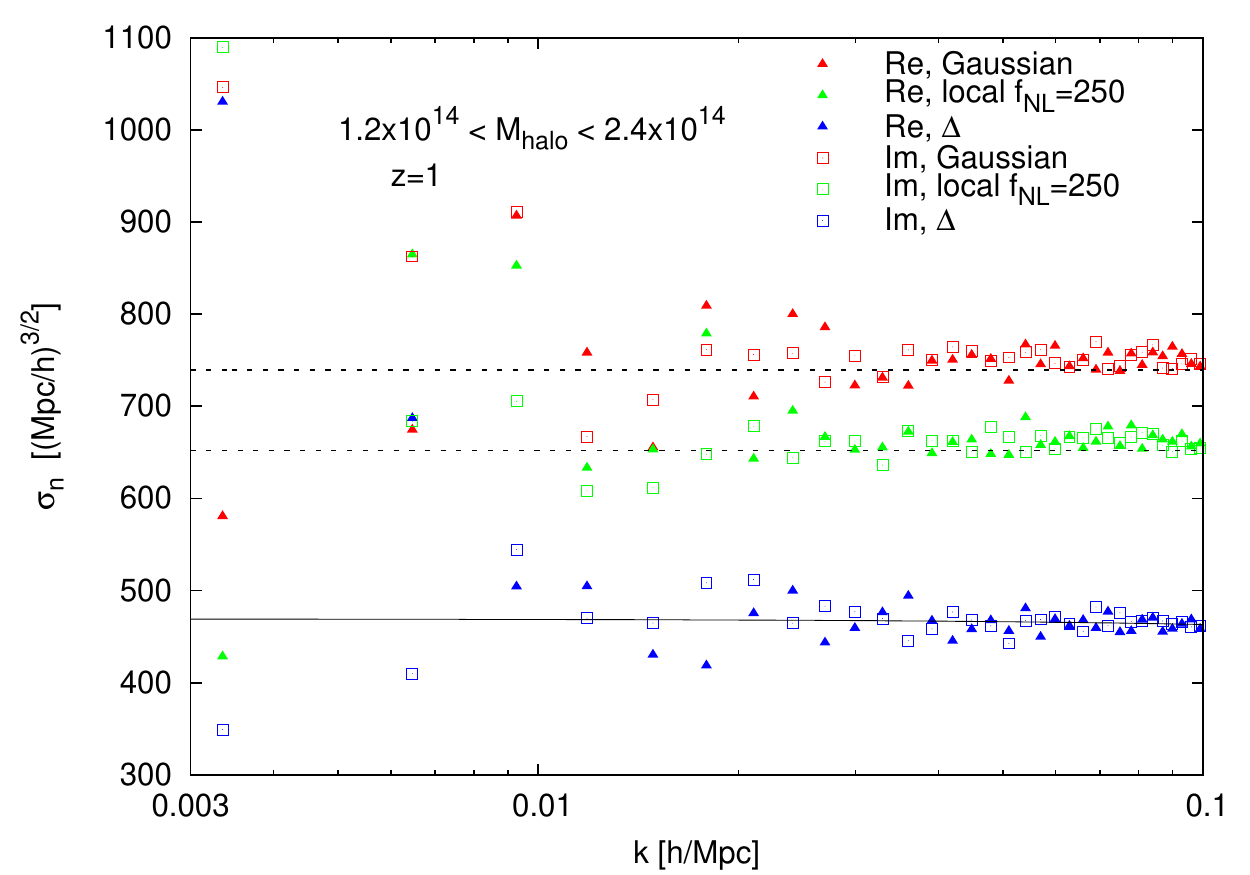}
\end{center}
\caption
{Variance of the distribution of $\operatorname{Re}(n \delta_m^*)/|\delta_{\rm m}|$ and $\operatorname{Im}(n \delta_m^*)/|\delta_{\rm m}|$ derived from N-body simulations for halos with mass of $1.2\times 10^{14} - 2.4\times 10^{14}\, \MSUN/h$ at $z=1$. Red and green symbols correspond to a Gaussian and non-Gaussian simulation of the local type with $\fnl=250$, respectively. The blue symbols show the noise derived from the distribution of $\Delta$ (see. Eq.~\ref{eq:Delta}). The dashed horizontal lines depict the Poisson noise prediction. The solid line shows the linear fit to the data.
}
\label{fig:noise}
\end{figure}
 
In Fig.~\ref{fig:noise}, we show the square root of the variance of the noise distribution for halos at $z=1$ with masses of $1.2\times 10^{14} - 2.4\times 10^{14}\, \MSUN/h$, measured from different simulations. Filled triangles and open squares depict the estimates derived from $\operatorname{Re}(n \delta_m^*)/|\delta_{\rm m}|$ and $\operatorname{Im}(n \delta_m^*)/|\delta_{\rm m}|$, respectively. The dashed lines show the Poisson noise prediction $\sigma_{\rm Poisson}=\sqrt{1/(2n_{\rm h})}$, where $n_{\rm h}$ is the halo number density. For the halos shown, the measured noise is in good agreement with the Poisson noise prediction. However, in general, depending on the halo mass and redshift, the actual noise can be smaller or larger than the Poisson noise and can also be scale-dependent (e.g., \cite{gil2010}). 

As the Gaussian and non-Gaussian simulations share the same realization of the Gaussian field, we expect that in the quantity
\begin{equation}\label{eq:Delta}
\Delta = \frac{\operatorname{Re}(\delta^{NG}_{\rm h} {\delta_m^{NG}}^*)}{|\delta^{NG}_{\rm m}|} - \frac{\operatorname{Re}(\delta^G_{\rm h} {\delta_m^{G}}^*)}{|\delta^{G}_{\rm m}|}
\end{equation}
the corresponding noise terms $\operatorname{Re}(n^{NG} {\delta_m^{NG}}^*)/|\delta^{NG}_{\rm m}|$ and $\operatorname{Re}(n^G {\delta^G_m}^*)/|\delta^G_{\rm m}|$ cancel each other partly. 
This is indeed what we find in the simulation data. The blue symbols in Fig.~\ref{fig:noise} represent $\sigma_n$ derived from the distribution of $\Delta$ in each $k$-bin. The solid line is a linear fit to the data.
When we fit the non-Gaussian halo bias models to the simulation data, we use this fit as our noise estimate. 
The resulting $\chi^2/\rm d.o.f.$ values are close to one.

\begin{figure}[htb]
\begin{center}
\includegraphics[angle=0,width=0.8\textwidth]{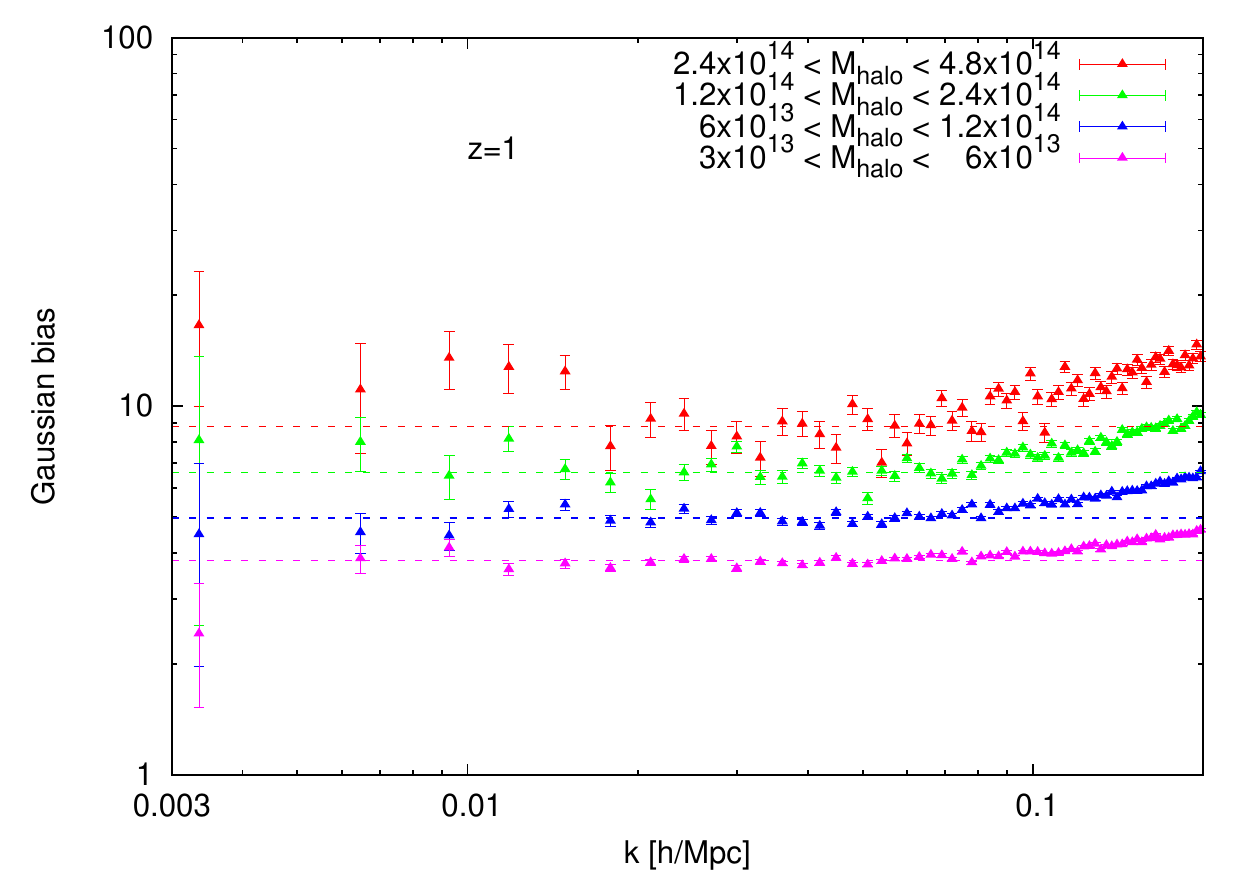}
\end{center}
\caption
{Gaussian halo bias for different halo mass bins measured from a Gaussian simulation at $z=1$. The dashed lines show the corresponding best fit of the linear bias.
}
\label{fig:lin_bias}
\end{figure}

Another ingredient in our fitting procedure of the non-Gaussian halo bias is the Gaussian linear bias, $b_1^G$. We estimate the linear bias by fitting the bias derived from the Gaussian simulations on large scales. In Fig.~\ref{fig:lin_bias}, the bias measurements for halos of different masses at redshift $z=1$ are shown. The dashed lines depict the corresponding  best fits of the linear bias. Only data points on large scales, $k\le0.07$, were included in the fitting. 

The remaining quantities needed for the fitting of the non-Gaussian bias (see Eq.~(\ref{eq:bias_fit})), like the growth function, the form factor, etc., are computed numerically for the given redshift, type of non-Gaussianity, and halo mass\footnote{For each halo mass bin, we compute the mass dependent $\mathcal{F}_M(k)$ and $\mathcal{M}_M(k)$ for a single mass value, $M_{\rm bin}$, given by 1.3 times the lower mass bound. We checked that this is a good choice by splitting the mass bin in ten smaller bins. The weighted average of $\mathcal{F}_M(k)/\mathcal{M}_M(k)$ is in good agreement with $\mathcal{F}_{M=M_{\rm bin}}(k)/\mathcal{M}_{M=M_{\rm bin}}(k)$.}.

Having all quantities at hand, we fit the non-Gaussian bias derived from the simulation data with the model given in Eq.~(\ref{eq:bias_fit}) out to wavenumbers $k<k_{\rm max}$. We use two different fitting ranges, a conservative $k_{\rm max}=0.03\,{\rm Mpc}/h$ and a more ambitious $k_{\rm max}=0.1\, {\rm Mpc}/h$. Both fitting ranges yield results consistent with each other. However, in the case of the equilateral and orthogonal shape, the errors on the fitting parameters are significantly smaller (approximately by a factor of ten and two, respectively) when the larger fitting range is used. For the local type of non-Gaussianity, the larger fitting range reduces the error by approximately $30\%$.

\bibliographystyle{JHEP}
\bibliography{ng2}

\providecommand{\href}[2]{#2}\begingroup\raggedright\begin{thebibliography}{10}

\bibitem{komatsu2011}
E.~{Komatsu}, K.~M. {Smith}, J.~{Dunkley}, C.~L. {Bennett}, B.~{Gold},
  G.~{Hinshaw}, N.~{Jarosik}, D.~{Larson}, M.~R. {Nolta}, L.~{Page}, D.~N.
  {Spergel}, M.~{Halpern}, R.~S. {Hill}, A.~{Kogut}, M.~{Limon}, S.~S. {Meyer},
  N.~{Odegard}, G.~S. {Tucker}, J.~L. {Weiland}, E.~{Wollack}, and E.~L.
  {Wright}, {\it {Seven-year Wilkinson Microwave Anisotropy Probe (WMAP)
  Observations: Cosmological Interpretation}},  {\em \apjs} {\bf 192} (Feb.,
  2011) 18--+, [\href{http://xxx.lanl.gov/abs/1001.4538}{{\tt
  arXiv:1001.4538}}].

\bibitem{maldacena2003}
J.~{Maldacena}, {\it {Non-gaussian features of primordial fluctuations in
  single field inflationary models}},  {\em Journal of High Energy Physics}
  {\bf 5} (May, 2003) 13--+,
  [\href{http://xxx.lanl.gov/abs/astro-ph/0210603}{{\tt astro-ph/0210603}}].

\bibitem{acquaviva2003}
V.~{Acquaviva}, N.~{Bartolo}, S.~{Matarrese}, and A.~{Riotto}, {\it
  {Gauge-invariant second-order perturbations and non-Gaussianity from
  inflation}},  {\em Nuclear Physics B} {\bf 667} (Sept., 2003) 119--148,
  [\href{http://xxx.lanl.gov/abs/astro-ph/0209156}{{\tt astro-ph/0209156}}].

\bibitem{bartolo_review}
N.~{Bartolo}, E.~{Komatsu}, S.~{Matarrese}, and A.~{Riotto}, {\it
  {Non-Gaussianity from inflation: theory and observations}},  {\em \physrep}
  {\bf 402} (Nov., 2004) 103--266,
  [\href{http://xxx.lanl.gov/abs/astro-ph/0406398}{{\tt astro-ph/0406398}}].

\bibitem{chen_review}
X.~{Chen}, {\it {Primordial Non-Gaussianities from Inflation Models}},  {\em
  Advances in Astronomy} {\bf 2010} (2010)
  [\href{http://xxx.lanl.gov/abs/1002.1416}{{\tt arXiv:1002.1416}}].

\bibitem{linde1997}
A.~{Linde} and V.~{Mukhanov}, {\it {Non-Gaussian isocurvature perturbations
  from inflation}},  {\em \prd} {\bf 56} (July, 1997) 535--+,
  [\href{http://xxx.lanl.gov/abs/astro-ph/9610219}{{\tt astro-ph/9610219}}].

\bibitem{lyth2003}
D.~H. {Lyth}, C.~{Ungarelli}, and D.~{Wands}, {\it {Primordial density
  perturbation in the curvaton scenario}},  {\em \prd} {\bf 67} (Jan., 2003)
  023503--+, [\href{http://xxx.lanl.gov/abs/astro-ph/0208055}{{\tt
  astro-ph/0208055}}].

\bibitem{babich2004}
D.~{Babich}, P.~{Creminelli}, and M.~{Zaldarriaga}, {\it {The shape of
  non-Gaussianities}},  {\em \jcap} {\bf 8} (Aug., 2004) 9--+,
  [\href{http://xxx.lanl.gov/abs/astro-ph/0405356}{{\tt astro-ph/0405356}}].

\bibitem{chen2007a}
X.~{Chen}, M.~{Huang}, S.~{Kachru}, and G.~{Shiu}, {\it {Observational
  signatures and non-Gaussianities of general single-field inflation}},  {\em
  \jcap} {\bf 1} (Jan., 2007) 2--+,
  [\href{http://xxx.lanl.gov/abs/hep-th/0605045}{{\tt hep-th/0605045}}].

\bibitem{chen2007b}
X.~{Chen}, R.~{Easther}, and E.~A. {Lim}, {\it {Large non-Gaussianities in
  single-field inflation}},  {\em \jcap} {\bf 6} (June, 2007) 23--+,
  [\href{http://xxx.lanl.gov/abs/astro-ph/0611645}{{\tt astro-ph/0611645}}].

\bibitem{holman2008}
R.~{Holman} and A.~J. {Tolley}, {\it {Enhanced non-Gaussianity from excited
  initial states}},  {\em \jcap} {\bf 5} (May, 2008) 1--+,
  [\href{http://xxx.lanl.gov/abs/0710.1302}{{\tt arXiv:0710.1302}}].

\bibitem{langlois2008}
D.~{Langlois}, S.~{Renaux-Petel}, D.~A. {Steer}, and T.~{Tanaka}, {\it
  {Primordial perturbations and non-Gaussianities in DBI and general multifield
  inflation}},  {\em \prd} {\bf 78} (Sept., 2008) 063523--+,
  [\href{http://xxx.lanl.gov/abs/0806.0336}{{\tt arXiv:0806.0336}}].

\bibitem{creminelli2003}
P.~{Creminelli}, {\it {On non-Gaussianities in single-field inflation}},  {\em
  \jcap} {\bf 10} (Oct., 2003) 3--+,
  [\href{http://xxx.lanl.gov/abs/astro-ph/0306122}{{\tt astro-ph/0306122}}].

\bibitem{alishahiha2004}
M.~{Alishahiha}, E.~{Silverstein}, and D.~{Tong}, {\it {DBI in the sky:
  Non-Gaussianity from inflation with a speed limit}},  {\em \prd} {\bf 70}
  (Dec., 2004) 123505--+, [\href{http://xxx.lanl.gov/abs/hep-th/0404084}{{\tt
  hep-th/0404084}}].

\bibitem{arkani-hamed2004}
N.~{Arkani-Hamed}, P.~{Creminelli}, S.~{Mukohyama}, and M.~{Zaldarriaga}, {\it
  {Ghost inflation}},  {\em \jcap} {\bf 4} (Apr., 2004) 1--+,
  [\href{http://xxx.lanl.gov/abs/hep-th/0312100}{{\tt hep-th/0312100}}].

\bibitem{meerburg2009}
P.~D. {Meerburg}, J.~P. {van der Schaar}, and P.~{Stefano Corasaniti}, {\it
  {Signatures of initial state modifications on bispectrum statistics}},  {\em
  \jcap} {\bf 5} (May, 2009) 18--+,
  [\href{http://xxx.lanl.gov/abs/0901.4044}{{\tt arXiv:0901.4044}}].

\bibitem{ashoorioon2010}
A.~{Ashoorioon} and G.~{Shiu}, {\it {A Note on Calm Excited States of
  Inflation}},  {\em ArXiv e-prints} (Dec., 2010)
  [\href{http://xxx.lanl.gov/abs/1012.3392}{{\tt arXiv:1012.3392}}].

\bibitem{ssz2010}
L.~{Senatore}, K.~M. {Smith}, and M.~{Zaldarriaga}, {\it {Non-Gaussianities in
  single field inflation and their optimal limits from the WMAP 5-year data}},
  {\em \jcap} {\bf 1} (Jan., 2010) 28--+,
  [\href{http://xxx.lanl.gov/abs/0905.3746}{{\tt arXiv:0905.3746}}].

\bibitem{fergusson2010_cmb}
J.~R. {Fergusson}, M.~{Liguori}, and E.~P.~S. {Shellard}, {\it {General CMB and
  primordial bispectrum estimation: Mode expansion, map making, and measures of
  $f_{NL}$}},  {\em \prd} {\bf 82} (July, 2010) 023502--+,
  [\href{http://xxx.lanl.gov/abs/0912.5516}{{\tt arXiv:0912.5516}}].

\bibitem{jorge2011}
P.~{Creminelli}, G.~{D'Amico}, M.~{Musso}, J.~{Nore{\~n}a}, and
  E.~{Trincherini}, {\it {Galilean symmetry in the effective theory of
  inflation: new shapes of non-Gaussianity}},  {\em \jcap} {\bf 2} (Feb., 2011)
  6--+, [\href{http://xxx.lanl.gov/abs/1011.3004}{{\tt arXiv:1011.3004}}].

\bibitem{planck}
{Planck Collaboration}, P.~A.~R. {Ade}, N.~{Aghanim}, M.~{Arnaud},
  M.~{Ashdown}, J.~{Aumont}, C.~{Baccigalupi}, M.~{Baker}, A.~{Balbi}, A.~J.
  {Banday}, and et~al., {\it {Planck Early Results: The Planck mission}},  {\em
  ArXiv e-prints} (Jan., 2011) [\href{http://xxx.lanl.gov/abs/1101.2022}{{\tt
  arXiv:1101.2022}}].

\bibitem{verde2010}
L.~{Verde}, {\it {Non-Gaussianity from Large-Scale Structure Surveys}},  {\em
  Advances in Astronomy} {\bf 2010} (2010)
  [\href{http://xxx.lanl.gov/abs/1001.5217}{{\tt arXiv:1001.5217}}].

\bibitem{desjacques_review}
V.~{Desjacques} and U.~{Seljak}, {\it {Primordial non-Gaussianity from the
  large-scale structure}},  {\em Classical and Quantum Gravity} {\bf 27} (June,
  2010) 124011--+, [\href{http://xxx.lanl.gov/abs/1003.5020}{{\tt
  arXiv:1003.5020}}].

\bibitem{verde2000}
L.~{Verde}, L.~{Wang}, A.~F. {Heavens}, and M.~{Kamionkowski}, {\it
  {Large-scale structure, the cosmic microwave background and primordial
  non-Gaussianity}},  {\em \mnras} {\bf 313} (Mar., 2000) 141--147,
  [\href{http://xxx.lanl.gov/abs/astro-ph/9906301}{{\tt astro-ph/9906301}}].

\bibitem{verde2001a}
L.~{Verde}, {\it {Non-gaussianity Versus Nonlinearity of Cosmological
  Perturbations}},  in {\em The Onset of Nonlinearity in Cosmology}
  ({J.~N.~Fry, J.~R.~Buchler, \& H.~Kandrup}, ed.), vol.~927 of {\em Annals of
  the New York Academy of Sciences}, pp.~54--69, 2001.
\newblock \href{http://xxx.lanl.gov/abs/astro-ph/0004341}{{\tt
  astro-ph/0004341}}.

\bibitem{verde2001b}
L.~{Verde}, R.~{Jimenez}, M.~{Kamionkowski}, and S.~{Matarrese}, {\it {Tests
  for primordial non-Gaussianity}},  {\em \mnras} {\bf 325} (July, 2001)
  412--418, [\href{http://xxx.lanl.gov/abs/astro-ph/0011180}{{\tt
  astro-ph/0011180}}].

\bibitem{baldauf2010}
T.~{Baldauf}, U.~{Seljak}, and L.~{Senatore}, {\it {Primordial non-Gaussianity
  in the Bispectrum of the Halo Density Field}},  {\em ArXiv e-prints} (Nov.,
  2010) [\href{http://xxx.lanl.gov/abs/1011.1513}{{\tt arXiv:1011.1513}}].

\bibitem{MVJ}
S.~{Matarrese}, L.~{Verde}, and R.~{Jimenez}, {\it {The Abundance of
  High-Redshift Objects as a Probe of Non-Gaussian Initial Conditions}},  {\em
  \apj} {\bf 541} (Sept., 2000) 10--24,
  [\href{http://xxx.lanl.gov/abs/astro-ph/0001366}{{\tt astro-ph/0001366}}].

\bibitem{LV}
M.~{Lo Verde}, A.~{Miller}, S.~{Shandera}, and L.~{Verde}, {\it {Effects of
  scale-dependent non-Gaussianity on cosmological structures}},  {\em \jcap}
  {\bf 4} (Apr., 2008) 14--+, [\href{http://xxx.lanl.gov/abs/0711.4126}{{\tt
  arXiv:0711.4126}}].

\bibitem{cayon2010}
L.~{Cay{\'o}n}, C.~{Gordon}, and J.~{Silk}, {\it {Probability of the most
  massive cluster under non-Gaussian initial conditions}},  {\em ArXiv
  e-prints} (June, 2010) [\href{http://xxx.lanl.gov/abs/1006.1950}{{\tt
  arXiv:1006.1950}}].

\bibitem{hoyle2010}
B.~{Hoyle}, R.~{Jimenez}, and L.~{Verde}, {\it {Too big, too early? Multiple
  High-Redshift Galaxy Clusters: implications}},  {\em ArXiv e-prints} (Sept.,
  2010) [\href{http://xxx.lanl.gov/abs/1009.3884}{{\tt arXiv:1009.3884}}].

\bibitem{enqvist2010}
K.~{Enqvist}, S.~{Hotchkiss}, and O.~{Taanila}, {\it {Estimating f\_NL and
  g\_NL from Massive High-Redshift Galaxy Clusters}},  {\em ArXiv e-prints}
  (Dec., 2010) [\href{http://xxx.lanl.gov/abs/1012.2732}{{\tt
  arXiv:1012.2732}}].

\bibitem{mortonson2011}
M.~J. {Mortonson}, W.~{Hu}, and D.~{Huterer}, {\it {Simultaneous falsification
  of {$\Lambda$}CDM and quintessence with massive, distant clusters}},  {\em
  \prd} {\bf 83} (Jan., 2011) 023015--+,
  [\href{http://xxx.lanl.gov/abs/1011.0004}{{\tt arXiv:1011.0004}}].

\bibitem{dalal2008}
N.~{Dalal}, O.~{Dor{\'e}}, D.~{Huterer}, and A.~{Shirokov}, {\it {Imprints of
  primordial non-Gaussianities on large-scale structure: Scale-dependent bias
  and abundance of virialized objects}},  {\em \prd} {\bf 77} (June, 2008)
  123514--+, [\href{http://xxx.lanl.gov/abs/0710.4560}{{\tt arXiv:0710.4560}}].

\bibitem{MV}
S.~{Matarrese} and L.~{Verde}, {\it {The Effect of Primordial Non-Gaussianity
  on Halo Bias}},  {\em \apjl} {\bf 677} (Apr., 2008) L77--L80,
  [\href{http://xxx.lanl.gov/abs/0801.4826}{{\tt arXiv:0801.4826}}].

\bibitem{VM}
L.~{Verde} and S.~{Matarrese}, {\it {Detectability of the Effect of
  Inflationary Non-Gaussianity on Halo Bias}},  {\em \apjl} {\bf 706} (Nov.,
  2009) L91--L95, [\href{http://xxx.lanl.gov/abs/0909.3224}{{\tt
  arXiv:0909.3224}}].

\bibitem{slosar2008}
A.~{Slosar}, C.~{Hirata}, U.~{Seljak}, S.~{Ho}, and N.~{Padmanabhan}, {\it
  {Constraints on local primordial non-Gaussianity from large scale
  structure}},  {\em \jcap} {\bf 8} (Aug., 2008) 31--+,
  [\href{http://xxx.lanl.gov/abs/0805.3580}{{\tt arXiv:0805.3580}}].

\bibitem{xia2010a}
J.~{Xia}, M.~{Viel}, C.~{Baccigalupi}, G.~{De Zotti}, S.~{Matarrese}, and
  L.~{Verde}, {\it {Primordial Non-Gaussianity and the NRAO VLA Sky Survey}},
  {\em \apjl} {\bf 717} (July, 2010) L17--L21,
  [\href{http://xxx.lanl.gov/abs/1003.3451}{{\tt arXiv:1003.3451}}].

\bibitem{xia2010b}
J.~{Xia}, A.~{Bonaldi}, C.~{Baccigalupi}, G.~{De Zotti}, S.~{Matarrese},
  L.~{Verde}, and M.~{Viel}, {\it {Constraining primordial non-Gaussianity with
  high-redshift probes}},  {\em \jcap} {\bf 8} (Aug., 2010) 13--+,
  [\href{http://xxx.lanl.gov/abs/1007.1969}{{\tt arXiv:1007.1969}}].

\bibitem{carbone2008}
C.~{Carbone}, L.~{Verde}, and S.~{Matarrese}, {\it {Non-Gaussian Halo Bias and
  Future Galaxy Surveys}},  {\em \apjl} {\bf 684} (Sept., 2008) L1--L4,
  [\href{http://xxx.lanl.gov/abs/0806.1950}{{\tt arXiv:0806.1950}}].

\bibitem{carbone2010}
C.~{Carbone}, O.~{Mena}, and L.~{Verde}, {\it {Cosmological parameters
  degeneracies and non-Gaussian halo bias}},  {\em \jcap} {\bf 7} (July, 2010)
  20--+, [\href{http://xxx.lanl.gov/abs/1003.0456}{{\tt arXiv:1003.0456}}].

\bibitem{cunha2010}
C.~{Cunha}, D.~{Huterer}, and O.~{Dor{\'e}}, {\it {Primordial non-Gaussianity
  from the covariance of galaxy cluster counts}},  {\em \prd} {\bf 82} (July,
  2010) 023004--+, [\href{http://xxx.lanl.gov/abs/1003.2416}{{\tt
  arXiv:1003.2416}}].

\bibitem{fedeli2010}
C.~{Fedeli}, C.~{Carbone}, L.~{Moscardini}, and A.~{Cimatti}, {\it {The
  clustering of galaxies and galaxy clusters: constraints on primordial
  non-Gaussianity from future wide-field surveys}},  {\em ArXiv e-prints}
  (Dec., 2010) [\href{http://xxx.lanl.gov/abs/1012.2305}{{\tt
  arXiv:1012.2305}}].

\bibitem{desjacques2009}
V.~{Desjacques}, U.~{Seljak}, and I.~T. {Iliev}, {\it {Scale-dependent bias
  induced by local non-Gaussianity: a comparison to N-body simulations}},  {\em
  \mnras} {\bf 396} (June, 2009) 85--96,
  [\href{http://xxx.lanl.gov/abs/0811.2748}{{\tt arXiv:0811.2748}}].

\bibitem{pillepich2010}
A.~{Pillepich}, C.~{Porciani}, and O.~{Hahn}, {\it {Halo mass function and
  scale-dependent bias from N-body simulations with non-Gaussian initial
  conditions}},  {\em \mnras} {\bf 402} (Feb., 2010) 191--206,
  [\href{http://xxx.lanl.gov/abs/0811.4176}{{\tt arXiv:0811.4176}}].

\bibitem{grossi2009}
M.~{Grossi}, L.~{Verde}, C.~{Carbone}, K.~{Dolag}, E.~{Branchini},
  F.~{Iannuzzi}, S.~{Matarrese}, and L.~{Moscardini}, {\it {Large-scale
  non-Gaussian mass function and halo bias: tests on N-body simulations}},
  {\em \mnras} {\bf 398} (Sept., 2009) 321--332,
  [\href{http://xxx.lanl.gov/abs/0902.2013}{{\tt arXiv:0902.2013}}].

\bibitem{nishimichi2010}
T.~{Nishimichi}, A.~{Taruya}, K.~{Koyama}, and C.~{Sabiu}, {\it {Scale
  dependence of halo bispectrum from non-Gaussian initial conditions in
  cosmological N-body simulations}},  {\em \jcap} {\bf 7} (July, 2010) 2--+,
  [\href{http://xxx.lanl.gov/abs/0911.4768}{{\tt arXiv:0911.4768}}].

\bibitem{wagner2010}
C.~{Wagner}, L.~{Verde}, and L.~{Boubekeur}, {\it {N-body simulations with
  generic non-Gaussian initial conditions I: power spectrum and halo mass
  function}},  {\em \jcap} {\bf 10} (Oct., 2010) 22--+,
  [\href{http://xxx.lanl.gov/abs/1006.5793}{{\tt arXiv:1006.5793}}].

\bibitem{kaiser1984}
N.~{Kaiser}, {\it {On the spatial correlations of Abell clusters}},  {\em
  \apjl} {\bf 284} (Sept., 1984) L9--L12.

\bibitem{MW1996}
H.~J. {Mo} and S.~D.~M. {White}, {\it {An analytic model for the spatial
  clustering of dark matter haloes}},  {\em \mnras} {\bf 282} (Sept., 1996)
  347--361, [\href{http://xxx.lanl.gov/abs/astro-ph/9512127}{{\tt
  astro-ph/9512127}}].

\bibitem{ST2001}
R.~K. {Sheth}, H.~J. {Mo}, and G.~{Tormen}, {\it {Ellipsoidal collapse and an
  improved model for the number and spatial distribution of dark matter
  haloes}},  {\em \mnras} {\bf 323} (May, 2001) 1--12,
  [\href{http://xxx.lanl.gov/abs/astro-ph/9907024}{{\tt astro-ph/9907024}}].

\bibitem{gao2005}
L.~{Gao}, V.~{Springel}, and S.~D.~M. {White}, {\it {The age dependence of halo
  clustering}},  {\em \mnras} {\bf 363} (Oct., 2005) L66--L70,
  [\href{http://xxx.lanl.gov/abs/astro-ph/0506510}{{\tt astro-ph/0506510}}].

\bibitem{gao2007}
L.~{Gao} and S.~D.~M. {White}, {\it {Assembly bias in the clustering of dark
  matter haloes}},  {\em \mnras} {\bf 377} (Apr., 2007) L5--L9,
  [\href{http://xxx.lanl.gov/abs/astro-ph/0611921}{{\tt astro-ph/0611921}}].

\bibitem{afshordi2008}
N.~{Afshordi} and A.~J. {Tolley}, {\it {Primordial non-Gaussianity, statistics
  of collapsed objects, and the integrated Sachs-Wolfe effect}},  {\em \prd}
  {\bf 78} (Dec., 2008) 123507--+,
  [\href{http://xxx.lanl.gov/abs/0806.1046}{{\tt arXiv:0806.1046}}].

\bibitem{mcdonald2008}
P.~{McDonald}, {\it {Primordial non-Gaussianity: Large-scale structure
  signature in the perturbative bias model}},  {\em \prd} {\bf 78} (Dec., 2008)
  123519--+, [\href{http://xxx.lanl.gov/abs/0806.1061}{{\tt arXiv:0806.1061}}].

\bibitem{taruya2008}
A.~{Taruya}, K.~{Koyama}, and T.~{Matsubara}, {\it {Signature of primordial
  non-Gaussianity on the matter power spectrum}},  {\em \prd} {\bf 78} (Dec.,
  2008) 123534--+, [\href{http://xxx.lanl.gov/abs/0808.4085}{{\tt
  arXiv:0808.4085}}].

\bibitem{jeong2009}
D.~{Jeong} and E.~{Komatsu}, {\it {Primordial Non-Gaussianity, Scale-dependent
  Bias, and the Bispectrum of Galaxies}},  {\em \apj} {\bf 703} (Oct., 2009)
  1230--1248, [\href{http://xxx.lanl.gov/abs/0904.0497}{{\tt
  arXiv:0904.0497}}].

\bibitem{sefusatti2009}
E.~{Sefusatti}, {\it {One-loop perturbative corrections to the matter and
  galaxy bispectrum with non-Gaussian initial conditions}},  {\em \prd} {\bf
  80} (Dec., 2009) 123002--+, [\href{http://xxx.lanl.gov/abs/0905.0717}{{\tt
  arXiv:0905.0717}}].

\bibitem{giannantonio2010}
T.~{Giannantonio} and C.~{Porciani}, {\it {Structure formation from
  non-Gaussian initial conditions: Multivariate biasing, statistics, and
  comparison with N-body simulations}},  {\em \prd} {\bf 81} (Mar., 2010)
  063530--+, [\href{http://xxx.lanl.gov/abs/0911.0017}{{\tt arXiv:0911.0017}}].

\bibitem{desimone2011}
A.~{De Simone}, M.~{Maggiore}, and A.~{Riotto}, {\it {Conditional Probabilities
  in the Excursion Set Theory. Generic Barriers and non-Gaussian Initial
  Conditions}},  {\em ArXiv e-prints} (Jan., 2011)
  [\href{http://xxx.lanl.gov/abs/1102.0046}{{\tt arXiv:1102.0046}}].

\bibitem{smith2011}
R.~E. {Smith}, V.~{Desjacques}, and L.~{Marian}, {\it {Nonlinear clustering in
  models with primordial non-Gaussianity: The halo model approach}},  {\em
  \prd} {\bf 83} (Feb., 2011) 043526--+,
  [\href{http://xxx.lanl.gov/abs/1009.5085}{{\tt arXiv:1009.5085}}].

\bibitem{valageas2010}
P.~{Valageas}, {\it {Mass function and bias of dark matter halos for
  non-Gaussian initial conditions}},  {\em \aap} {\bf 514} (May, 2010) A46+,
  [\href{http://xxx.lanl.gov/abs/0906.1042}{{\tt arXiv:0906.1042}}].

\bibitem{reid2010}
B.~A. {Reid}, L.~{Verde}, K.~{Dolag}, S.~{Matarrese}, and L.~{Moscardini}, {\it
  {Non-Gaussian halo assembly bias}},  {\em \jcap} {\bf 7} (July, 2010) 13--+,
  [\href{http://xxx.lanl.gov/abs/1004.1637}{{\tt arXiv:1004.1637}}].

\bibitem{desjacques2010}
V.~{Desjacques} and U.~{Seljak}, {\it {Signature of primordial non-Gaussianity
  of ${\phi}^{3}$ type in the mass function and bias of dark matter haloes}},
  {\em \prd} {\bf 81} (Jan., 2010) 023006--+,
  [\href{http://xxx.lanl.gov/abs/0907.2257}{{\tt arXiv:0907.2257}}].

\bibitem{bartolo2005}
N.~{Bartolo}, S.~{Matarrese}, and A.~{Riotto}, {\it {Signatures of primordial
  non-Gaussianity in the large-scale structure of the universe}},  {\em \jcap}
  {\bf 10} (Oct., 2005) 10--+,
  [\href{http://xxx.lanl.gov/abs/astro-ph/0501614}{{\tt astro-ph/0501614}}].

\bibitem{yoo2009}
J.~{Yoo}, A.~L. {Fitzpatrick}, and M.~{Zaldarriaga}, {\it {New perspective on
  galaxy clustering as a cosmological probe: General relativistic effects}},
  {\em \prd} {\bf 80} (Oct., 2009) 083514--+,
  [\href{http://xxx.lanl.gov/abs/0907.0707}{{\tt arXiv:0907.0707}}].

\bibitem{wands2009}
D.~{Wands} and A.~{Slosar}, {\it {Scale-dependent bias from primordial
  non-Gaussianity in general relativity}},  {\em \prd} {\bf 79} (June, 2009)
  123507--+, [\href{http://xxx.lanl.gov/abs/0902.1084}{{\tt arXiv:0902.1084}}].

\bibitem{yoo2010}
J.~{Yoo}, {\it {General relativistic description of the observed galaxy power
  spectrum: Do we understand what we measure?}},  {\em \prd} {\bf 82} (Oct.,
  2010) 083508--+, [\href{http://xxx.lanl.gov/abs/1009.3021}{{\tt
  arXiv:1009.3021}}].

\bibitem{bartolo2010}
N.~{Bartolo}, S.~{Matarrese}, and A.~{Riotto}, {\it {The Gauge-Invariant Bias
  of Dark Matter Haloes with Primordial non-Gaussianity}},  {\em ArXiv
  e-prints} (Nov., 2010) [\href{http://xxx.lanl.gov/abs/1011.4374}{{\tt
  arXiv:1011.4374}}].

\bibitem{shandera2010}
S.~{Shandera}, N.~{Dalal}, and D.~{Huterer}, {\it {A generalized local ansatz
  and its effect on halo bias}},  {\em ArXiv e-prints} (Oct., 2010)
  [\href{http://xxx.lanl.gov/abs/1010.3722}{{\tt arXiv:1010.3722}}].

\bibitem{becker2010}
A.~{Becker}, D.~{Huterer}, and K.~{Kadota}, {\it {Scale-Dependent
  Non-Gaussianity as a Generalization of the Local Model}},  {\em ArXiv
  e-prints} (Sept., 2010) [\href{http://xxx.lanl.gov/abs/1009.4189}{{\tt
  arXiv:1009.4189}}].

\bibitem{schmidt2010}
F.~{Schmidt} and M.~{Kamionkowski}, {\it {Halo clustering with nonlocal
  non-Gaussianity}},  {\em \prd} {\bf 82} (Nov., 2010) 103002--+,
  [\href{http://xxx.lanl.gov/abs/1008.0638}{{\tt arXiv:1008.0638}}].

\bibitem{fergusson2010_lss}
J.~R. {Fergusson}, D.~M. {Regan}, and E.~P.~S. {Shellard}, {\it {Rapid
  Separable Analysis of Higher Order Correlators in Large Scale Structure}},
  {\em ArXiv e-prints} (Aug., 2010)
  [\href{http://xxx.lanl.gov/abs/1008.1730}{{\tt arXiv:1008.1730}}].

\bibitem{creminelli2006}
P.~{Creminelli}, A.~{Nicolis}, L.~{Senatore}, M.~{Tegmark}, and
  M.~{Zaldarriaga}, {\it {Limits on non-Gaussianities from WMAP data}},  {\em
  \jcap} {\bf 5} (May, 2006) 4--+,
  [\href{http://xxx.lanl.gov/abs/astro-ph/0509029}{{\tt astro-ph/0509029}}].

\bibitem{EFTI}
C.~{Cheung}, A.~L. {Fitzpatrick}, J.~{Kaplan}, L.~{Senatore}, and
  P.~{Creminelli}, {\it {The effective field theory of inflation}},  {\em
  Journal of High Energy Physics} {\bf 3} (Mar., 2008) 14--014,
  [\href{http://xxx.lanl.gov/abs/0709.0293}{{\tt arXiv:0709.0293}}].

\bibitem{smith2006}
K.~M. {Smith} and M.~{Zaldarriaga}, {\it {Algorithms for bispectra:
  forecasting, optimal analysis, and simulation}},  {\em ArXiv Astrophysics
  e-prints} (Dec., 2006) [\href{http://xxx.lanl.gov/abs/astro-ph/0612571}{{\tt
  astro-ph/0612571}}].

\bibitem{CAMB}
A.~{Lewis}, A.~{Challinor}, and A.~{Lasenby}, {\it {Efficient Computation of
  Cosmic Microwave Background Anisotropies in Closed Friedmann-Robertson-Walker
  Models}},  {\em \apj} {\bf 538} (Aug., 2000) 473--476,
  [\href{http://xxx.lanl.gov/abs/astro-ph/9911177}{{\tt astro-ph/9911177}}].

\bibitem{zeldovich}
Y.~B. {Zel'Dovich}, {\it {Gravitational instability: An approximate theory for
  large density perturbations.}},  {\em \aap} {\bf 5} (Mar., 1970) 84--89.

\bibitem{2LPT}
R.~{Scoccimarro}, {\it {Transients from initial conditions: a perturbative
  analysis}},  {\em \mnras} {\bf 299} (Oct., 1998) 1097--1118,
  [\href{http://xxx.lanl.gov/abs/astro-ph/9711187}{{\tt astro-ph/9711187}}].

\bibitem{sirko}
E.~{Sirko}, {\it {Initial Conditions to Cosmological N-Body Simulations, or,
  How to Run an Ensemble of Simulations}},  {\em \apj} {\bf 634} (Nov., 2005)
  728--743, [\href{http://xxx.lanl.gov/abs/astro-ph/0503106}{{\tt
  astro-ph/0503106}}].

\bibitem{springel2005}
V.~{Springel}, {\it {The cosmological simulation code GADGET-2}},  {\em \mnras}
  {\bf 364} (Dec., 2005) 1105--1134,
  [\href{http://xxx.lanl.gov/abs/astro-ph/0505010}{{\tt astro-ph/0505010}}].

\bibitem{knollmann2009}
S.~R. {Knollmann} and A.~{Knebe}, {\it {AHF: Amiga's Halo Finder}},  {\em
  \apjs} {\bf 182} (June, 2009) 608--624,
  [\href{http://xxx.lanl.gov/abs/0904.3662}{{\tt arXiv:0904.3662}}].

\bibitem{scoccimarro2011}
R.~{Scoccimarro} and {collaborators} {\em (in preparation)}.

\bibitem{MR}
M.~{Maggiore} and A.~{Riotto}, {\it {The Halo Mass Function from Excursion Set
  Theory. III. Non-Gaussian Fluctuations}},  {\em \apj} {\bf 717} (July, 2010)
  526--541, [\href{http://xxx.lanl.gov/abs/0903.1251}{{\tt arXiv:0903.1251}}].

\bibitem{norena2011}
G.~{D'Amico}, M.~{Musso}, J.~{Nore{\~n}a}, and A.~{Paranjape}, {\it {An
  improved calculation of the non-Gaussian halo mass function}},  {\em \jcap}
  {\bf 2} (Feb., 2011) 1--+, [\href{http://xxx.lanl.gov/abs/1005.1203}{{\tt
  arXiv:1005.1203}}].

\bibitem{yadav2010}
A.~P.~S. {Yadav} and B.~D. {Wandelt}, {\it {Primordial Non-Gaussianity in the
  Cosmic Microwave Background}},  {\em Advances in Astronomy} {\bf 2010} (2010)
  [\href{http://xxx.lanl.gov/abs/1006.0275}{{\tt arXiv:1006.0275}}].

\bibitem{chen2010a}
X.~{Chen} and Y.~{Wang}, {\it {Large non-Gaussianities with intermediate shapes
  from quasi-single-field inflation}},  {\em \prd} {\bf 81} (Mar., 2010)
  063511--+, [\href{http://xxx.lanl.gov/abs/0909.0496}{{\tt arXiv:0909.0496}}].

\bibitem{chen2010b}
X.~{Chen} and Y.~{Wang}, {\it {Quasi-single field inflation and
  non-Gaussianities}},  {\em \jcap} {\bf 4} (Apr., 2010) 27--+,
  [\href{http://xxx.lanl.gov/abs/0911.3380}{{\tt arXiv:0911.3380}}].

\bibitem{gil2010}
H.~{Gil-Mar{\'{\i}}n}, C.~{Wagner}, L.~{Verde}, R.~{Jimenez}, and A.~F.
  {Heavens}, {\it {Reducing sample variance: halo biasing, non-linearity and
  stochasticity}},  {\em \mnras} {\bf 407} (Sept., 2010) 772--790,
  [\href{http://xxx.lanl.gov/abs/1003.3238}{{\tt arXiv:1003.3238}}].

\end{thebibliography}\endgroup

\end{document}